\documentstyle[12pt]{article}
\newcommand{\beq}{\begin{equation}}
\newcommand{\eeq}{\end{equation}}
\newcommand{\gsim}{\lower.7ex\hbox{$\;\stackrel{\textstyle>}
{\sim}\;$}}
\newcommand{\lsim}{\lower.7ex\hbox{$\;\stackrel{\textstyle<}
{\sim}\;$}}

\begin{document}
\begin{titlepage}
\renewcommand{\thefootnote}{\fnsymbol{footnote}}

\begin{center} \Large
{\bf Theoretical Physics Institute}\\
{\bf University of Minnesota}
\end{center}
\begin{flushright}
TPI-MINN-95/31-T\\
UMN-TH-1413-95\\
October, 1995
\end{flushright}
\vspace{.3cm}
\begin{center} \Large
{\bf  Lectures on \\
Heavy
Quarks  in Quantum Chromodynamics}
\footnote{An extended version of the lectures  given at
Theoretical Advanced Study Institute {\em QCD and Beyond},
University of Colorado, Boulder, Colorado, June 1995.}
\end{center}
\vspace*{.3cm}
\begin{center} {\Large
Mikhail Shifman } \\
\vspace{0.4cm}
{\it  Theoretical Physics Institute, University of Minnesota,
\\
Minneapolis, MN 55455}\\
\vspace*{.4cm}

{\Large{\bf Abstract}}
\end{center}

\vspace*{.2cm}
A pedagogical introduction
to the heavy quark theory is given. It is explained that various
expansions in
the inverse heavy quark mass $1/m_Q$
 present a version of the Wilson operator product expansion in QCD.
A systematic approach is developed and many practically interesting
problems are considered. I show how the
$1/m_Q$ expansions  can be built using the background field
technique and how they work in particular applications.  Interplay
between perturbative and nonperturbative aspects of the heavy
quark theory is discussed.

\end{titlepage}
%\addtocounter{footnote}{-2}
\renewcommand{\theequation}{1.\arabic{equation}}
\setcounter{equation}{0}

 \section{Lecture 1. Heavy Quark Symmetry}

The statement that Quantum Chromodynamics (QCD) is {\em the}
theory of
hadrons has become  common place.  It is a very strange theory,
 since
many
questions concerning dynamics of the quarks and gluons at large
distances  --
however simple they might seem -- remain unanswered or, at best,
understood only at a qualitative level. Progress in the direction of the
quantitative description of the hadronic properties is slow -- every
step
bringing us closer to such a description is painfully difficult. At the
same time
new results, even modest, have a special weight for obvious
reasons -- QCD, unlike many other trendy theories in the modern
high energy
physics, definitely has a direct relation to Nature and will stay with
us
forever.

Every hadron in a sense is built from quarks and/or gluons. I say ``in
a sense"
because  these are no ordinary building blocks. The number of
degrees of
freedom  fluctuates and is not fixed; this we know for sure.
At large
distances we have to deal with a genuine strongly coupled field
theory, and, as
usual, the strong coupling creates  complicated structures which can
not be
treated by perturbative methods. Then we feel helpless and are
ready to use
every opportunity, no matter where it comes from,  if only  it gives
the
slightest
hope of getting a solid quantitative approach based on QCD.

QCD has two faces,  two components -- hard and soft.  The hard
component is
the realm of perturbative QCD. Not much will be said in these
lectures about
this aspect.  Instead, we will concentrate on the soft component.
Many years
ago, at the dawn of the QCD era, it was noted \cite{NOSVVZ} that
heavy quarks
are, probably, the best probe of the soft component of the gluon
fields out of
all probes we have at our disposal. The developments we
witnessed in
recent
years  confirm this conclusion.

The dynamics of  soft degrees of freedom in QCD is the realm of
non-perturbative phenomena. Having said this I hasten to add that
there is an
element of luck -- transition from the perturbative regime to the
non-perturbative one is very abrupt in QCD. In a sense the gauge
coupling
constant is abnormally small. I do not mean here the conventional
logarithmic
suppression of the running constant but, rather, the fact that $b$, the
first
coefficient in the Gell-Mann-Low function, is numerically large. This
fact
allows us to forget, in the first approximation, about perturbative
effects and
focus on non-perturbative ones in a wide range of problems. It is
more exact to say that we
will
concentrate on studying the soft degrees of freedom, but due to the
fortunate
circumstance of ``abnormal" smallness of $\alpha_s (\mu)/\pi$  for
as low
normalization point as $\mu \sim$1 GeV, all effects due to the soft
degrees of
freedom
are essentially non-perturbative. I will elucidate the precise meaning
of this
statement later.

It would be great if we could just switch off -- by adjusting some
parameter
-- all hard processes in QCD without changing its soft component.
Then we
would be left with the confining dynamics in a clean and
uncontaminated
form; formulation of the theory would be much easier. The only
parameter
which might do the job is $b$. If we could tend
$b\rightarrow\infty$
with $\Lambda_{\rm QCD}$ fixed, the hard gluons would be
suppressed by
powers of $1/b$
while the soft component would presumably remain unaltered or
almost
unaltered. Unfortunately, nobody knows how to make the
enhancement of
$b$ parametric. (The limit of the large number of colors, $N_c
\rightarrow\infty$, does not work since, although $b$ is definitely
proportional to $N_c$ in this case, the perturbative expansion for all
planar graphs goes in $N_c/b$, not in $1/b$ \cite{tHooft}.) Therefore,
we will have
to rely on the  numerical enhancement of $b$.  In the first lectures
I will merely assume that the hard gluon exchanges are non-existent.
Later on, at the very end,  we will return to this issue and will
briefly discuss the
impact of  hard gluons.

The purpose of these lectures is mainly pedagogical -- the coverage
of
the topic is neither chronological nor  comprehensive. Technically
sophisticated issues and calculations are avoided whenever possible;
instead I discuss particularly illuminating problems, in a simplified
setting.
The readers interested in specific advanced applications (e.g.
combining the $1/m_Q$ expansions with the chiral perturbation
theory \cite{CHPTH})   are referred to the original publications and
the review
papers
\cite{HQETRev} summarizing a wealth of results obtained in the
heavy quark
theory after 1990. The presentation of the heavy quark theory below
as a rule
does not  follow the standard pattern and is, rather, complementary
with
respect to the more traditional  reviews \cite{HQETRev}. We try to
emphasize
that the heavy quark theory and the heavy quark expansion is
nothing else than a version of the Wilson operator product expansion
(OPE) \cite{Wilson}, an aspect
which
usually remains fogged.

\subsection{Why heavy quarks?}

The quark-gluon dynamics is governed by the QCD Lagrangian
$$
{\cal L} =-\frac{1}{4} G_{\mu\nu}^a G_{\mu\nu}^a
+\sum_{q}\bar q i\not\!\!{D} q
 +\sum_{Q}\bar Q (i\not\!\!{D} - m_Q) Q =
$$
\begin{equation}
{\cal L}_{\rm light}
+ \sum_{Q}\bar Q (i\not\!\!{D} - m_Q) Q
\label{lagr}
\eeq
where $G_{\mu\nu}^a$ is the gluon field strength tensor,
the light quark fields ($u, d$ and $s$) are generically denoted by $q$
and are
assumed, for simplicity, to be massless while the heavy quark fields
are
generically denoted by $Q$. To qualify as a heavy quark $Q$ the
corresponding
mass term $m_Q$ must be much larger than $\Lambda_{\rm QCD}$.
The charmed quark $c$ can be called heavy only with
some
reservations and, in discussing the heavy quark theory, it is more
appropriate
to keep in mind $b$ quarks. The hadrons to be considered are
composed from
one heavy quark $Q$, a light antiquark $\bar q$, or diquark $qq$,
and
a gluon
cloud which can also contain  light quark-antiquark pairs. The role of
the
cloud is, of course, to keep all these objects together, in a colorless
bound state
which  will be generically denoted by $H_Q$.

Quite naturally in the heavy quark theory, the gamma matrices used
are those of the standard representation,
\beq
\gamma^0 = \left(
\begin{array}{cc}
1 & 0 \\
0 & -1
\end{array}\right) \,
, \,\, \vec\gamma = \left(
\begin{array}{cc}
0 & \vec\sigma \\
-\vec\sigma & 0
\end{array}\right)\,  , \,\,
\gamma^5 = \left(
\begin{array}{cc}
0 & -1 \\
-1 & 0
\end{array}\right)\, .
\label{gamma}
\eeq
With these definitions of the gamma matrices the left-handed spinor
has the form $\psi_L = (1+\gamma_5)\psi$.

The light component of $H_Q$, its light cloud, \footnote{In some
papers
devoted to the subject the light cloud is referred to as `brown muck'.
I think it
is absolutely unfair with respect to the soft components of the
quark and gluon fields to call them `brown muck' only because we
are not
smart enough to fully understand the corresponding dynamics.}
 has a complicated structure -- the soft modes of the light fields are
strongly
coupled and strongly fluctuate. Basically, the only fact which we
know for sure
is that the light cloud is indeed light; typical frequencies are of order
of
$\Lambda_{\rm QCD}$. One can try to visualize the light cloud as a
soft
medium. The heavy quark $Q$ is then  submerged in this medium. If
the hard
gluon exchanges are discarded the momentum which the heavy
quark
can borrow from the light cloud is of order of $\Lambda_{\rm QCD}$,
and the
corresponding uncertainty in the energy of the heavy quark is of
order $\Lambda_{\rm QCD}^2/m_Q$. Since
these quantities are much smaller than $m_Q$ this means, in
particular,
that the heavy quark-antiquark pairs can not play a role. In other
words,
 the field-theoretic (second-quantized) description of the heavy
quark
becomes redundant, and under the circumstances it is perfectly
sufficient to
treat one
single heavy quark $Q$ within quantum mechanics, which is
infinitely
simpler, of course, than any field theory. Moreover, one can
systematically
expand in $1/m_Q$. Thus, in the limit
$m_Q/\Lambda_{\rm QCD}\rightarrow\infty$ the heavy quark
component of
$H_Q$ becomes easily manageable allowing one to use the heavy
quark
as a probe of the light cloud dynamics. The special advantages of this
limit in
QCD were first emphasized by Shuryak \cite{Shuryak}.

\subsection{Descending Down}

 In  field theory
one has to
specify  the normalization point $\mu$ where all operators
are
defined; in particular, the gauge coupling constant $g$ and the quark
mass $m_Q$ are functions of
$\mu$.
The original QCD Lagrangian (\ref{lagr}) is formulated at very short
distances, or, which is the same, at a high normalization point $\mu =
M_0$ where $M_0$ is the mass of an ultraviolet
regulator. In other words, the
normalization point is assumed to be much
higher
than all mass scales in the theory, $\mu\gg m_Q$. Constructing an
effective theory intended for  description of  the
low-energy
properties of the heavy flavor hadrons we must evolve the
Lagrangian from the original high scale $M_0$ down to a
normalization point
$\mu$ lying below  the heavy quark masses $m_Q$. By evolving
down I mean that we  integrate out, step by step,
all high-frequency modes in the theory thus calculating the
Lagrangian ${\cal L}(\mu )$ describing dynamics of the soft modes,
with characteristic frequencies less than $\mu$.  The hard
(high-frequency) modes determine the coefficient  functions
in ${\cal L}(\mu )$ while the contribution of the soft modes
is hidden in the matrix elements of (an infinite set) of
operators appearing in ${\cal L}(\mu )$. This approach, which in the
context of QCD was put forward by K. Wilson long ago, has become
common. It is widely recognized and exploited in countless
applications -- from the ancient problem of the $K$ meson decays to
fresh trends in the lattice calculations \cite{Lepage}. The peculiarity
of the heavy quark theory is due to the fact that the {\em in} and
{\em out} states
we deal with contain heavy quarks. Therefore, although we do
integrate out the  field fluctuations with the frequencies down to
$\mu$ the heavy quark fields themselves are not integrated out
since we will be interested in physics in the sector with the $Q$
charge
$\neq 0$. The effective Lagrangian ${\cal L}(\mu )$ acts in this
sector.

If QCD was solved we could include in our explicit calculation
of the effective Lagrangian  all modes, descending down to $\mu =0$.
The Lagrangian
obtained in this way would be built in terms of the
fields of physical mesons and baryons, not in terms of quarks and
gluons, since the latter become irrelevant degrees of freedom in the
infrared limit $\mu\rightarrow 0$.  This Lagrangian would give us
the
full set of all conceivable amplitudes and  would, thus, represent the
final answer for the theory. There would be no need for any further
calculations -- one would just pick up the amplitude of interest
and compare it with experimental data.

This picture is quite Utopian, of course. The real QCD is not solved
in the closed form,
and in doing explicit calculations of the
coefficients in the effective Lagrangian one can not put $\mu = 0$.
The lower the value of $\mu$ the larger part of dynamics is
accounted for in the explicit calculation.
Therefore, we would like to have $\mu$ as low as possible; definitely
$\mu \ll m_Q$. The heavy quark can be treated as a non-relativistic
object moving in the soft background field only provided the latter
condition is met. On the other hand,
to keep theoretical control over the explicit calculations of the
coefficient functions we must stop
at some $\mu\gg \Lambda_{\rm QCD}$, so that $\alpha_s (\mu
)/\pi$ is still a sufficiently small expansion parameter. In practice
this means
that the best choice (which we will always stick to) is $\mu\sim$
several units times $\Lambda_{\rm QCD}$.  All coefficients in
the effective Lagrangian obtained in this way will be functions of
$\mu$.

Since
$\mu$ is an auxiliary parameter
predictions for physical quantities must be $\mu$ independent, of
course. The $\mu$ dependence of the coefficients must be canceled
by that coming from the physical matrix elements
of the operators in ${\cal L}(\mu )$.
However, in calculating in the hard and soft domains (i.e.
above $\mu$ and below $\mu$) we make different approximations,
so that the exact $\mu$ independence of the physical quantities can
be lost.
Since the transition from the hard to soft physics is very steep one
may hope that our predictions will be very insensitive   to the
precise choice of $\mu$ provided that
$\mu\sim$ several units times $\Lambda_{\rm QCD}$. Below, if not
stated to the contrary we will assume that the normalization point
$\mu$
is chosen in this way.

In descending from $M_0$ down to $\mu$ the
form of the
Lagrangian (\ref{lagr}) changes, and a series of operators of higher
dimension  appears.
It
is important that all these operators are
Lorentz scalars. For instance, the heavy quark part of the
Lagrangian takes the form
\begin{equation}
{\cal L}_{\rm heavy}=\sum_Q \left\{\bar Q (i\not\!\!D
-m_Q)Q +
\frac{c_G}{2m_Q}\bar Q (i/2)\sigma_{\mu\nu}G_{\mu\nu} Q\; +
\sum_{\Gamma ,\;q}
\frac{d_{Qq}^{(\Gamma )}}{m_Q^2} \bar Q \Gamma Q \bar q \Gamma
q \right\} + {\cal O}\left(\frac{1}{m_Q^3}\right)
\label{N2}
\end{equation}
where $c_G$ and $d_{Qq}^{(\Gamma )}$ are coefficient functions,
$G_{\mu\nu}\equiv g
G_{\mu\nu}^a t^a $ and $t^a$ is the color generator, (${\rm
Tr}\,t^at^b=\delta^{ab}/2$); below we will
often use the short-hand notation
$i\sigma G=i\sigma_{\mu\nu}G_{\mu\nu}=i\gamma_{\mu}
\gamma_{\nu} G_{\mu\nu}$.
The sum over the light quark flavors is shown explicitly as well as
the
sum over
possible structures $\Gamma$ of the four-fermion operators. All
masses and
couplings, as well as the  coefficient functions $c_G$ and
$d^{(\Gamma )}$,
depend on the normalization point. For example, the coefficient
$c_G$ in the leading logarithmic approximation can be written as
\begin{equation}
c_G(\mu)=\left(\frac{\alpha_s(\mu)}{\alpha_s(m_Q)}\right)^{-
\frac{3}{b}}-1
\, ,\;\;\;\;b=11-\frac{2}{3}n_f\, ,
\label{N3}
\end{equation}
where $n_f$ is the number of the light flavors.  The power $-3/b$
was first calculated in Ref. \cite{Falk}. In Sect. 5.3 I will
explain
how to derive Eq. (\ref{N3}).

The operators of dimension five and higher in Eq. (\ref{N2})
 are due to the  contribution of hard gluons,
with offshellness from $\mu$ up to $M_0$.  Since we agreed that
in this lecture we will ignore the existence of such gluons,
we will forget about these operators for the time being. Does this
mean that what remains from the Lagrangian (\ref{N2}) contains no
$1/m_Q$ terms?

The answer to this question is negative.  The $1/m_Q$ expansion is
generated by the first (``tree-level") term in the Lagrangian
(\ref{N2}),
\begin{equation}
{\cal L}_{\rm heavy}^0 = \bar Q (\not\!\!{\cal P} -m_Q)Q\, .
\label{lagrzero}
\end{equation}
Although the field $Q$ in this Lagrangian is normalized at a low point
$\mu$   the field $Q$ carries a hidden large parameter, $m_Q$;
isolating this parameter  opens the way to the $1/m_Q$
expansion. Indeed,
the interaction of the heavy quark with the light degrees of freedom
enters through ${\cal P}_\mu = i D_\mu$, where
$$
D_\mu = \partial_\mu - igA_\mu^at^a\, .
$$
The background gluon field $A_\mu$ is weak if
measured in
the scale $m_Q$, which means, of course,  that there is a large
``mechanical"
part
in the $x$ dependence of $Q(x)$, known from the very beginning
\cite{KSSV},
\beq
Q(x) = e^{-im_Qt}{\tilde Q} (x)
\label{tildeq}
\eeq
where ${\tilde Q} (x)$ is a ``rescaled" bispinor field which, in the
leading approximation, carries no
information
about
the heavy quark mass. It describes  a residual
motion
of the heavy quark inside the heavy hadron \cite{HQET} with
typical
momenta
of order
$\Lambda_{\rm QCD}$. Remnants of the heavy quark mass
appear in $\tilde Q$ only at the level of $1/m_Q$ corrections.

Equation (\ref{tildeq}) is written in the rest
frame of
$H_Q$.  In the arbitrary frame one singles out the factor
$\exp (-im_Qv_\mu x_\mu )$ where $v_\mu$ the
four-velocity of
the heavy hadron,
$$
v_\mu = p_\mu/M_{H_Q} \, .
$$

The covariant momentum operator ${\cal P}_\mu$ acting on the
original filed
$Q$, when acting on
the rescaled field $\tilde Q$, is substituted by the operator
$m_Qv_\mu + \pi_\mu$,
\beq
iD_\mu Q(x) = e^{-im_Qv_\mu x_\mu}\left( m_Q v_\mu + i
D_\mu\right) \tilde Q (x)
\equiv e^{-im_Qv_\mu x_\mu}\left( m_Q v_\mu + \pi_\mu
\right) \tilde Q (x)
\, .
\eeq
Below we will consistently use different letters,
${\cal P}_\mu$ and $\pi_\mu$ for the momentum operators
$iD_\mu$ acting
on
$Q$ and $\tilde Q$, respectively. If not stated to the contrary, we will
 use the rescaled
field
$\tilde Q$, {\em omitting} the  tilde in all expressions where there is
no risk of
confusion \footnote{Whenever one sees an expression containing
$\pi$'s one may be sure that it refers to the rescaled fields $\tilde Q$
even if the tildes are not written out explicitly}.
In the local colorless operators bilinear in the heavy quark fields it
does not
matter whether the original field $Q$ or the rescaled one is used,
since, say,
$$
\bar QQ = \bar{\tilde Q}\tilde Q ,\;   \bar Q {\cal P}_\mu Q =
\bar{\tilde Q}\pi_\mu
\tilde Q ,...
$$
and so on.  Using these distinct notations for the momentum operator
is convenient since all expressions written in terms of
$\pi_\mu$ and $\tilde Q$ do not contain implicitly the large
parameter
$m_Q$.

I pause here to make a reservation. The rescaled field $\tilde Q$
is a {\em four-component} Dirac bispinor, not a two component
non-relativistic spinor which is usually introduced in the heavy
quark effective theory (HQET) \cite{HQET}. HQET is a formalism
invented in
the very beginning of the 90's \cite{HQET} which is very often used
in connection with the heavy quark physics \cite{HQETRev}. It  is
convenient in a
range of problems but can be quite misleading in some other
problems.
I prefer to discuss the heavy quark expansions directly and
systematically in
{\em full QCD} in the
framework of
the Wilson OPE. In many instances the careful reader will certainly
recognize a significant overlap, but the Wilson language, being more
general, seems to  give a better understanding and command over
the $1/m_Q$ expansions. Moreover, some issues can not be
addressed in the framework of HQET at all.

The Dirac equation $(\not\!\!{\cal P} -m_Q)Q = 0$ in terms of  the
rescaled field can be written as follows:
\beq
\frac{1-\gamma_0}{2} Q = \frac{\not\!\!{\pi} }{2m_Q} Q\, ,
\label{DEt1}
\eeq
and
\beq
\pi_0 Q = -\frac{\pi^2 +(i/2)\sigma G}{2m_Q} Q\, .
\label{DEt2}
\eeq
The last equation is actually the squared Dirac equation,
$$
\frac{1}{2m_Q}\left( \not\!\!{\cal P} + m_Q\right)
\left( \not\!\!{\cal P} - m_Q\right) Q = \frac{1}{2m_Q}
\left( {\cal P}^2 +\frac{i}{2} \sigma G - m_Q^2\right) Q=
0\, .
$$
In deriving Eq. (\ref{DEt2}) we used the fact that
\beq
[{\cal P}_\mu , {\cal P}_\nu ] = [\pi_\mu , \pi_\nu ] =
ig  G^a_{\mu\nu}t^a\, .
\label{commutator}
\eeq

Armed with this knowledge one can easily  obtain the $1/m_Q$
expansion of ${\cal L}_{\rm heavy}^0$, up to terms $1/m_Q^2$,
$$
{\cal L}_{\rm heavy}^0 =
\bar Q (i\not\!\!D -m_Q)Q\;=\;\bar Q
\frac{1+\gamma_0}{2}\left(1+\frac{(\vec\sigma\vec\pi)^2}{8m_Q^2}
\right)\left[
\pi_0-\frac{1}{2m_Q}(\vec\pi\vec\sigma)^2\;-\right.
$$
\beq
\left.
-\;\frac{1}{8 m_Q^2}\,\left(-(\vec D\vec E)+2\vec\sigma\cdot\vec
E\times\vec\pi\right)\,
\right]\left(1+\frac{(\vec\sigma\vec\pi)^2}{8m_Q^2}
\right)\frac{1+\gamma_0}{2}\,Q
\;+\;{\cal O}\left(\frac{1}{m_Q^{3}}\right)\, ,
\label{HQLm}
\eeq
where $\vec\sigma$ denote the Pauli matrices and
$$
(\vec\pi\vec\sigma )^2 ={\vec\pi}^2 +\vec\sigma \vec B\, ,
$$
$\vec E$
and $\vec B$ denote the background chromoelectric and
chromomagnetic fields, respectively. The coupling constant $g$ and
the color matrix $t^a$ are included in the definition of these fields.
The derivation of this Lagrangian is a good home exercise.
I encourage everyone to obtain Eq. (\ref{HQLm})
by using the commutation relation (\ref{commutator})
and the properties of the gamma matrices.
Those who will have problems with getting Eq. (\ref{HQLm})
should consult Chapter 4 of Bjorken and Drell \cite{BD} or Sect. 33
of the Landau-Lifshitz course \cite{LL}
from where this Lagrangian  follows immediately.  It is worth noting
that
\beq
{\cal L}_{\rm heavy}^0 \equiv
\varphi^+(\pi_0-{\cal H}_Q\,)
\varphi
\label{ham}
\end{equation}
where
\begin{equation}
\varphi=\left(1+\frac{(\vec\sigma\vec\pi)^2}{8m_Q^2}
\right)\frac{1+\gamma_0}{2}\,Q
\label{18a}
\end{equation}
and ${\cal H}_Q$ is a non-relativistic Hamiltonian, through second
order in $1/m_Q$,
\beq
{\cal H}_Q\,=\,\frac{1}{2m_Q}\,({\vec\pi}^2 +
\vec\sigma \vec B)\,+\,
\frac{1}{8m_Q^2}\,\left(-(\vec D\vec E)+
2\vec\sigma\cdot\vec E\times\vec\pi\right)
\label{hamil}
\end{equation}
well-known (in the Abelian case) from the text-book expressions
\cite{BD,LL}.  Equation (\ref{18a}) is merely
the Foldy-Wouthuysen transformation which is necessary to keep
the
term linear in
$\pi_0$ in its canonic form.

\subsection{$m_Q\rightarrow\infty$; The heavy quark symmetry}

Let us first neglect all $1/m_Q$ corrections altogether. In this limit
$m_Q$ drops out from ${\cal L}^0_{\rm heavy}$,
\beq
{\cal L}^0_{\rm heavy} = \bar Q
\frac{1+\gamma_0}{2}\pi_0 Q \, .
\label{Lzerore}
\eeq
This expression takes place in the rest frame of $H_Q$; in the
arbitrary frame \cite{HQET}
\beq
{\cal L}^0_{\rm heavy} = \bar Q
\frac{1+\not\!{v}}{2}\pi_\mu v_\mu Q \, .
\label{Lzero}
\eeq
In the limit $m_Q\rightarrow\infty$ the masses of all $Q$-containing
hadrons
become equal to that of the heavy quark $Q$,
$$
M_{H_Q} = m_Q +{\cal O}(\Lambda_{\rm QCD})\, .
$$
The mass splittings between different hadrons are generically of
order $\Lambda_{\rm QCD}\ll m_Q$.  Soon, we will relate these
mass splittings to the expectation values of certain operators.

The assertion that all $Q$-containing hadrons
are degenerate to the zeroth order in $m_Q$ is trivial.
This ``degeneracy" by no means implies that the internal
structure of all $Q$-containing hadrons is the same. A little less
trivial is the fact that there exist hadrons whose masses are
degenerate to much better accuracy, ${\cal O}(m_Q^{-1})$,
and whose internal structure is, indeed, identical in the limit
$m_Q \rightarrow\infty$.

Since all effects due to the heavy quark spin are, obviously,
proportional to
$1/m_Q$, in this limit the heavy quark spin becomes irrelevant,
see Eqs. (\ref{HQLm}), (\ref{Lzero}).
Correspondingly, there emerges a symmetry between  the states
 which differ only by the spin orientation of the heavy quark. The
pseudoscalar and vector mesons of the type $B$ and $B^*$ (both
are the ground state $S$ wave mesons)
present an example of such spin family. In the limit $m_Q
\rightarrow\infty$ their masses must be degenerate up to terms
${\cal O}(m_Q^{-1})$, and the light clouds of $B$ and $B^*$
coincide.
If there is more than one heavy quark, say $Q_1$
and $Q_2$,
 the theory is symmetric with respect to the interchange
$Q_1\leftrightarrow
Q_2$ even if their masses are not close to each other (in physical
applications
we, of course, keep in mind $b$ and $c$). Indeed, the heavy quark
$Q_i$ plays
the role of the static force center
inside $H_{Q_i}$; the light cloud is flavor-blind and does not notice
the substitution of $Q_1$ by $Q_2$ provided that the four-velocities
of both quarks are the same. Notice that at this level the
four-velocity of
the heavy quark coincides with that of the heavy hadron. (Only when
higher order corrections in $1/m_Q$ are taken into account
the difference between the four-velocities becomes important and
the
symmetry $Q_1\leftrightarrow
Q_2$ is violated. At the level of $1/m_Q$ also the spin
symmetry is not valid any more.) If the hard gluon effects
are
neglected the interaction with the light cloud can not change the
heavy quark
four-velocity;
therefore, this quantity is conserved in the strong interactions
\cite{HQET}.
(This conservation is, of course, destroyed by the hard gluons
which can easily carry away a finite fraction of the heavy quark
momentum.)

The symmetry connecting $Q_1$ and $Q_2$
 emerges
in the limit $m_{Q_{1,2}}\rightarrow\infty$ even if the masses
of the heavy quarks are not close to each other. What is important
is that both  must be much larger than $\Lambda_{\rm QCD}$.
We encounter here  a situation which is conceptually close
to the problem of the isotopic symmetry of the strong interactions.
Everybody knows that the strong amplitudes are isotopically
invariant
with the accuracy up to a few percent, and, at the same time,
the masses of the $d$ and $u$ quarks are not too close
to each other, $m_d/m_u\sim 2$.  It is not the proximity of these
masses which counts, but the fact that the both masses are much
less than the QCD scale $\Lambda_{\rm QCD}$.

Usually the existence of an internal symmetry implies a degeneracy
of the
spectrum. For instance, the isotopic symmetry mentioned above,
apart from certain relations between the scattering amplitudes,
predicts
that the proton and neutron masses are the same, up to small
corrections
due to the symmetry breaking effects.  The heavy quark symmetry
does not manifest itself as a degeneracy in the spectrum --
the $D$ and $B$ masses are very far from each other. One has to
subtract the
mechanical part
of the heavy quark mass in order to see that all dynamical
parameters
are insensitive to the substitution $Q_1\leftrightarrow Q_2$
in the limit $m_{Q_{1,2}}\rightarrow\infty$ \cite{Dodik}. Perhaps,
this is the
reason why it was discovered so late.

To elucidate the issue of the heavy quark symmetry
let us consider a practical problem, semileptonic
decay of the $B$ meson induced by the weak $b\rightarrow c$
transition.
The initial $B$ meson decays into an electron-neutrino pair plus the
$D$
meson.  Since we do not now discuss  the $1/m_Q$ corrections we
may
make no
distinction between
the four-velocities of the quark $Q$ and the hadron $H_Q$, and
between their
masses. Assume that the $B$ meson is at rest. Furthermore, let us
assume that
the four-momentum $q$ carried away by the lepton pair is maximal,
$q^2=(M_B-M_D)^2$. This means that the $D$ meson produced is also
at rest --
the hadronic system experiences no recoil. The corresponding regime
is sometimes called the point of zero recoil.

In this regime the $B\rightarrow D$ transition  form factor is exactly
unity!
More exactly,
\beq
\langle D |\bar c \gamma_0 b|B\rangle =(2M_B2M_D)^{1/2}
\times\mbox{unity}\; \mbox {(at zero recoil)}
\label{BD}
\eeq
where the square root factors are due to the relativistic
normalization of our
amplitudes. By the same token
\beq
\langle D^* |\bar c \gamma_i \gamma_5 b|B\rangle
=i (2M_B2M_D)^{1/2}
D^*_i\times\mbox{unity}\; \mbox {(at zero recoil)}
\label{BD*}
\eeq
where $D^*_i$ is the polarization vector of $D^*$.
As
well-known, the
exact relations of this type always reflect an underlying symmetry.
They can
never
emerge accidentally because only a symmetry can protect the form
factors
from renormalizations.

It is very easy to understand why Eqs. (\ref{BD}) and (\ref{BD*})
take place.
Indeed, the space-time picture is very transparent. The $b$ quark at
rest is
surrounded by its light cloud, the latter being the eigenstate of the
problem
of color interaction with a static force center. At time zero the weak
current
instantaneously substitutes the $b$ quark by $c$; the charmed quark
is also at
rest, and since the color interactions are flavor-blind the same light
cloud
continues to be the eigenstate, this time with the $c$ quark as the
static center. If instead of the field-theoretic light cloud we had a
quantum-mechanical problem one could say that the overlap integral
for these
identical wave functions is 1. The light cloud will feel the substitution
$b\rightarrow c$ only to the extent the heavy quark momentum
inside the
heavy meson does not vanish exactly -- this effect is, of course,
suppressed
by powers of $1/m_Q$.  As we will see later corrections
in the right-hand side of Eqs. (\ref{BD}) and (\ref{BD*}) are actually
of order
$1/m_Q^2$; there are no linear corrections in $1/m_Q$. In the
$B\rightarrow
D^*$ transition generated by the axial-vector current the current,
additionally,
changes the orientation of the heavy quark spin. As was already
mentioned,
all effects related to the heavy quark spin are suppressed by
$1/m_Q$;
$D$ and $D^*$ are in the same multiplet, and the $B\rightarrow D^*$
transition
is governed by the same symmetry. This symmetry allows one to
rotate
arbitrarily four states,
$$
b \;\mbox{spin up},\; b\;\mbox{spin down},\;
c\;\mbox{spin up},\; c\;\mbox{spin down};
$$
therefore, we obviously deal here with  an SU(4) invariance.

The symmetry relations (\ref{BD}) and (\ref{BD*}) were first derived
in Refs.
\cite{NW,VS}. Shortly after it was realized \cite{IW} that the actual
symmetry
is much stronger -- the SU(4) invariance takes place for any given
value of
$v_\mu$, the four-velocity of
the recoiling $c$ quark, not necessarily at the point of zero recoil or
close to it.
Thus, many
different form factors connecting $(B,B^*)$ and $(D,D^*)$ can be
expressed in
terms of one  function depending only on the velocity of the
recoiling hadron (in the rest frame of the decaying hadron). The
universal
form factor is called the {\em Isgur-Wise function}.

\subsection{The Isgur-Wise function}

Now we are finally ready to discuss a very elegant observation due
to Isgur
and Wise \cite{IW}.
Let us consider now the  amplitudes induced by the transition $
\bar c \Gamma b $  off the zero recoil point. Here $\Gamma$ is any
Lorentz
matrix; of special interest are, of course, the vector and the
axial-vector cases,
$$
\Gamma = \gamma_\mu\;\mbox{or}\;\gamma_\mu\gamma_5 ;
$$
the weak decays of the $B$ meson are induced by the $V-A$
currents. The
physically measurable amplitudes are
$\langle D|\bar c \Gamma b|B\rangle$ and $\langle D^*|\bar c
\Gamma
b|B\rangle$; for completeness one can also consider the amplitudes of
the
type $\langle D|\bar c \Gamma b|B^*\rangle$, or $\langle D|\bar c
\Gamma
b|B^*\rangle$, or $\langle B|\bar b \Gamma b|B\rangle$ -- this adds
nothing new. The four-velocity of the
particle $H_Q$
is defined as
\beq
v_\mu = \frac{(p_{H_Q})_\mu}{M_{H_Q}};
\label{v}
\eeq
the four-velocities of the initial particles will be denoted by $v$
while those
of the final particles by $v'$. It is obvious that $v^2=1$, and,
additionally, in
the rest frame $v=\{1,0,0,0\}$. In the most general case the
amplitude
$\langle D|\bar c\gamma_\mu b|B\rangle$ can be expressed in terms
of two
form factors,
the amplitude
$\langle D^*|\bar c\gamma_\mu\gamma_5 b|B\rangle$ in terms of
three form
factors and the amplitude $\langle D^*|\bar c\gamma_\mu
b|B\rangle$ in
terms of one form factor. The heavy quark symmetry tells us that in
the
limit $m_Q\rightarrow\infty$ these six functions, {\em a priori}
independent,
reduce to one and the same function which depends only on the
scalar product
$vv'$. Specifically,
\beq
\langle D|\bar c\gamma_\mu b|B\rangle =
\sqrt{M_BM_D}\left[ \left( v+v'\right)_\mu \right] \xi (y)\, ;
\label{1}
\eeq
\beq
\langle D^*|\bar c\gamma_\mu\gamma_5 b|B\rangle =
\sqrt{M_BM_D}\,  i \left[ D_\mu^* (1+vv') -
\left( D_\alpha^* v_\alpha\right)v'_\mu\right] \xi (y)\, ;
\label{2}
\eeq
\beq
\langle D^*|\bar c\gamma_\mu b|B\rangle = \sqrt{M_BM_D}\left[
- \epsilon_{\mu\nu\lambda\sigma}D_\mu^*v'_\lambda v_\sigma
\right] \xi (y)\, .
\label{3}
\eeq
where
$$
y = vv'
$$
and $\xi (y)$ is the Isgur-Wise function, $D_\mu^*$ is the
polarization vector of $D^*$.
The Isgur-Wise function is independent of the heavy quark masses.
The square root $\sqrt{M_BM_D}$ reflects the relativistic
normalization of the states.
The symmetry relations (\ref{BD})
and (\ref{BD*}) imply that the normalization of the Isgur-Wise
function at
zero recoil is fixed,
\beq
\xi (y=1) = 1 .
\label{norm}
\eeq

Perhaps, it is worth noting that the phases in Eqs. (\ref{2}) and
(\ref{3})  differ from what you might see in the literature. They are,
of course, a matter of convention and reflect the definition of the
states. The definition I follow is in accord with the standard
relativistic convention, see Eq. (\ref{genIW}) and Sect. 3.5.

The fact that a  large set of  form factors degenerate into a single
function
depending only on $y$ might seem a miracle; but after   the
assertion is
made,
with the knowledge you already have, it should be not  difficult to
understand
why it happens. Indeed, let us turn again to the space-time picture
described
above. A $b$ quark at rest, surrounded by the light cloud,
instantaneously
converts into a $c$ quark. This time the four-momentum carried
away by the
lepton pair is not maximal; therefore, the $c$ quark is not at rest.
This force
center flies away with the velocity $\vec v'$. But the light cloud
stays intact.
So, the question is: ``what is the amplitude for the flying $c$ quark
and the
cloud at rest to form a $D$ or $D^*$ meson?" We can look at this
process in
another way. After the $b\rightarrow c$ transition happened let us
proceed
to the rest frame of $c$. In this reference frame the $c$ quark
produced is at
rest, but the cloud, as a whole, moves away with the velocity $-\vec
v'$. It is
clear that this system -- the static charmed force center plus a
moving light
cloud -- has a projection on $D$ or $D^*$. The amplitude
{\it per se}, with the kinematic structures {\em excluded}, can
depend
only on
$|\vec v'|$ -- there is no preferred orientation in the space, and the
direction
of $\vec v'$ is irrelevant. Using covariant notations one can say that
the
amplitude depends only on $vv'$  since in the $B$ rest frame $vv' =
\sqrt{(1+{\vec v}'^2 )}$. There is simply no place for the dependence
on the
heavy quark masses, apart from the overall normalization factors
appearing
because we stick to the relativistic normalization of states
\footnote{Warning:
an additional dependence on the heavy quark masses may emerge if
we
include the hard gluon exchanges neglected so far. For details see Ref.
\cite{FaG}}.

Since the heavy quark spin is irrelevant in the limit
$m_Q\rightarrow\infty$ to warm up let us consider a toy model
where the
heavy quarks are deprived of their spins from the very beginning.
In other words, I replace the genuine spin-1/2 heavy quarks of QCD
by spin-0 color triplets with the same mass. We will turn to this
toy model more than once below.

In QCD, $B$ and $B^*$ form a multiplet which includes 4 states:
the total angular momentum of the light cloud (1/2)
combines with the heavy quark spin (1/2) to produce either
spin-0 state ($B$) or three spin-1 states ($B^*$). In our toy model
the analog of this ground-state multiplet is obviously a baryon of
spin 1/2; let us denote the corresponding field by $N^\alpha_Q$,
where $\alpha$ is the spinorial index. The current generating the
transition $N_b\rightarrow N_c$ has the form $c^\dagger b$,
where the fields $b$ and $c$ are assumed to be scalar now.
At first sight the amplitude $\langle N_c|c^\dagger b|N_b\rangle$
might contain four different kinematic structures,
$$
\bar N_c N_b\, , \,\,\,
\bar N_c \not\!{v}N_b\, , \,\,\,
\bar N_c {\not\!{v}}' N_b\, , \,\,\,
\bar N_c \sigma_{\mu\nu}N_b v_\mu v_\nu ' \, ,
$$
where I list only P-even structures, of course. On mass shell they all
reduce to the first one, however; for instance, $\not\!\!{v}N_b =
N_b$.
Hence
$$
\langle N_c|c^\dagger b|N_b\rangle = \bar N_c N_b \xi (y)\, .
$$

Returning to  real QCD what
remains to be done is to work out the consequences of  spin.
The most concise general formula can be written in terms of the
matrices
\beq
{\cal M} = B(i\gamma_5) + B_\mu \gamma_\mu\; \mbox{and}
\;
{\cal M}' = D^*(i\gamma_5) + D^*_\mu \gamma_\mu\, ,
\label{m}
\eeq
where $B_\mu$ and $D_\mu$ are the polarization vectors of $B^*$
and
$D^*$, respectively. The (heavy quark) spin independence of the
strong
interactions at
$m_Q\rightarrow\infty$ manifests itself in the fact that the
couplings of the
ground state pseudoscalar to $i\gamma_5$ and the ground state
vector to
$\gamma_\mu$ are the same, see Sect. 3.4.  Now, the whole set of
the
transition
amplitudes can be expressed by one compact cute formula
\footnote{Strictly speaking, for the outgoing particles one must use
$\overline{\cal M}' =\gamma_0 ({\cal M}')^\dagger\gamma_0$.
With our conventions, however, $\overline{\cal M}' = {\cal M}'$.}
,
\beq
\frac{1}{\sqrt{2M_D2M_B}}\langle H_c (v') | \bar c \Gamma b | H_b
(v)
\rangle = \frac{1}{2}
 \mbox{Tr} \{ {{\cal M}'} \frac{1+{\not\!v}\, '}{2}
\Gamma \frac{1+{\not\!v}}{2} {\cal M}\}\, \xi (y=vv') .
\label{genIW}
\eeq
Completing the trace we recover Eqs. (\ref{1}) -- (\ref{3}).

Equation (\ref{genIW})  can be derived in many different
ways. Originally it was obtained in Ref. \cite{FaG} (see also \cite{JBJ}).
In Sect. 3.5 we
will discuss one
 of the possible derivations -- perhaps, not the simplest,
but very instructive.  Before we will be able to do that it is necessary
to make a digression and study some elements of the background
field
technique.

Equations (\ref{1}) -- (\ref{3}) are valid not only in the space-like
domain (the form factor kinematics) but also in the time-like
domain. The latter assertion calls for an immediate  reservation,
though.
The heavy quark symmetry implies that $M_B = M_{B^*}$.
In the real world this equality is not exact: the  heavy quark
symmetry  is violated by small
$1/m_Q$ terms. This small violation can be strongly enhanced in the
near threshold domain, $E\approx 2M_B$ ,where the symmetry
breaking parameter
turns out to be  of order one \cite{IWviol}. Indeed, let us consider a
kinematical
point above the
threshold of the $B\bar B$ production but below
$B\bar B^*$. In this domain all form factors describing three
amplitudes
$$
\langle B\bar B |J_\mu |0\rangle\, ,\,\,\,
\langle B\bar B^*| J_\mu |0\rangle\, ,\,\,\,
\langle B^*\bar B^*| J_\mu |0\rangle\, ,
$$
(here $J_\mu$ is some heavy quark current, say, $J_\mu =\bar
b\gamma_\mu b$)
have imaginary parts associated with the normal thresholds due to
the intermediate state
$B\bar B$. On the other hand, there is no contribution to the
imaginary part from the intermediate state
$B\bar B^*$ and $B^*\bar B^*$. In the
pseudoscalar meson $B$ the spin of the heavy quark $Q$
is rigidly correlated with that of the light cloud. Hence, the
spin independence of the heavy quark interaction is totally lost in
the imaginary part in this point. In particular, in the amplitude
$\langle B^*\bar B^*| J_\mu |0\rangle$ a kinematic structure
forbidden by the
Isgur-Wise formula appears. An even  more pronounced effect of the
heavy quark symmetry violation takes place in the anomalous
thresholds generated by the pion exchange which can start
parametrically much below the
the normal thresholds depending on the interplay
between $M_{B^*}^2-M_B^2$ and the pion mass \cite{Ball}.

\subsection{The mass formula}

To complete our first encounter with the basics of the
heavy quark theory we will now derive a $1/m_Q$ expansion
for the masses of the $Q$-containing hadrons.

It is intuitively  clear that the heavy hadron mass can be expanded
in terms
of that of the heavy quark as follows
\beq
M_{H_Q} = m_Q +\bar\Lambda + {\cal O}(m_Q^{-1})
\label{mf}
\eeq
where $\bar\Lambda$ is a constant, of order of $\Lambda_{\rm
QCD}$,
which depends on the light quark content and the quantum numbers
of $H_Q$
but is independent of $m_Q$. (It first appeared in Ref. \cite{Luke}.
Later on we will see that this expression is not as trivial as it might
naively
seem and requires thoughtful definitions of all parameters involved.
In particular, since the quarks are never observed  as isolated
objects,
one may ask what the quark mass $m_Q$ actually means. In due
time we will
return to this question, of course. For the time being we agreed to
disregard hard gluon exchanges; then $m_Q$ is just the mass
parameter in
the
Lagrangian (\ref{lagr}).

Formally Eq. (\ref{mf}) can be most easily derived by analyzing the
trace of the energy-momentum tensor in QCD,
\beq
\theta_{\mu\mu} = m_Q\bar Q Q +\frac{\beta (\alpha_s)}{4\alpha_s}
G_{\mu\nu}^a G_{\mu\nu}^a
\label{emt}
\eeq
where $\beta (\alpha_s)$ is the Gell-Mann-Low function.  For
simplicity we
assume the light quarks to be massless; introduction of the light
quark masses changes only technical details at intermediate stages of
our analysis. If the mass term of the light quarks is set equal to zero
the light quark fields do not appear explicitly in the
trace of the energy-momentum tensor. The expression (\ref{emt})
contains two terms: the first one is a mechanical part while the
second term is the famous trace anomaly of QCD \cite{aemt} (for a
review
see e.g. Ref. \cite{shifman}).

Furthermore, as well-known, for any given one-particle state
the expectation value of the trace of the energy-momentum tensor
reduces to the mass of the state.
Then, the hadron mass can be
expressed in terms of two expectation values,
\begin{equation}
M_{H_Q} = \frac{1}{2M_{H_Q}}\langle H_Q | m_Q\bar QQ
|H_Q\rangle +  \frac{1}{2M_{H_Q}}\langle H_Q |
\frac{\beta (\alpha_s)}{4\alpha_s}G^2
|H_Q\rangle
\label{mass}
\end{equation}
where
the relativistic normalization of the states is implied
$$
\langle H_Q|H_Q\rangle = 2M_{H_Q}V
$$
in the rest frame; $V$ is the normalization volume. We will always
use only the relativistic normalization of
states
which will routinely result in the factors $(2M_{H_Q})^{-1}$ in all
expressions.

Let us discuss the expectation values of the operators in Eq.
(\ref{mass})
in turn. The first one is explicitly proportional to $m_Q$. To be more
quantitative we must determine the matrix element of the heavy
quark
density $\bar QQ$. To this end it is convenient to use an argument
suggested  in Ref. \cite{BUV} which will show us that the expectation
value of
$\bar QQ$ is very close to unity;  as a matter of fact, with our present
accuracy
it is just equal to unity. The second expectation value reduces to
$\bar\Lambda$.

Indeed, in the rest frame of $H_Q$ a typical momentum of $Q$ is of
order $\Lambda_{\rm QCD}$, i.e. the heavy quark is very slow. This
means
that the lower components of the bispinor field $Q$ are small
compared to the
upper ones and, hence, the scalar density of the heavy quark is close
to its
vector charge, $\bar QQ \approx \bar Q\gamma_0 Q$. The difference
is only
due to the lower components. The vector charge, however, just
measures the
number of the heavy quarks inside $H_Q$; therefore, its matrix
element is
exactly unity.

It is instructive to do the simple derivation outlined above in
some detail.
Combining  the equations of motion, (\ref{DEt1}) and
(\ref{DEt2}),
it is easy to get that
\beq
\frac{1-\gamma_0}{2} Q =-\frac{1}{2m_Q}\vec\pi\vec\gamma
\frac{1+\gamma_0}{2} Q +{\cal O}(m_Q^{-2}) \, ,
\label{1-g}
\eeq
which implies, in turn,
\beq
\bar Q Q = \bar Q\gamma_0 Q -\frac{1}{2m_Q^2}
\bar Q\left( {\vec\pi}^2 +\vec\sigma\vec B\right) Q +\mbox{higher
orders}
\label{QQQQ}
\eeq
where  $\vec B$ is the
chromomagnetic field, $B_i=\epsilon_{ijk}G_{jk}$. Equation
(\ref{QQQQ}) is the
desired result
demonstrating that
\beq
\frac{1}{2M_{H_Q}}\langle H_Q |\bar QQ |H_Q\rangle
= \frac{1}{2M_{H_Q}}\langle H_Q |\bar Q\gamma_0Q |H_Q\rangle
+{\cal O}(m_Q^{-2}) = 1 +{\cal O}(m_Q^{-2})\, .
\label{QQ}
\eeq
The  matrix element
of the vector charge (appropriately normalized) is set equal to unity,
as was discussed above.

This  digression has been undertaken merely to familiarize
the reader
 with
the basics of the $1/m_Q$ expansion in QCD.  As our
understanding  progresses the level of the explanatory remarks will
be
reduced  so that in the subsequent lectures many derivations of a
more technical nature will be suggested as an exercise.

Thus, we have established that the first expectation value in Eq.
(\ref{mass})
produces $m_Q$ in the expansion for the heavy hadron mass. The
second
expectation
value which also has the dimension of mass obviously does not scale
with
$m_Q$ in the
limit $m_Q\rightarrow\infty$, so one can define
\footnote{This expression relating $\bar\Lambda$ to the expectation
value of the gluon anomaly operator was obtained in Ref.
\cite{BSUV}. Some subtleties left aside in the derivation presented
here are discussed in detail in this paper. It is instructive to compare
Eq. (\ref{defl}) with a similar expression for the nucleon mass,
$$
M_N = \frac{1}{2M_N}\langle N |
\frac{\beta (\alpha_s)}{4\alpha_s}G^2
|N\rangle\, ,
$$
known from  ancient times \cite{GAN}.
}
\beq
\bar\Lambda = \frac{1}{2M_{H_Q}}\langle H_Q |
\frac{\beta (\alpha_s)}{4\alpha_s}G^2
|H_Q\rangle_{m_Q\rightarrow\infty} .
\label{defl}
\eeq
Thus, the parameter $\bar\Lambda$ of the heavy quark theory is, in
a sense,
similar to the gluon condensate \cite{SVZ}. The latter is the
expectation value
of the same gluon operator over the vacuum state. In the case of
$\bar\Lambda$ the gluon operator is averaged over the lowest state
of the
system with the given (unit) value of the heavy quark charge. The
lowest
state is, of course, the ground state pseudoscalar meson, $B$.
Generally
speaking, $H_Q$ can be any $Q$-containing hadron. $B$ mesons are
most
interesting from the point of view of  applications; of practical
interest also are
$Q$-containing baryons which are the lowest-lying states in the
given channel with the baryon quantum numbers.
Therefore, strictly speaking, unlike the gluon condensate, there exist
many
different parameters $\bar\Lambda$, one for every channel
considered.
Usually we will tacitly assume that $\bar\Lambda$ is defined with
respect
to the $B$ mesons.

Both expectation values,
$$
\frac{1}{2M_{H_Q}}\langle H_Q | m_Q\bar QQ
|H_Q\rangle\;\mbox{and}\; \frac{1}{2M_{H_Q}}\langle H_Q |
\frac{\beta (\alpha_s)}{4\alpha_s}G^2
|H_Q\rangle \, ,
$$
have $1/m_Q$ corrections which show up at the level
${\cal O}(m_Q^{-1})$ in Eq. (\ref{mf}). Later on we will derive
the expansion for $M_{H_Q}$ which takes into account these terms
${\cal
O}(m_Q^{-1})$.

The $1/m_Q$ corrections in the expectation value of the gluon
anomaly are due to the fact that in our approach the states
$|H_Q\rangle$ are physical heavy flavor states, rather than the
asymptotic states corresponding to $m_Q=\infty$ which are usually
considered within HQET. Instead of working with these fictitious
states I prefer to explicitly keep track of all $1/m_Q$ corrections,
both in the operators and in the definition of the states, appealing
directly to the Wilsonean operator product expansion.

\newpage

\section{Lecture 2. Basics of the Background Field Technique}

\renewcommand{\theequation}{2.\arabic{equation}}
\setcounter{equation}{0}

The essence of our approach is separation
of all  momenta into two classes -- hard and soft. For the time being
we will continue to pretend that the role of the gluon degrees of
freedom  reduces to a soft gluon medium.  This is an ideal situation
where the gluons can be treated as a background field. A powerful
method allowing one to put calculations in the background fields on
an industrial basis was developed by Schwinger in electrodynamics
many years ago. In the eighties it was adapted to QCD. We will  be
unable to submerge in all details of  this technique, and  will, rather,
present  some basic elements in particular examples.
The review paper \cite{NSVZ} is recommended for further education.
This lecture will be rather technical -- its primary goal is to teach
how the heavy quark mass expansions can be constructed in a
systematic way in different problems.

The starting point of the method  is
decomposition of  fields into  two parts -- the
quantum part and the background  one.  The propagation
of quanta is described by the correlation functions of the quantum
part of the fields considered in the  external field. Later on
the external field is to be considered as a fluctuating field of the light
cloud, but this stage
need not concern us at the moment.

Let us start with a brief review of the Schwinger method, as it can be
applied in QCD. We introduce the coordinate and momentum
operators, $X_\mu$ and $p_\mu$, respectively, $[p_\mu , X_\nu ] = i
g_{\mu\nu}, \,\,  [p_\mu , p_\nu ] = [X_\mu , X_\nu ] = 0$.
Moreover, introduce a formal set of states $|x)$ which are
the eigenstates of the coordinate operator $X_\mu$,
\begin{equation}
X_\mu | x) =x_\mu |x).
\label{13}
\end{equation}
Please, note that $|x)$ has nothing to do with the field-theoretic
eigenstates, e.g. $|H_Q\rangle$. To emphasize this fact
the use the regular bracket ) in the notation instead of the angle one,
which is reserved for the field-theoretic
eigenstates.

Then define the covariant momentum operator ${\cal
P}_\mu$ satisfying
the following commutation relations
\begin{equation}
[{\cal P}_\mu , X_\nu ] = i g_{\mu\nu}, \,\,\,
[{\cal P}_\mu , {\cal P}_\nu ] = i gt^a G_{\mu\nu}^a ,
\label{14}
\end{equation}
where $t^a$ are the generators of the color group, $G_{\mu\nu}^a$
is the external field.

The algebra (\ref{14}) is the basic tool of the Schwinger formalism.
We will  expand the Green functions in the background field, and in
each  order of the expansion we will need to use only this algebra.

In the coordinate basis ${\cal P}_\mu$ acts as a
covariant derivative, namely
\begin{equation}
(y|{\cal P}_\mu |x) =
\left( i\frac{\partial}{\partial x_\mu} +gt^aA_\mu^a (x)\right)
\delta (x-y)
\label{15}
\end{equation}
if
\begin{equation}
(y|x) = \delta (x-y).
\label{16}
\end{equation}

Now we can write formal expressions for the Green functions. For
instance, for the quark Green function (mass $m_q$) describing
propagation from the point 0 to the point $x$ we have
\begin{equation}
S(x,0) = (x| \frac{1}{\not\!\!{\cal P} -m_q} |0)\, .
\label{17}
\end{equation}

Eq. (\ref{17}), rather obvious by itself, is readily verified by applying
the
Dirac operator to both sides of Eq. (\ref{17}).  Furthermore, it can be
identically rewritten as follows
\begin{equation}
S(x,0) = (x| (\not\!\!{\cal P} +m_q)\,
\frac{1}{{\cal P}^2 -m_q^2 + (i/2)
G_{\mu\nu}\sigma_{\mu\nu}} |0).
\label{18}
\end{equation}
where
$$
G_{\mu\nu}\equiv gt^a G_{\mu\nu}^a.
$$
Please, note that the ordering is important here since ${\cal P}_\mu$
does not commute with ${\cal P}_\nu$ of $G_{\mu\nu}$.

If we are aimed at calculating the coefficient functions in the Born
approximation we need nothing else -- Eq. (\ref{18}) is just
systematically
expanded in powers of the background field by using the
commutation relations (\ref{14}).

Observe that one can always shift ${\cal P}_\mu$ by a c-number
vector due to the fact that
\begin{equation}
{\rm e}^{iqX}{\cal P}_\mu {\rm e}^{-iqX} = {\cal P}_\mu + q_\mu .
\label{19}
\end{equation}
Hence, the Fourier transformed propagator reduces to
$$
\int d^4 x {\rm e}^{iqx} S(x,0) = \int d^4 x \, (x|
{\rm e}^{iqX} \frac{1}{\not\!\!{\cal P} -m_q}{\rm e}^{-iqX} |0)
=\int d^4 x (x|\frac{1}{\not\!\!{\cal P} +\not\! q-m_q}|0)\, .
$$
 This simple trick allows one to readily develop  the
expansion
sought for. Indeed, assume that $q$ is large (hard momentum) and
${\cal P}$ represents soft modes and is small in this sense.  Then we
can expand in ${\cal P}$,
\begin{equation}
\int d^4 x {\rm e}^{iqx} S(x,0) \rightarrow \frac{1}{\not\!\! q - m_q}
-
\frac{1}{\not\!\! q - m_q} \not\!\!{\cal P}\frac{1}{\not\!\! q - m_q}
+...
\label{20}
\end{equation}
Next, we transpose ${\cal P}$ to the right-most (left-most) position
and act on the states using the equations of motion.

It may seem that so far we got almost nothing compared to the
standard Feynman graph calculations.  Let us demonstrate the
efficiency of the background field technique in a few examples.

\subsection{Inclusive decay of the heavy quark -- toy model}

One of the most important practical problems in the heavy quark
theory is the description of the inclusive decays of heavy flavors.
The semileptonic and radiative decays of the $B$ mesons
$B\rightarrow X_c l\nu$ and $B\rightarrow X_s \gamma$ are
particular examples. Both are two key elements of the ongoing
experimental efforts, in quest of new physics. Needless to say that
a reliable QCD-based theory of such decays is badly needed. In this
section we start discussing basics of such a theory.

Since this is our first exercise, for pedagogical reasons, it seems
reasonable to ``peel off''
all inessential technicalities, like the quark
spins, and  resort to a simplified model.  In this toy model we will
consider
the inclusive decay of a spinless heavy quark into a spinless lighter
quark  plus a photon.
Of course, our photon is also a toy photon. We will assume it to be
scalar and the corresponding field will be denoted by $\phi$.

The Lagrangian describing the transition of a heavy quark $Q$
into a lighter quark $q$ and a ``photon" has  the form
\begin{equation}
 {\cal L}_\phi = h \bar{Q} \phi q \; + \; {\rm h.c.} \; ,
\label{Qqphi}
\end{equation}
where $h$ is the coupling constant and
$\bar{Q}=Q^{\dagger}$.
The masses of the quarks
$Q$ and $q$ are both {\em large} (and I remind that they are both
spinless).  Moreover, to further simplify the problem
we will
analyze a special limit (the so called small velocity or SV
limit  suggested in Ref. \cite{VS})
in which
\begin{equation}
\Lambda_{\rm QCD}\ll m_Q - m_q\ll m_Q \, .
\label{SV}
\end{equation}
The field $\phi$
carries color charge zero;
the  reaction $Q\rightarrow q+\phi$ could be considered  a toy model
for the
radiative
decays of the type $B\rightarrow X_s\gamma$ where $X_s$ is an
arbitrary inclusive hadronic state containing the $s$ quark produced
in the
$b$ quark decay.

It is very easy to calculate  the total  width for the {\em free} quark
decay
$Q\rightarrow q +\phi$,
\begin{equation}
\Gamma_{\rm free\,\,\, quark} (Q\rightarrow q\phi )
=\frac{h^2E_0}{8\pi m_Q^2}\equiv
\Gamma_0
\label{gamma0}
\end{equation}
where
\begin{equation}
E_0 = \frac{m_Q^2-m_q^2}{2m_Q} \, .
\label{E0}
\end{equation}
This free quark expression is valid for the total inclusive
probability in the asymptotic limit
when $m_Q\rightarrow\infty$. We are interested, however,  in the
preasymptotic corrections proportional to powers of $1/m_Q$.

First of all we must formulate what object we must deal with in
order
to be able to calculate these corrections systematically. Upon
reflection one concludes that it can not be the decay amplitude
$Q\rightarrow q\phi$ itself. Instead we must consider
the $Q\rightarrow Q$ forward ``scattering" amplitude depicted on Fig.
1. By scattering I mean that $Q$ scatters off the $\phi$ quantum
and off the background gluon field which is not shown on Fig. 1
explicitly but is implied. It is implied that all quark lines,
$Q$ and $q$, are submerged into this soft-gluon background field.
Through the
optical theorem the imaginary part of the amplitude of Fig. 1 is
related to the inclusive probability of the $Q\rightarrow q\phi$
transition. More specifically, if we introduce the {\em transition
operator}
\begin{equation}
\hat T = i\int d^4 x \,{\rm e}^{-iqx} T\{ \bar Q(x) q(x) \, , \,\bar
q (0)
Q (0)\} .
\label{trans}
\end{equation}
then the energy spectrum of the $\phi$ particle  in the inclusive
decay is
obtained from
$\hat T$ in the following way:
\begin{equation}
\frac{d\Gamma}{dE} = \frac{h^2 E}{4\pi^2M_{H_Q}}
  {\rm Im}\, \langle H_Q|\hat T |H_Q \rangle \, .
\label{spectrum}
\end{equation}
Here as usual $H_Q$ denotes a hadron built from the heavy quark
$Q$ and
the light cloud (including the light antiquark), $q$ in the exponent is
the four-momentum carried away by $\phi$ and $E$ is the energy of
the $\phi$ quantum.

Equation (\ref{spectrum}) immediately translates the $1/m_Q$
expansion for the transition operator in the $1/m_Q$ expansion for
the inclusive decay rate.  The fact that the transition operator
must be the primary object of the analysis in all problems of this
type was realized in Refs. \cite{KSSV,1,chay}.

Now we use what we have already learnt about the background field
technique to write the transition operator in the form
$$
\hat T = - \int d^4 x \,{\rm e}^{-iqx} (x|\bar Q\frac{1}{{\cal P}^2-
m_q^2}Q|0) =
$$
\begin{equation}
- \int d^4 x (x|\bar Q\frac{1}{(P_0 - q +\pi )^2-
m_q^2}Q|0)\, ,
\label{born}
\end{equation}
where
$$
(P_0)_\mu  = m_Qv_\mu \, .
$$
The Green function of the quark $q$  differs
from  the Green function given in  Eq. (\ref{18}) in an obvious way
since we assume for the time being that  our
quarks $Q$ and $q$ have spin zero, and, correspondingly, instead of
Eq. (\ref{18}) referring to the spinor quarks we have
\begin{equation}
S(x,0) = (x| \frac{1}{{\cal P}^2 -m_q^2 } |0)\, .
\label{GFS}
\end{equation}
In the second line of Eq. (\ref{born}) we proceeded to the rescaled
fields $\tilde Q$
which singles out the mechanical part of the momentum operator.

One more thing which will be needed is the equation of motion for
the scalar field $Q$ substituting the Dirac equation. Starting from
$({\cal P}^2 -m_Q^2)Q = 0$ we obviously get
\beq
\pi_0\tilde Q = - \frac{\pi^2}{2m_Q}\tilde Q\, .
\label{eoms}
\eeq

Finally we are ready to begin constructing the $1/m_Q$ expansion.
Since $\pi$ is of order $\Lambda_{\rm QCD}$ while $P_0-q$ scales as
$m_Q$, in the
leading approximation $\pi$ in the denominator of Eq. (\ref{born})
can be neglected altogether. Then, obviously,
\beq
\hat T^{(0)} =\bar QQ\frac{1}{m_q^2 - k^2}
\eeq
where
$$
k= P_0 - q\, .
$$
We see that the leading operator
 appearing in the expansion is
$\bar Q Q$ and it  has  dimension 2 (let us recall that the scalar $Q$
field
has
dimension $1$ in contrast to the real quark fields of dimension
$3/2$, which
leads in particular to different normalization factors in the matrix
elements). Taking the imaginary part we conclude that in the leading
approximation
\beq
\frac{1}{\pi} {\rm Im}\, \langle H_Q|\hat T^{(0)}|H_Q\rangle
=
\frac{\langle \bar Q Q\rangle}{2m_Q}\delta (E-E_0 )\, .
\label{imazero}
\eeq
Here and below  I will use a very convenient short-hand notation,
$$
\langle \bar Q Q\rangle \equiv
\langle H_Q|\bar Q Q|H_Q \rangle\, .
$$

The delta function in the imaginary part is characteristic of a
two-body decay. As a matter of fact, combining Eq. (\ref{imazero})
with the general expression  (\ref{spectrum}) and approximating
$\langle \bar Q Q\rangle $ by unity -- which can and must be
done in the leading order in $1/m_Q$ -- we get the delta-function
spectrum of the
the free quark decay.  Integrating over the energy we then arrive at
the free quark decay width (\ref{gamma0}).

Although this little achievement is quite gratifying and shows that
we are on the right track the real $1/m_Q$ expansion begins
when the preasymptotic terms switch on.  To this end the
terms with $\pi$ in the denominator of Eq. (\ref{born}) must be
kept,
and then the expansion in $(k\pi +\pi^2)$  must be carried
out. The general term of this expansion is
\beq
\hat T = \frac{1}{m_q^2-k^2}\sum_{n=0}^\infty
\bar Q \left( \frac{2m_Q\pi_0 +\pi^2- 2q\pi }{m_q^2-k^2}\right)^n Q
\, .
\label{S1}
\eeq
In Lecture 4 where the theory of the end point spectrum will
be presented we will need the whole sum.  At the moment our
purpose is
more limited -- we are aimed at getting the first
correction in the total decay width. This task does not require
the infinite sum; only two terms, with $n=1$ and $n=2$, are relevant.
Both terms are especially simple.

Indeed, if $n=1$ the combination $2m_Q\pi_0 +\pi^2$
in the numerator acting on $Q$ is nothing else than the equation of
motion, and can be dropped, see Eq. (\ref{eoms}). We can further
discard the $\vec q\vec\pi$ part -- since the $H_Q$ spin is assumed
to be zero there is no preferred orientation and, hence,
$\langle\bar Q\vec\pi Q\rangle = 0$. In this way we arrive at
\beq
\langle H_Q|{\hat T}^{(1)}|H_Q\rangle =-\frac{q_0}{m_Q}\,
\frac{\langle{\vec\pi}^2\rangle}{(m_q^2-k^2)^2}
\label{S3}
\eeq
plus terms of higher order in $1/m_Q$.
Here the same equation of motion (\ref{eoms}) was applied to
eliminate $\pi_0$ in favor of $\pi^2$ which, in the given order
in $1/m_Q$, coincides with $-{\vec\pi}^2$. The notation  is even
more concise than previously, namely $\langle{\vec\pi}^2\rangle$
stands for $\langle H_Q|\bar Q{\vec\pi}^2 Q|H_Q\rangle$.
You will often see similar short-hand below.

I pause here to make a side remark. The physical meaning of the
matrix element
$\langle{\vec\pi}^2\rangle$ is quite transparent -- it merely
represents the average value of the square of the momentum of the
heavy quark $Q$ inside the heavy hadron $H_Q$. This quantity is
of order $\Lambda_{\rm QCD}^2$ .
This is one of the most important parameters of the heavy quark
theory, along with $\bar\Lambda$.   Note the gap in dimensions of
the operators
appearing in the expansion. The dimension-2 operator $\bar QQ$
is followed by dimension-4 operator $\bar Q{\vec\pi}^2 Q$.
No relevant operator of dimension 3 exists. Due to this reason
the contribution of $\hat T^{(1)}$ in the total width
is ``unnaturally" suppressed by two powers of the inverse
heavy quark mass, not one power as one would expect
{\em apriori}.  The observation of the dimension gap was
first made in Ref. \cite{chay} in the context of HQET; it is crucial in
phenomenological applications.

Let us return now to the construction of the $1/m_Q$ expansion,
and consider the term with $n=2$ in the sum (\ref{S1}).  One of two
factors  $(2m_Q\pi_0 +\pi^2)$ can be applied to the right, the other
one to the left. The difference between $\pi$ applied to the right and
to the left is a total derivative which vanishes anyway in the forward
matrix element $\langle H_Q|...|H_Q\rangle$. This simple observation
implies
that  the combination $(2m_Q\pi_0 +\pi^2)$ in the numerator again
vanishes by virtue of the equation of motion and we are left with
\beq
\langle H_Q|{\hat T}^{(2)}|H_Q\rangle = \frac{4}{3}{\vec q}^2
\frac{1}{(m_q^2-k^2)^3}
\langle {\vec\pi}^2\rangle \, ,
\label{S4}
\eeq
where I have singled out and retained only the spin-0 part of
the operator $\bar Q\pi_i\pi_j Q \rightarrow
(1/3)\delta_{ij}\bar Q{\vec\pi}^2 Q$ for the reasons explained above.

We are almost done. The imaginary parts of
$\hat T^{(1)}$ and $\hat T^{(2)}$ are expressible in terms of  the
first and  second derivatives of the delta function, and
after
some simple algebra it is not difficult to get
$$
\frac{1}{\pi}{\rm Im}\,{\langle\hat T\rangle}
=\left( \frac{\langle\bar QQ\rangle }{2m_Q}
-\frac{\langle\bar Q {\vec\pi}^2 Q\rangle}{12m_Q^3}\right)
\delta (E-E_0 )\; -
$$
\begin{equation}
-\;\frac{E_0 \langle\bar Q {\vec\pi}^2 Q\rangle}{12m_Q^3}\delta '(E-
E_0) +
\frac{E_0^2 \langle\bar Q {\vec\pi}^2 Q\rangle}{12m_Q^3}\delta ''(E-
E_0) + ...
\label{ImT}
\end{equation}
where operators of higher dimension are ignored;
I have taken into account that $q_0 = E$ and ${\vec q}^2= E^2$
and played a little with the delta functions.

The expansion of ${\rm Im}\, {\hat T}$
into local operators  generates more and more
singular terms at the point where the $\phi$ spectrum would be
concentrated in the free quark approximation. You should not be
surprised by this circumstance which will be elucidated in every
detail in due time. What is important is that the physical spectrum is
a smooth function of  $E$. One
could derive a smooth spectrum by summing up the  infinite set of
operators in Eq. (\ref{S1}) -- this will be the subject of
Lecture 4.
There is
no need to carry out this summation now, however, since we are
interested only in the
integral characteristics of the type of the total probability. As
far
as such integral characteristics  are concerned, the expansion in Eq.
(\ref{ImT})
is perfectly legitimate.

At first, we calculate the total width by substituting eq.
(\ref{ImT})
into eq. (\ref{spectrum}) and integrating over $E$,
\begin{equation}
\Gamma = \int dE \frac{d\Gamma}{dE}=\Gamma_0
\frac{m_Q}{M_{H_Q}}\langle\bar QQ\rangle
\label{tgamma}
\end{equation}
where the integration runs from 0 to the physical boundary
$E_0^{phys}$, expressed in terms of the hadron masses
\begin{equation}
E_0^{phys} = \frac{M_{H_Q}^2 -M_{H_q}^2}{2M_{H_Q}}\, .
\label{E0phys}
\end{equation}
The power correction
proportional to
$\langle\bar Q {\vec\pi}^2 Q\rangle /m_Q^2$ which might have
appeared cancels in the total width! Is this cancellation
unexpected? No, we could have anticipated it on general grounds.
Indeed, the total width $\Gamma$ is a Lorentz scalar, and, quite
naturally, the $1/m_Q$ expansion for this quantity must run over
the Lorentz scalar operators;
$\bar QQ$ is Lorentz scalar while $\bar Q {\vec\pi}^2 Q$ is not.
The fact that there are no explicit $1/m_Q$ corrections in Eq.
(\ref{tgamma}) does not mean that they are absent in
$\Gamma$ at all. They could appear through
$\langle\bar Q  Q\rangle m_Q/M_{H_Q}$. Hence,
our next task is to find the expansion for $\langle\bar Q Q\rangle $
in the toy model at hand. To solve the problem we will use the very
same idea as in Sect. 1.5;
the only difference is the form of the heavy quark current.
For the scalar quarks the
current whose diagonal
matrix
element counts the number of quarks is
$\bar Q \, i {\stackrel{\leftrightarrow}{D}}_\mu Q$. Hence, in the rest
frame of
$H_Q$
we have:
\beq
\frac{1}{2M_{H_Q}} \langle H_Q|\bar Q \,i
{\stackrel{\leftrightarrow}{D}}_0 Q
|H_Q\rangle= 1 \; \; .
\label{shqc}
\eeq
Passing to the rescaled fields
we arrive at the  relation
$$
1= \frac{1}{M_{H_Q}} \langle H_Q|m_Q\bar Q Q +\bar Q \pi_0 Q|H_Q
\rangle =
$$
\beq
 \frac{m_Q}{M_{H_Q}} \langle H_Q|\bar Q Q|H_Q\rangle
+\frac{1}{2M_{H_Q}m_Q}
\langle H_Q|\bar Q {\vec\pi}^2Q|H_Q \rangle,
\label{shqc1}
\eeq
where the second line is due to the equation of motion. Equation
(\ref{shqc1})
leads us to the result sought for,
\begin{equation}
 \frac{m_Q}{M_{H_Q}}\langle H_Q|\bar Q Q|H_Q \rangle =
\left( 1
-\frac{\mu_\pi^2}{2m_Q^2}   + ... \right) ;
\label{QQ1}
\end{equation}
I have introduced here  the standard
notation for the expectation value of ${\vec\pi}^2$,
\begin{equation}
\left(\mu_\pi^2 \right)_{\rm toy\,\,\,  model} =
\frac{1}{2M_{H_Q}}\langle H_Q|2m_Q\bar
Q{\vec\pi}^2Q|H_Q\rangle
\;\;.
\label{mupi}
\end{equation}
As was already mentioned, $\mu_\pi^2$ is a crucial parameter of
the heavy quark theory. Its definition in QCD will be slightly
different from that in our toy model, but the physical meaning
will be the same.

Plugging in Eq. (\ref{QQ1}) in (\ref{tgamma}) we finally arrive at the
desired expression,
\beq
\Gamma = \Gamma_0\left( 1-\frac{\mu_\pi^2}{2m_Q^2}\right)\ .
\label{theorem}
\eeq
The inclusive width coincides with
the parton-model result up to  terms of order $1/m_Q^2$.
There is no correction of order $1/m_Q$! Moreover, the term
$1/m_Q^2$ is calculable and
its physical meaning  is quite
transparent: it reflects the time dilation for the moving quark
inside the heavy hadron at rest
(the Doppler effect). The
coefficient $(-1/2)$ in front of
$\langle{\vec\pi}^2\rangle /m_Q^2$ could, therefore,
have been guessed from the very
beginning, without explicit calculations,  were we a little bit smarter.

This situation is quite general, it takes place not only in the toy
model at hand but in  real QCD as well.
The absence of the correction of order $1/m_Q$ in the total
inclusive
widths (say, the semileptonic width of the $B$ mesons,
or $\Gamma (B\rightarrow X_s\gamma )$, and so on)  is called the
CGG/BUV  theorem  \cite{chay,BUV}.

I hasten to add, though, that the
 absence of the $1/m_Q$ correction
is {\em not} merely a consequence of the dimension gap in the
set of the relevant operators, as it is sometimes stated in the
literature. Indeed, let me give a counterexample.
Let us calculate the average energy of the $\phi$ particle, or, more
exactly,
the first moment of the spectrum,
\begin{equation}
I_1 =\int_0^{E_0^{phys}} dE \, (E_0^{phys} - E)\,
\frac{1}{\Gamma_0}
\frac{d\Gamma}{dE}\, .
\label{I1def}
\end{equation}
In the parton model the $\phi$ spectrum is a pure delta function
and,
consequently, $I_1$ vanishes. The heavy quark expansion does
generate a non-vanishing result, a $1/m_Q$ effect. To see that this is
indeed the case we integrate the theoretical spectrum (\ref{ImT})
which yields us
\begin{equation}
I_1 = \Delta - \frac{\mu_\pi^2E_0^{phys}}{2m_Q^2}
\label{I1res}
\end{equation}
where $\Delta$ is defined as follows
\begin{equation}
\Delta = E_0^{phys} - E_0 ,
\label{delta}
\end{equation}
and the parameters $E_0$ and $E_0^{phys}$ are given in   Eqs.
(\ref{E0}) and (\ref{E0phys}). Now, invoking what we have already
learnt in
Sect. 1.5 about the heavy hadron masses we find that
\beq
\Delta = \frac{1}{2}
v_0^2\bar\Lambda +{\cal O}(\Lambda_{\rm QCD}^2)
\eeq
where
$$
v_0= \frac{M_{H_Q}-M_{H_q}}{M_{H_Q}} \, .
$$
In the SV limit $v_0$ is small and coincides with the velocity
of the final hadron produced in the transition
$Q\rightarrow q\phi$.

With all these definitions
\beq
I_1 = \frac{1}{2}
v_0^2\bar\Lambda +{\cal O}(\Lambda_{\rm QCD}^2)
\label{I1e}
\eeq
i.e. the preasymptotic correction in the first moment is of the order
of
$\Lambda_{\rm QCD}$, not $\Lambda_{\rm QCD}^2$. The reason for
the occurrence of a ``wrong" power of the QCD parameter is that
the leading correction term in the $1/m_Q$ expansion  in this
particular quantity
is unrelated to any local operator. As we will see later on such a
situation is not rare in the heavy quark theory.
The sum rule   (\ref{I1e}) is
just a version of Voloshin's optical sum rule \cite{Volopt}, while that
of Eq. (\ref{theorem}) can be interpreted in terms of the  Bjorken
sum rule \cite{Bj}. We will dwell on the both sum rules in the real
QCD in Sect. 3.6.

I apologize for this little waterfall of new letters and definitions
and hope that a simple picture behind our results is not
overshadowed.
Notice that in the SV limit $E_0^{phys} - E$ reduces to the
excitation energy of the final hadron produced in the decay. The
factor $E_0^{phys} - E$ in the integrand
eliminates the ``elastic" peak, so
that the integral is saturated only by the inelastic contributions.
Say, in the $b\rightarrow c\phi$ transition the contribution
of $B\rightarrow D\phi$ is eliminated, only the excited $D$ mesons
survive
in the first moment. Since the excitation energies are of order of
$\Lambda_{\rm QCD}$, or $\bar\Lambda$, the prediction (\ref{I1e})
means that the probabilities of the inelastic transitions
$B\rightarrow$ excited $D$'s are all proportional to $v^2$.
This is in full agreement with the theorem \cite{VS} discussed in
Sect. 1.3 -- that in the point
of  zero recoil the only transition that can occur is the elastic
$B\rightarrow D$ transition, with the unit probability. Away from
the point of the zero recoil (but in the SV limit) the inelastic
transitions are generated. However, Eq. (\ref{theorem}) shows,
that up to small corrections ${\cal O}(\Lambda_{\rm QCD}^2/m_Q^2)$
which can be neglected if we are interested only in the linear in
$\Lambda_{\rm QCD}$ effects, the total probability remains unity.
In other words, the total probability is just reshuffled:
a small $v^2$ part is taken away from the elastic transition
and is given to the inelastic transitions. The QCD analog of this
assertion is the essence of the Bjorken sum rule \cite{Bj}.

It is quite evident that the series of such sum rules can readily be
continued further. For the
next moment, for instance, we get
\begin{equation}
I_2 =\int_0^{E_0^{phys}} dE \, (E_0^{phys} - E)^2
\,\frac{1}{\Gamma_0}\frac{d\Gamma}{dE}\;
= \;\Delta^2
+\frac{\mu_\pi^2E_0^2}{3m_Q^2}\;\;.
\label{I2}
\end{equation}
Analyzing this sum rule in the SV limit one obtains, in principle,
additional information, not included in the results of Refs.
\cite{VS,Volopt,Bj}. It is worth emphasizing that in Eqs.
(\ref{tgamma}), (\ref{I1res}) and (\ref{I2}) we have collected all
terms
through order $\Lambda_{\rm QCD}^2$, whereas
those of order $\Lambda_{\rm QCD}^3$ are systematically
omitted. Predictions for higher moments would require
calculating terms ${\cal O}(\Lambda_{\rm QCD}^3)$ and higher.

Concluding this part let me suggest to you an exercise which will
show
whether the technology introduced above is well understood by you.
Try to repeat in  real QCD, with the quark spins switched on,
everything we have done in the toy model. Of particular interest to
us will be the transition operator
\begin{equation}
\hat T = i\int d^4 x \,{\rm e}^{-iqx} T\{ \bar Q(x)\Gamma_\mu q(x) \,
, \,\bar
q (0)\Gamma_\nu
Q (0)\} .
\label{transex}
\end{equation}
where $\Gamma_\mu$ is either $\gamma_\mu$ or
$\gamma_\mu\gamma_5$. This transition operator is relevant for
the semileptonic $b$ to $c$ decays. To facilitate the task consider
special kinematics: (i) zero recoil (the vanishing spatial momentum of
the lepton
pair,
$\vec q = 0$; (ii) small velocity limit ${\vec q}^2\ll m_Q^2$. To
further facilitate the task limit yourself to the spatial components
of $\Gamma_\mu$.
If you still have problems go over this lecture again and consult the
original works \cite{koyrakh,k2,k3}. The full answer for the
transition
operator (\ref{transex}) is given, for instance, in Appendix of Ref.
\cite{koyrakh}.

\subsection{The Fock-Schwinger gauge}

In some situations (especially when one deals with  massless quarks)
a variant of the
background field technique based on the so called Fock-Schwinger
gauge
for the external filed turns out to very  efficient (for a
review and extensive list of references see \cite{NSVZ}). The gauge
condition on the background gluon  field has the form
\begin{equation}
x_\mu A_\mu (x) = 0 .
\label{26}
\end{equation}
What is remarkable in this condition is that in
this gauge the gauge four-potential can be
represented as an expansion which runs only over the gauge
covariant quantities, the gluon field strength tensor and its covariant
derivatives,
\begin{equation}
A_\mu (x) = \frac{1}{2}x_\rho G_{\rho\mu}(0) +\frac{1}{3}
\frac{1}{1!}x_\alpha x_{\rho} D_\alpha G_{\rho\mu}(0) +...\, .
\label{27}
\end{equation}
This expression implies, in particular, that $A(0) = 0$. It is worth
noting that the gauge condition (\ref{26}) singles out the origin and,
hence, breaks the translational invariance. The latter is restored only
in the final answer for the gauge invariant amplitudes.

It is rather easy to show (see \cite{NSVZ}) that the
massless quark Green function in the coordinate space is
\begin{equation}
S(x,0) =\frac{1}{2\pi^2}\frac{\not\!\! x}{x^4}
-\frac{1}{8\pi^2}\frac{x_\alpha}{x^2}\tilde{G}_{\alpha\phi}
\gamma_\phi\gamma_5 + ... , \,\,\,\, \tilde{G}_{\alpha\phi}
=\frac{1}{2}
\epsilon_{\alpha\phi\mu\nu}G_{\mu\nu} .
\label{28}
\end{equation}
One can also  construct a similar  expansion for the Green function in
the momentum space $S(q)$.

If the quark is not massless, $m_q\neq 0$, the expansion of the
Green function in the background field becomes much more
cumbersome. Although we will hardly need it in full I quote it here
for the
sake of completeness,
$$
S(x,0) =\frac{1}{2\pi^2}\frac{\not\!\! x}{x^4}\{ -\frac{1}{2}m^2x^2K_2
(m\sqrt{-x^2})\}-
\frac{1}{8\pi^2}\frac{x_\alpha}{x^2}\tilde{G}_{\alpha\phi}
\gamma_\phi\gamma_5\left\{ \frac{-x^2mK_1(m\sqrt{-
x^2})}{\sqrt{-x^2}}\right\}
$$
\begin{equation}
-\frac{im^2}{4\pi^2}\frac{K_1(m\sqrt{-x^2})}{\sqrt{-x^2}}
\frac{m}{16\pi^2}G_{\rho\lambda}\sigma^{\rho\lambda}
K_0(m\sqrt{-x^2}) +...\, .
\label{29}
\end{equation}
Here $K$ is the McDonald function. This result was obtained in Ref.
\cite{BB}. Further education on the Fock-Schwinger gauge technique
can be obtained from Ref. \cite{NSVZ}.
The best way to master those aspects which are most common in the
heavy quark
theory is merely to play with this tool kit. Let us see how it works
in the  calculation  of
the  $1/m_Q^2$  correction in
the total probability of the semileptonic decay of the heavy quark in
the real QCD.  Unlike Sect. 2.1 we will address directly  the total
width bypassing the stage of  the spectrum. The first
calculation of the power correction in
$\Gamma (B\rightarrow
X_c l\nu )$ along these lines was carried out in Ref. \cite{BUV}
(see also \cite{BSUV1}).

\subsection{The $1/m_Q$ corrections to the semileptonic inclusive
width in QCD}

In this section I will describe probably the most elegant  application
of
the ideas developed above -- calculation of the leading correction in
the total semileptonic widths. In the toy model considered in
the previous section it was established that the $1/m_Q$ correction
was absent, and the first non-trivial correction $1/m_Q^2$ was
associated with
the matrix element of $\langle \bar QQ\rangle$. As a matter
of fact, through the heavy quark expansion, we managed to express
it in terms of the matrix element $\langle \bar
Q{\vec\pi}^2Q\rangle$.

When we take the quark spins into account a new dimension-5
Lorentz scalar operator appears,
$\langle \bar Q (i/2) \sigma GQ\rangle$. On general grounds one may
expect that
the main lesson abstracted from the toy model -- the absence of the
$1/m_Q$ term -- persists but the $1/m_Q^2$ correction will receive a
contribution from the operator $\bar Q (i/2) \sigma GQ$.
The conclusion will be confirmed by the analysis presented below.
You will see how efficiently the Fock-Schwinger technique is in this
case.

Thus,  let us proceed to calculation of  semileptonic widths. The
final
quark mass $m_q$ is arbitrary -- we do not assume the SV limit
now, nor is
any other constraint  imposed on $m_q$. The
weak Lagrangian responsible for the semileptonic decays has the
following generic form
\begin{equation}
{\cal L} =\frac{G_F}{\sqrt{2}}V_{Qq} (\bar q\Gamma_\mu Q)(\bar l
\Gamma_\mu \nu )\, , \,\,\,  \,  \Gamma_\mu = \gamma_\mu
(1+\gamma_5)\, ,
\label{74}
\end{equation}
where $l$ is a charged lepton, electron for definiteness. The mass of
the charged lepton will be neglected. Moreover, $G_F$ and $V_{Qq}$
are constants irrelevant for our purposes.

As usual, at the first stage we construct the transition operator
$\hat T(Q\rightarrow X\rightarrow Q)$,
\begin{equation}
\hat T = i\int d^4x T\{ {\cal L}(x) {\cal L}(0)\} =
\sum_i C_i {\cal O}_i
\label{75}
\end{equation}
 describing a diagonal amplitude
with the heavy quark $Q$ in the initial and final state (with identical
momenta). The lowest-dimension operator in the expansion of
$\hat T(Q\rightarrow X\rightarrow Q)$ is $\bar Q  Q$, and the
complete
perturbative prediction -- the spectator
model -- corresponds to the perturbative calculation of the
coefficient of this operator. For the time being we are not interested
in perturbative calculations. Our task is the analysis of the influence
of the soft modes in the  gluon field manifesting
themselves  as a series of higher-dimension operators in $\hat T$.

At the second stage we average $\hat T$ over the hadronic state of
interest, say, $B$ mesons. At this stage the non-perturbative
large distance dynamics enters through matrix elements of the
operators of dimension  5 and higher.

Finally, the imaginary part of $\langle H_Q|T|H_Q\rangle $ presents
the $H_Q$
semileptonic width sought for,
\begin{equation}
\Gamma = \frac{1}{M_{H_Q}}{\rm Im }\langle H_Q|T|H_Q\rangle\,  .
\label{76}
\end{equation}

The diagram determining the transition operator is depicted on Fig.
2.
The lepton
propagators are, of course, free -- they  do not feel
the background gluon field.  Thick lines refer to the initial quark
$Q$.  Although the gluon field is not shown one
should understand that the lines corresponding to $Q$ and $q$ are
submerged into a soft gluon background.

In  the Fock-Schwinger gauge the line corresponding to the final
quark $q$ (Fig. 2)
remains free, and the only source of the dimension-5 operators is the
external line corresponding to $ Q$ (or $\bar Q$).  Let us elaborate
this point in more detail.

If we do not target corrections higher than $1/m_Q^2$  it is sufficient
to use the expression for
the quark Green function given in Eq. (\ref{28}) or (\ref{29}).  The
particular form
is
absolutely inessential; the only important point is the chiral structure
of the vertices in the weak Lagrangian and the fact that the leptons
are massless.

The currents in the weak Lagrangian (\ref{74}) are left-handed.
Therefore,
the
Green function of the quark $q$ is sandwiched between
$\Gamma_\mu$ and $\Gamma_\nu$.  This means that
$1+\gamma_5$
projectors annihilate the part of the Green function with the even
number of the $\gamma$ matrices. Then the only potential
contribution
is associated with the first line in eq. (\ref{29}).

The non-perturbative term is the one containing ${\tilde
G}_{\alpha\phi}$.  This term vanishes, however,  after convoluting it
with the
lepton part. Indeed, the lepton loop (with the massless leptons) has
the form
\begin{equation}
L_{\mu\nu} = -\frac{2}{\pi^4}\frac{1}{x^8}(2x_\mu x_\nu - x^2
g_{\mu\nu}).
\label{77}
\end{equation}
Here I  take  the product of two massless fermion propagators in
the coordinate space, with the appropriate $\gamma$ matrices
inserted, and do the trace.
Actually we need to know only the last bracket.  Now, convoluting it
with the $\tilde G$ term from the quark Green function we get
$$
(\Gamma_\mu x_\alpha {\tilde G}_{\alpha\phi}\gamma_\phi
\Gamma_\nu )(2x_\mu x_\nu - x^2 g_{\mu\nu})\equiv 0,
$$
q.e.d. \cite{BUV, BS}.

Thus, if one uses the Fock-Schwinger gauge the only source for the
$1/m_Q$  correction in the total semileptonic widths (at the level up
to $1/m_Q^2$) is through the
equations of motion for $Q$. Here is how it works.

The expression for the amplitude corresponding to the diagram of
Fig. 2
can be generically written as follows:
\begin{equation}
{\cal A} = \int d^4x\bar Q (0) F(x) Q(x),
\label{78}
\end{equation}
where a function $F(x)$ incorporates the lepton loop and the $q$
quark Green function. It may
include Lorentz and color matrices, etc. Now, let us single out the
large, mechanical part of the motion of the heavy quark,
$$
Q(x) = {\rm e}^{-iP_0 x}{\tilde Q}(x),
$$
Then
$$
{\cal A} = \int d^4x \left\{ {\bar{\tilde  Q}} (0) F(x)
{\rm e}^{-iP_0 x}[{\tilde Q}(0)
+x_\mu \partial_\mu {\tilde Q}(0) +\frac{1}{2}x_\mu x_\nu
\partial_\mu \partial_\nu {\tilde Q}(0) + ...]\right\} =
$$
$$
\left\{ {\bar{\tilde  Q}} (0)[\tilde F (P_0){\tilde Q}(0)
+ i \frac{\partial}{\partial P_{0\mu}}\tilde F (P_0) \partial_\mu
{\tilde Q}(0)
\right.
$$
$$
\left. +i^2 \frac{1}{2}\frac{\partial^2}{\partial P_{0\mu}\partial
P_{0\nu}}
\tilde F (P_0)\partial_\mu \partial_\nu {\tilde Q}(0) + ...]\right\} =
$$
\begin{equation}
{\bar{\tilde  Q}} (0)\tilde F (P_0+i\partial ){\tilde Q}(0)
=\bar Q (0) \tilde F(i\partial ) Q(0),
\label{80}
\end{equation}
where $\tilde F $ is the Fourier transform of $F(x)$.

Our next goal is to convert $i\partial$ in the covariant derivative and
then use the equation of motion, $i\not\!\! D Q = m_Q Q$. More
exactly, we start from the expressions of the form
\begin{equation}
\bar Q (0) \not\! p (p^2)^k Q(0),\,\,\, p_\mu = i\partial_\mu ,
\label{81}
\end{equation}
rewrite $p_\mu$ in terms of ${\cal P}_\mu = iD_\mu$ plus terms
with the gluon field strength tensor (in the Fock-Schwinger gauge)
and then substitute $\not\!\!{\cal P}$ acting on $Q$ by $m_Q$.
Expressions (\ref{81}) appear in Im$\hat T$. If the final $q$ quark is
massless, $m_q = 0$, the only relevant power is $k=2$. Switching
on the quark mass, $m_q\neq 0$, brings in other values of $k$ as
well. (Warning: in the procedure sketched above all operators $p$ in
Eq. (\ref{81}) should be considered as acting either only to the right
or only to the left. I will assume they act to the right. We can not
make some of them act to the right and others to the left and neglect
full derivatives. Question: do you understand why?)

Since we focus now on $\bar Q\sigma G Q$ it is sufficient to keep
only the
terms linear in the gluon field strength tensor; the terms with
derivatives
of $G_{\mu\nu}$ are to be neglected as well. In this approximation in
the Fock-Schwinger gauge $A_\mu =(1/2)x_\rho G_{\rho\alpha}$.

Furthermore,
\begin{equation}
p^2  = {\cal P}^2 - 2Ap
\label{82}
\end{equation}
where we neglected the terms quadratic in $A$ and used the fact
that $[p_\mu , A_\mu] =0$.  Equation  (\ref{82}) should be
substituted in
Eq. (\ref{81}),
and then we start transposing $Ap$ trying to put it in the left-most
position, next to $\bar Q (0)$.  If $Ap$ appears in this position the
result is zero since $A(0) =0$. Notice that
$$
[p^2, Ap]\propto G_{\alpha\beta}p_\alpha p_\beta = 0,
$$
so, one can freely transpose $Ap$ through $p^2$. In this way we
arrive at
\begin{equation}
\bar Q (0) \not\!\! p (p^2)^k Q(0) = \bar Q (0) \left[ \not\!\! {\cal P}
({\cal P}^2)^k -2k\not\!\! {\cal P} (A{\cal P})({\cal P}^2)^{k-1}
\right] Q(0) \, .
\label{83}
\end{equation}
Moreover, with our accuracy the last term reduces to
$$
\gamma_\alpha [{\cal P}_\alpha , A_\mu ]{\cal P}_\mu
({\cal P}^2)^{k-1}=
\gamma_\alpha\frac{i}{2}G_{\alpha\mu}{\cal P}_\mu
({\cal P}^2)^{k-1}
$$
\begin{equation}
=-\frac{i}{8} (\not\!\! {\cal P} \sigma G - \sigma G \not\!\! {\cal P})
{\not\!\! {\cal P}}^{2k-2}
\label{84}
\end{equation}
and, hence,  using the equations of motion we conclude that
\begin{equation}
\bar Q (0) \not\!\! {\cal P} (A{\cal P})({\cal P}^2)^{k-1}Q(0) = 0.
\label{85}
\end{equation}
As a result, the $\sigma G$ terms emerge only due to the fact that
$$
{\cal P}^2 = {\not\!\! {\cal P}}^2 -\frac{i}{2}\sigma G,
$$
and the final expression is as follows:
\begin{equation}
\bar Q (0) \not\!\! p (p^2)^k Q(0)\rightarrow
m_Q^{2k+1}\bar Q (0) Q(0) -\frac{ik}{2}\bar Q (0)\sigma G Q(0)
m_Q^{2k-1}.
\label{86}
\end{equation}

After these explanatory remarks  the procedure of
calculating the leading $1/m_Q^2$ correction  in the total
semileptonic
width should be perfectly clear. Let us summarize it in the form of a
prescription.

(i) Calculate the semileptonic width in the parton model. The result
has the form
\begin{equation}
\frac{G_F^2|V_{Qq}|^2}{192\pi^3} m_Q^5 F_0(\frac{m_q^2}{m_Q^2})
\label{87}
\end{equation}
where $F_0(m_q^2/m_Q^2) $ is the phase space factor, a function of
the ratio $m_q^2/m_Q^2$ well-known in the literature
(see Eq. (\ref{89}) below).

(ii) Then construct the expression for $\Gamma$ including the
${\cal O}(m_Q^{-2})$
corrections in the following way \cite{BUV}:
$$
\Gamma  = \frac{G_F^2|V_{Qq}|^2}{192\pi^3} m_Q^5
\left\{ \frac{1}{2M_{H_Q}}\langle
H_Q|\bar QQ|H_Q\rangle  F_0(\rho )
\right.
$$
\begin{equation}
\left.
+\frac{\mu_G^2}{m_Q^2} \,
(\rho\frac{d}{d\rho} -2)F_0(\rho )\right\}
\label{88}
\end{equation}
where
\beq
\mu_G^2 = \frac{1}{2M_{H_Q}} {\langle H_Q|\bar Q (i/2)\sigma G
Q|H_Q\rangle}
\label{muG}
\eeq
and
$$
\rho = \frac{m_q^2}{m_Q^2} \, .
$$

A few comments are in order concerning this beautiful expression for
the total semileptonic width.  The expansion contains two Lorentz
scalar operators, $\bar QQ $ and $\bar Q (i/2)\sigma G Q$,
of dimension 3 and 5, respectively. The fact that only the Lorentz
scalars contribute is obvious since $\Gamma$ is a Lorentz invariant
quantity. We observe here the very same gap in dimensions
mentioned previously in the context of the toy model -- there is no
operator of dimension 4 \cite{chay}. The only element still needed to
complete the derivation is the matrix element $\langle \bar
QQ\rangle$. Fortunately, the corresponding heavy quark expansion
has been already built, see  Eq. (\ref{QQQQ}).

Borrowing the explicit expression for $F_0 (\rho )$ from textbooks (it
is
singled out below in the braces) and assembling all other pieces
together we finally get
$$
\Gamma =\frac{G_F^2|V_{Qq}|^2}{192\, \pi^3} m_Q^5\times
$$
\begin{equation}
\left[ \left( 1 +\frac{\mu_G^2-\mu_\pi^2}{2m_Q^2}\right)
\left\{1 - 8\rho +
 8\rho^3 - \rho^4 -
     12\rho^2\ln {\rho}\right\}  -
 2\frac{\mu_G^2}{m_Q^2} (1-\rho )^4 \right]\, ,
\label{89}
\end{equation}
where $\mu_\pi^2$ is defined in QCD as follows
\beq
\mu_\pi^2 = \frac{1}{2M_{H_Q}}
\langle H_Q|\bar Q {\vec\pi}^2
Q|H_Q\rangle \, .
\label{mupid}
\eeq
This result   is due to Bigi {\em et al.} \cite{BUV}. The
absence of the $1/m_Q$ correction is a manifestation of the
CGG/BUV theorem.

If the mass of the final charged lepton is
non-negligible, the property of no soft gluon emission from the $q$
quark
line (in the Fock-Schwinger gauge) is lost. The expansion for
$\Gamma (B\rightarrow X_c \tau\nu_\tau)$ in this case is much
more cumbersome; it was
constructed in Ref.
\cite{koyr1}.

Dimension-5 operators are responsible for the {\em leading}
non-perturbative corrections in the total semileptonic widths. To
assess
the convergence of the expansion it may be instructive to have an
idea of the higher order terms in the expansion.
The $1/m_Q^3$ terms were estimated in Ref. \cite{BDS}. This is a
rather
messy and time-consuming analysis, and it is hardly in order to
comment on it in this lecture. Surprising though it is, from
what we
already know it is practically trivial to find the coefficient of the
dimension-7 operator
\begin{equation}
\bar Q  G_{\mu\nu}G_{\mu\nu} Q ,
\label{96}
\end{equation}
with two gluon field strength tensors fully contacted over the
Lorentz indices for massless final quarks (say, $b\rightarrow u$
transition).  Of course, this is a purely academic exercise, for many
reasons. In particular, because it is only one of a rather large number
of
dimension-7 operators. Since we do not know their matrix elements
anyhow, it seems to be meaningless to carry out full classification
and calculate all coefficients. The operator (\ref{96}) is chosen since
one can
at least use factorization for a rough estimate of the corresponding
matrix element and since we get its coefficient essentially for free.

The point is that the massless quark propagator (in the
Fock-Scwinger gauge) does not contain
$G_{\mu\nu}G_{\mu\nu}$ term at all (see ref. \cite{shur1}).  This
fact implies
that the only source of the operator (\ref{96}) is the same as in the
case of
$\bar Q\sigma GQ$. It is not difficult to get that
$$
\bar Q(0) \not\!\! p p^4 Q(0) = \bar Q(0)\not\!\!{\cal P} {\cal P}^4
Q(0) + \frac{1}{2} \bar Q(0) G_{\mu\nu}G_{\mu\nu} Q(0) =
$$
\begin{equation}
\bar Q(0){\not\!\!{\cal P}}^5Q(0) +\bar Q(0) G_{\mu\nu}G_{\mu\nu}
Q(0) ,
\label{97}
\end{equation}
where we omitted structures of the type
$[G_{\mu\nu}G_{\mu\alpha}
-(1/4) g_{\nu\alpha}G_{\mu\rho}G_{\mu\rho}]$.

\subsection{Digression}

This section is intended for curious readers -- those who are anxious
to find out where and how else, beyond the theory of the $H_Q$
states,  the background field technique can be used to obtain
interesting predictions.  Here I will discuss an estimate of the mass
splittings between the levels of the highly excited quarkonium states.
This part can be safely omitted
in first reading since it is unrelated to the remainder of these
lectures.

The quarkonium states to be considered below consist of one quark
$Q$, one antiquark $\bar Q$ and the soft gluon cloud connecting
them together.  To begin with we will assume that $m_Q$ is large,
$m_Q\gg\Lambda_{\rm QCD}$ (later on this assumption will be
relaxed). The $Q\bar Q$ mesons can have different quantum
numbers. We will analyze the excited $S$ and $P$ wave states
with the quantum numbers $0^-$ and $0^+$, respectively.
The naive quark model language is used to name  the states; this
does not mean, of course, that we accept any of the dynamical
assumptions of the naive quark model. It would be more accurate to
say that the mesons of interest are produced from the vacuum by
the currents
\beq
J_P = \bar Q i\gamma_5 Q\,\,\, \mbox{and} \,\,\, J_S =\bar QQ\, .
\label{currents}
\eeq

The central object of our analysis is the difference between the
two-point functions in the pseudoscalar and scalar channels. In terms
of
the Green functions in the background field this difference takes the
form (see Fig. 3)
$$
\Pi_P - \Pi_S = i\int dx e^{iqx}\left(
\langle vac|T\{J_P(x)J_P(0)\}|vac\rangle -
\langle vac|T\{J_S(x)J_S(0)\}|vac\rangle \right) =
$$
$$
i\mbox{Tr}\, \left\{
i\gamma_5\frac{1}{\not\!\!{\cal P} +(\not\! q/2) -m_Q}
i\gamma_5\frac{1}{\not\!\!{\cal P} -(\not\! q/2) -m_Q}\,
-\right.
$$
\beq
\left. \frac{1}{\not\!\!{\cal P} +(\not\! q/2) -m_Q}\, \,
\frac{1}{\not\!\!{\cal P} -(\not\! q/2) -m_Q}\right\}\, .
\label{corre}
\eeq
A comment is in order here concerning the trace operation in this
expression. It implies not only the trace over the Lorentz and color
indices, as usual, but  also the trace in the momentum space
substituting $\int d^4p/(2\pi)^4$ in the conventional Feynman
integral.
With the help of  Eq. (\ref{18}) the difference $\Pi_P - \Pi_S $ can be
identically rewritten as
\beq
\Pi_P - \Pi_S =
-2m_Q^2 i \mbox{Tr}\, \left\{\frac{1}{D_+}\, \frac{1}{D_-}
\right\}\,
\label{diftr}
\eeq
where
$$
D_\pm = {\left[ {\cal P} \pm (q/2)\right]^2 -m_Q^2 +(i/2)\sigma G}\, .
$$

We  continue to ignore  hard gluons assuming that
the only role of the gluon field is to provide a soft cementing
background. This is certainly an idealization, but let us see how far
one
can go within the framework of this simplified picture.
Neglecting hard gluons means, in particular, that we will be unable to
analyze the low-lying levels of heavy quarkonium where an essential
role is played by the short-distance Coulomb interaction.
Each
$\sigma G$ insertion in the denominator is of order of
$\Lambda_{\rm QCD}^2$, i.e. does not scale with the external
momentum $q$ when $q$ is large. Let us expand Eq. (\ref{diftr})
in $\sigma G$ and take the trace over the Lorentz indices. Then the
first order term drops out; the first surviving term is bilinear in $G$,
$$
\Pi_P - \Pi_S =
-8m_Q^2 i \mbox{Tr}\, \left\{\frac{1}{D_+}\, \frac{1}{D_-}
\right. +
$$
$$
\frac{1}{2}\frac{1}{D_+}
G_{\alpha\beta}\frac{1}{D_+}
G_{\alpha\beta}\frac{1}{D_+}\,
\frac{1}{D_-}
+ \frac{1}{2}\frac{1}{D_-}
G_{\alpha\beta}\frac{1}{D_-}
G_{\alpha\beta}\frac{1}{D_-}\,
\frac{1}{D_+}
$$
\beq
\left.
\frac{1}{2}\frac{1}{D_+}
G_{\alpha\beta}\frac{1}{D_+}\,
\frac{1}{D_-}G_{\alpha\beta}
\frac{1}{D_-}\right\}+ ...\, .
\label{difsc}
\eeq

A closer look at this expression reveals some peculiar features. First
of all, one can interpret each term as a certain correlation function
in the theory where the quark $Q$ is scalar, not spinor.
Take, for instance, this first line.  It is nothing else than the
two-point function of the $L=0$ quarkonium in the scalar QCD (i.e.
QCD with the scalar quarks; $L$ is the total angular momentum of the
meson). The current producing the scalar
quarkonium from the vacuum in the scalar QCD is $Q^\dagger Q$ (Fig.
4).
The second and the third line, together, represent the four-point
function of the type depicted on Fig. 5. The current denoted by
the dashed line on this figure is $Q^\dagger G_{\alpha\beta} Q$;
the momentum flowing through this line vanishes. Two insertions of
$G$ imply that this four-point function is explicitly proportional to
$\Lambda_{\rm QCD}^4$.

Now, let us examine Eq. (\ref{difsc}) in the complex $q^2$ plane.
At some positive values of $q^2$ the two-point function of Fig. 4
has simple poles corresponding to positions of the $L=0$ quarkonium
levels in the scalar QCD. The four-point function of Fig. 5 has double
and single
poles at the very same values of $q^2$ and single poles at some
other values of $q^2$ corresponding to the production of $L=1$
states in the scalar QCD. The latter are produced
from the $L=0$ states by applying to them the current
$Q^\dagger G_{\alpha\beta} Q$.

On the other hand, the original difference $\Pi_P - \Pi_S$ in real
QCD has only single poles at the positions of the $S$ and $P$ wave
states. These positions are shifted compared to the levels in the
scalar QCD. Expanding in the shift one generates  double poles.
The $L=1$ pole -- the partner to the $P$ wave meson states in
$\Pi_P - \Pi_S $ appears only in the four-point function of Fig. 5.
{}From this figure it is quite clear that the residue of the $L=1$ pole
scales as $\Lambda_{\rm QCD}^4/(\Delta(M^2))^2$ where
$\Delta(M^2)$ is a characteristic  splitting between the $L=0$
and $L=1$ states. On the other hand, the residue of the
$P$ wave meson in $\Pi_P - \Pi_S$, on general grounds, scales
as ${\vec p}^2/M^2$ where $\vec p $ is a characteristic quark
momentum, and I assume that $\Lambda_{\rm QCD}\ll |\vec p|\ll
M$. Equating these two estimates we find that
$$
\Delta M \sim \Lambda_{\rm QCD}^2/|\vec p| \, .
$$
One may observe, with satisfaction, that this is exactly the
characteristic level splitting (between radial or orbital excitations)
for two heavy quarks interacting through a string (``linear
potential"). What is remarkable is that in no place our estimate
invokes any reference  to the linear potential or other models. It was
based only on
some general features of QCD. For me this is a strong evidence that
a string-like picture should take place in QCD, at least, approximately,
for high excitations.

What changes if the quarks are light or even massless,
$m_Q\rightarrow\infty$? The only difference is that now
the residues of the $P$ wave mesons in $\Pi_P - \Pi_S $
are of the same order as those of the $S$ wave mesons
for highly excited states, which implies that
$$
\Lambda_{\rm QCD}^4/(\Delta(M^2))^2\sim 1\, ,
$$
or
$$
\Delta(M^2)\sim \Lambda_{\rm QCD}^2\, .
$$
In other words, we got the linear Regge trajectories, at least for
highly excited states. Moreover, this analysis makes clear a
potentially important point -- the empirical observation  that even
the
lowest states in every channel lie on the linear Regge trajectories
looks like a numerical coincidence and can not be exact.

\newpage
\section{Lecture 3.  Classic Problems with Heavy Quarks}

\renewcommand{\theequation}{3.\arabic{equation}}
\setcounter{equation}{0}

The number of problems  successfully solved within the
heavy quark expansion is quite large.  Even a brief review of the
main
applications is beyond the scope of these lectures.
Some issues, however, are quite general and are important in a
variety of applications. Everybody, not only the heavy quark
practitioners, should know them.  In this lecture we will discuss
several such topics -- the scaling  of the heavy meson coupling
constants,  some properties of the Isgur-Wise function, and, finally,
analysis of  corrections violating the heavy quark symmetry at the
point of zero recoil.  We begin, however, from a systematic
classification of all {\em local} operators which appear in the heavy
quark expansion up to the level ${\cal O}(m_Q^{-3})$. Some terms of
order $1/m_Q^3$ in particular heavy quark expansions are actually
not expressible in terms of the local operators and are, rather,
related
to non-local correlation functions. A full classification of such
correction also exists \cite{Mannel,BSUV}, but we will not go into
details only marginally mentioning them here and there. The
interested reader is referred to the original publications
\cite{Mannel,BSUV}.

\subsection{Catalogue of relevant operators}

The local operators in the heavy quark expansion are bilinear in the
heavy quark field. They are certainly gauge invariant, and in many
instances, when the expansion is built for scalar quantities, the
operators must be Lorentz scalars. As in any operator product
expansion in QCD they can be ordered according to their dimension.
We will limit ourselves here to dimension 6 and lower.
This leaves us with quite a few possibilities listed below. We start
with the Lorentz scalar operators. The only appropriate operators are
\beq
\bar QQ\, ,\,\,\,  \frac {i}{2} \bar Q\sigma_{\mu\nu}G_{\mu\nu}Q
\, , \,\,\, \mbox{and} \,\,\, \bar Q \Gamma Q \bar q \Gamma q
\label{list}
\eeq
where $\Gamma$ stands here for a combination of $\gamma$ and
color matrices. All other structures that might come to one's mind
reduce to those listed above and full derivatives by virtue of
the equations of motion.

(Exercise: prove that this is the case, for instance, for the operators
$\bar Q D^2Q$ and $ \bar QG_{\mu\nu} \gamma_\mu D_\nu Q $.)

(i) The only operator of dimension 3 is $\bar Q Q$. This operator is
related to the heavy quark current $\bar Q\gamma_0 Q$
plus terms suppressed by powers of $1/m_Q$. The leading term of
this expansion has been already discussed, see Eq. (\ref{QQQQ}).
Actually it is not difficult to continue this expansion one step further.
The following relation is {\em exact}:
$$
\bar Q\,Q\;=\;\bar Q\gamma_0Q\;+\;2\bar Q\left(\frac{1-
\gamma_0}{2}\right)^2 Q
\;=\;\bar Q\gamma_0Q\;-\;2\bar
Q\,\frac{\stackrel{\leftarrow}{\not\!\pi}}{2m_Q}\cdot
\frac{\stackrel{\rightarrow}{\not\!\pi}}{2m_Q}\,Q\;=
$$
\begin{equation}
=\;\bar Q\gamma_0Q\;+\;\bar
Q\frac{\not\!\pi^2}{2m_Q^2}
Q\;+\;\mbox{a total derivative }\, ;
\label{5m}
\end{equation}
in the first relation above the operators
$\stackrel{\leftarrow}{\pi}_\mu$
act on the $\bar Q$ field. Keeping in mind that we always  consider
only the forward matrix
elements, with the zero momentum transfer, we can drop all terms
with total
derivatives. Applying now the equations of motion
(\ref{DEt1}) and (\ref{DEt2})
 generates the  $1/m_Q$ expansion for
the scalar density,
$$
\bar Q\,Q\;=\;\bar Q\,\gamma_0\,Q\;
+\;\frac{1}{2 m_Q^2}
\bar Q\,(\pi^2+\frac{i}{2}\sigma G)\,Q\;=\;
\bar Q\,\gamma_0\,Q\;
-\;\frac{1}{2 m_Q^2}
\bar Q\,(\vec\pi\vec\sigma)^2 Q\;-\;
$$
\begin{equation}
-\;\frac{1}{4 m_Q^3}\,\bar Q\left(-(\vec D \vec
E)\,+\,2\vec\sigma\cdot\vec
E\times\vec\pi\right) Q
\;+\;{\cal O}\left(\frac{1}{m_Q^{4}}\right)\;\;.
\label{N9}
\end{equation}
Here $E_i=G_{i0}$ is the chromoelectric field, and its covariant
derivative is
defined as \footnote{Note that in our notations $\vec
D=
-\partial/\partial \vec x\,-\,i\vec A$, therefore $(\vec D\vec E)
=-{\rm div}\, {\bf E}$ in the Abelian case.} $D_jE_k=-i[\pi_j,E_k]$;
we have omitted the term
$\bar Q ([\pi_k,[\pi_0,\pi_i]]-[\pi_i,[\pi_0,\pi_k]])Q$ using the Jacobi
identity.
Moreover,
$$
\vec D \vec E=g^2 t^aJ_0^a
$$
by virtue of the QCD
equation of
motion (here $J_\mu^a=\sum_q \bar q \gamma_\mu t^a q$ is the
color quark
current).  Therefore the first of the
$1/m_Q^3$ terms can be rewritten as a four-fermion
operator.

(ii) As has been already mentioned, no operators of dimension 4
exist.

(iii) Dimension five. There is only one  such operator,
\begin{equation}
{\cal O}_G =\bar Q \frac{i}{2}\sigma_{\mu\nu}G_{\mu\nu} Q ,
\label{40}
\end{equation}
where $\sigma_{\mu\nu} =(1/2)[\gamma_\mu ,\gamma_\nu ]$.
Again, it can be expanded in the powers of  $1/m_Q$,
\begin{equation}
{\cal O}_G = -\bar Q \vec\sigma \vec B Q
-\frac{1}{2m_Q} \bar Q\left(-(\vec D \vec
E)\,+\,2\vec\sigma\cdot\vec
E\times\vec\pi\right) Q
\;+\;{\cal O}\left(\frac{1}{m_Q^{2}}\right).
\label{41}
\end{equation}
where $\vec B$ is the chromomagnetic field, $\vec B =\vec \nabla
\times\vec A = - (1/2)
\epsilon_{ijk}G_{jk}$.

(iv)
Dimension 6 four-quark operators $\sum_i \bar Q\Gamma_i Q
\bar q\Gamma_i q $. Generally speaking, the matrix $\Gamma_i$ can
be any Lorentz matrix  ($1,\gamma_5,\gamma_\mu ,
\gamma_\mu\gamma_5$ or $\sigma_{\mu\nu}$) or any of the
above multiplied by $t^a$.  Of course, in specific problems only a
subset of these matrices may appear.
The four-quark operators  differ by the chiral properties
of the light quark field $q$. Some of them  carry non-zero
chirality
(they are non-singlet with respect to $SU(N_f)_L\times SU(N_f)_R $).
Hence, they do not  show up in the transitions associated with  the
weak currents of the $V-A$ type.

Further remarks will concern operators that are spatial scalars but
not Lorentz scalars. They appear in the low-energy effective
Lagrangian (\ref{HQLm}) and  in the expansions of the type
(\ref{N9}) and
(\ref{41}).  The most important operator from the class
is
\beq
{\cal O}_\pi =\bar Q{\vec\pi}^2 Q\, ,
\label{kinop}
\eeq
which we have already encountered more than once.

Dimension 4 operators are all reducible to those of dimension 5 and
higher. For instance,
$$
\bar Q \vec\gamma \vec\pi  Q = \frac{1}{m_Q} \bar Q ({\vec\pi}^2
- {i \over 2} \sigma G) Q + {\cal O}(m_Q^{-2}),
$$
$$
\bar Q \vec\gamma \vec\pi\gamma_0  Q = {\cal O}(m_Q^{-2}),
$$
$$
\bar Q \pi_0 Q = \frac{1}{2m_Q} \bar Q ({\vec\pi}^2
- {i \over 2} \sigma G) Q + {\cal O}(m_Q^{-2}).
$$

At the level of dimension 6 only one additional operator emerges
(apart from the four-fermion operators), namely,
\beq
\bar Q \vec\sigma\cdot\vec E\times\vec\pi  Q\, .
\label{addop}
\eeq
At first sight it might seem that one could build extra operators of
dimension 6, from the gluon fields, e.g.
$$
\bar Q \pi_iE_i Q \,\,\, {\rm or}\,\,\,
\bar Q\sigma_i\epsilon_{ijk}(D_jE_k)Q.
$$
Actually they are reducible to operators of higher
dimension via the equations of motion. Indeed,
using the fact that
$$
E_i = -i[\pi_0 \pi_i]
$$
one can rewrite
$$
\bar Q \pi_i E_i Q = -i\bar Q\pi_i[\pi_0\pi_i]Q =
-i\bar Q (\pi_i\pi_0\pi_i -\pi_i^2\pi_0)Q =
$$
$$
-i\bar Q([\pi_i\pi_0]\pi_i
+\pi_0\pi_i^2 - \pi_i^2\pi_0)Q
=-i\bar Q([\pi_i\pi_0]\pi_i
+\frac{1}{2m_Q} (\vec\sigma\vec\pi )^2\pi_i^2
-\frac{1}{2m_Q}\pi_i^2 (\vec\sigma\vec\pi )^2 )Q
$$
$$
-\frac{i}{2}\bar Q [\pi_i[\pi_0\pi_i]]Q
+ \,\, {\rm dimension \,\, seven} =
$$
\begin{equation}
=-i\bar Q ({\rm div}\vec E) Q + \,\, {\rm dimension \,\, seven}.
\label{48}
\end{equation}
This is a four-fermion operator.  By the same token,
$$
\bar Q\sigma_i\epsilon_{ijk}(D_jE_k)Q =
\bar Q\sigma_iD_0B_iQ=
$$
\begin{equation}
-i\bar Q\sigma_i[\pi_0B_i]Q=
-\frac{i}{2m_Q}\bar Q [(\vec\sigma\vec\pi )^2,\vec\sigma\vec B]Q
\label{49}
\end{equation}
which is obviously of the next order in $1/m_Q$ (a dimension-seven
operator).

\subsection{Extracting/determining the matrix elements}

Construction of the operator product expansion is only the first step
in any  theoretical analysis. The heavy quark expansion
must be converted into predictions for the physical quantities.
To this end it is necessary to take the matrix elements of the
operators involved in the expansion. The latter carry all information
about the large distance dynamics responsible for the hadronic
structure, in all its peculiarity. These matrix elements in our
QCD-based approach play the same role as the wave functions in the
non-relativistic quark models.

In this section we will summarize
what is
known about the matrix elements of the operators from the list
presented
above.

(i) The most favorable situation takes place  at the level of dimension
three. Indeed, the only Lorentz  scalar operator of dimension 3 is
$\bar QQ$ which has a nice expansion (\ref{N9}).
The operator $\bar Q\gamma_0 Q$ is the time component of the
conserved
current, measuring the number of the quarks $Q$ in $H_Q$.
Therefore, both
for mesons and baryons
\begin{equation}
\frac{1}{2M_H}\langle H_Q|\bar Q\gamma_0 Q|H_Q\rangle = 1 .
\label{54}
\end{equation}
(As usual, we stick to the
rest frame of $H_Q$; in the case of baryons  averaging over
the baryon spin is implied).

(ii) The status of two operators of dimension 5 is different. Let us
consider first ${\cal O}_G$ whose matrix elements are expressible in
terms
of experimentally measurable quantities.

To  leading order in $1/m_Q$ the parameter  $\mu_G^2$ defined
in Eq. (\ref{muG}) reduces to \footnote{
Some authors prefer a different nomenclature \cite{FN}. The
expectation values of the chromomagnetic and kinetic energy
operators are sometimes called $\lambda_2$ and $\lambda_1$.}
\begin{equation}
\mu_G^2 =
\frac{1}{2M_{H_Q}}
\langle H_Q|{\cal O}_G|H_Q\rangle =
-\frac{1}{2M_{H_Q}}
\langle H_Q|\bar Q\vec\sigma\vec B Q |H_Q\rangle \, .
\label{55}
\end{equation}
For pseudoscalar mesons this quantity can be related to the
measured hyperfine
mass splittings. Indeed, $\bar Q\vec\sigma\vec B Q$ is the leading
spin-dependent operator in the
heavy quark Hamiltonian (\ref{hamil}). Hence
\begin{equation}
\mu_G^2 (B_Q) = \frac{3}{4} (M_{B^*}^2-M_B^2)
\label{56}
\end{equation}
where $B^*$ and $B$ are generic notations for the
vector and pseudoscalar mesons, respectively, and the limit
$m_Q\rightarrow\infty$ is implied. Assuming that the $b$ quark
already
belongs to this asymptotic limit one estimates $\mu_G^2$ from
the measured B meson masses,
$$
\mu_G^2\approx 0.35 \,\,\,  \mbox{GeV}^2\, .
$$

Furthermore, in the baryon family four baryons are expected
to decay weakly and are, thus, long-living states:  $\Lambda_Q,\,\,\,
\Sigma_Q,\,\,\, \Xi_Q$
and $\Omega_Q$. In the first three of them the total angular
momentum of the light cloud is zero; hence the chromomagnetic field
has no vector to be aligned with, and
\begin{equation}
\mu_G^2 (\Lambda_Q) = \mu_G^2 (\Sigma_Q) = \mu_G^2 (\Xi_Q)=0.
\label{57}
\end{equation}
In the case of $\Omega_Q$ the total angular momentum of the light
cloud is
1. Hence,
\begin{equation}
\mu_G^2 (\Omega_Q) = \frac{2}{3} \left( M_{\Omega_Q^{3/2}}^2-
M_{\Omega_Q^{1/2}}^2\right)
\label{58}
\end{equation}
where the superscripts 3/2 and 1/2 mark the spin of the baryon.
Although the mass splitting on the right-hand side of Eq. (\ref{58})
is in principle measurable, it is not known at present, and if one
wants
to get an estimate one has to resort to quark models or lattice
calculations. Both approaches are not mature enough at the moment
to give reliable predictions for this quantity and I suggest we wait
until experimental measurements appear.

Let us proceed now to the discussion of the matrix element of the
operator
${\cal O}_\pi$. The physical meaning of this matrix element is the
average
kinetic energy (more exactly, the spatial momentum squared) of the
heavy
quark $Q$ inside $H_Q$. This operator is spin-independent, and it is
much harder to   extract   $\mu_\pi^2$  (the parameter $\mu_\pi^2$
is defined in Eq. (\ref{mupid})) from
phenomenology, although such an extraction is possible, in principle
(see Ref. \cite{BSUV} and Lectures 4 and 5 for details). Since the
phenomenological
analysis has not been carried out yet  one has to rely on
theoretical estimates. Several calculations of $\mu_\pi^2$ within
the QCD sum rules yield \cite{VBE,mupie}
$$
\mu_\pi^2 (B) = \frac{1}{2M_B}\langle B|{\cal O}_\pi |B\rangle =
0.5\pm 0.1 \,\,\, \mbox{GeV}^2 \, .
$$
A remarkable model-independent lower bound on
$\mu_\pi^2 (B)$ exists in the literature,
\beq
\mu_\pi^2 (B) > \mu_G^2 (B) \approx 0.35 \,\,\,  \mbox{GeV}^2
\label{lb}
\eeq
The quantum-mechanical derivation of this inequality due to
Voloshin
(see Ref. \cite{HQETRev}) is straightforward.  Indeed, start from the
square of the Hermitean operator $(\vec\sigma\vec\pi )^2$ and
average it
over the $B$ meson state. It is obvious then that
$\langle (\vec\sigma\vec\pi )^2\rangle >0$. Using the fact that
$(\vec\sigma\vec\pi )^2 ={\vec\pi}^2+\vec\sigma\vec B $ we
immediately arrive at Eq. (\ref{lb}).  A field-theoretic derivation of
the same result can be found in Ref. \cite{BSUV}. It is remarkable
that the inequality   (\ref{lb}) almost saturates the QCD sum rule
estimate quoted above. Another lower bound on $\mu_\pi^2 (B)$,
obtained from a totally different line of reasoning,
is discussed in Sect. 3.6. It turns out to be close to Eq. (\ref{lb})
numerically.

It is plausible that $\mu_\pi^2$ for mesons and baryons is different
--
there is no reason why they should coincide.
The task of estimating $\mu_\pi^2$ for baryons remains open.

These parameters, $\mu_\pi^2$ and $\mu_G^2$, along
with $\bar\Lambda$, are most important in applications.
In most applications one deals with the expectation values
over the $B$ meson state. Therefore, let us agree that
$\mu_\pi^2 ,\, \mu_G^2$, and  $\bar\Lambda$, with no
subscripts or arguments, are defined with respect to the $B$ mesons.
This is a standard convention.
In a few cases when these quantities are defined with respect
to other heavy flavor hadrons we will mark them by the
corresponding subscripts or indicate with parentheses.

(iii) Operators of dimension 6 are studied to a much lesser extent
than
those of
dimension 5. Perhaps, the least favorable is the situation with the
operator
${\cal O}_E$ given in Eq. (\ref{addop}).
Let us
parametrize its matrix element  as follows:
\begin{equation}
\frac{1}{2M_H} \langle H_Q|\bar Q \vec\sigma\times \vec E\vec\pi
Q|H_Q\rangle
=\mu_E^3.
\label{64}
\end{equation}
This operator in the
heavy quark Hamiltonian is responsible for the spin-orbit
interactions and
consequently generates the spin-orbit splittings  between the
masses  of the ground states and the orbital excitations. Hence, in the
non-relativistic limit (non-relativistic with respect to the spectator
light quark)  $\mu_E^3$ vanishes for the
$S$ wave states. Of course, the non-relativistic approximation with
respect to the
light quark is very bad. The estimate of $\mu_E^3$ existing in the
literature \cite{BDS} is so rough that it, probably, does not deserve to
be discussed here.

As for the four-quark operators  the only
method of
estimating their matrix elements which does not rely heavily
on the most primitive (and hence totally unreliable) quark models is
the old
idea of factorization applicable only in mesons but -- alas -- not in
baryons.

First of all let us observe that each of the four-quark operators exists
in two variants differing by the
color flow. One can always rearrange the operators, using the Fierz
identities,
in the form
\begin{equation}
{\cal O}_{4q} = \bar Q\Gamma  q\bar q\Gamma  Q
\label{66}
\end{equation}
and
\begin{equation}
{\tilde{\cal O}}_{4q} = \bar Qt^a\Gamma  q\bar
qt^a\Gamma  Q\, .
\label{67}
\end{equation}
Take for definiteness $\Gamma  =\gamma_\mu
\gamma_5$. (Other $\gamma$ matrices can also appear, of
course.) In
the first
operator color is transferred from the initial heavy to the final  light
quark
and from the initial light to the final heavy quark. The second
operator is
essentially color-exchange.
Now, if we are interested in the matrix elements over the meson
states
we can simply factorize the currents appearing in Eqs. (\ref{66}) and
(\ref{67}) (i.e.
saturate by the vacuum intermediate state),
\begin{equation}
\frac{1}{2M_B}\langle
 B_Q|{\cal O}_{4q}|B_Q\rangle
 =
\frac{1}{2}M_Bf_B^2,\,\,\,
\frac{1}{2M_B}\langle B_Q|{\tilde{\cal O}}_{4q}|B_Q\rangle
 = 0
\label{68}
\end{equation}
where $f_B$ is the pseudoscalar decay constant,
\begin{equation}
\langle 0|\bar Q \gamma_\mu\gamma_5 q |B_Q \rangle = i f_B
p_\mu
\label{69}
\end{equation}
As we will discuss shortly,  in the limit $m_Q\rightarrow\infty$ the
combination
$M_Bf_B^2$ scales as $m_Q^0$ (modulo  logarithmic corrections) so
that the
right-hand side of Eq. (\ref{68})
is the cube of a typical hadronic mass, as it should be.

Factorization in Eq. (\ref{68}) is justified by $1/N_c$ arguments.
Indeed,
corrections to Eq. (\ref{68}) are of the order of $N_c^{-1}$.

Thus, from the whole  set of the four-quark operators we can say
something about the meson expectation values of those operators
which are reducible to
\begin{equation}
(\bar Q \Gamma q)(\bar q\Gamma Q)
\label{70}
\end{equation}
where $\Gamma$ stands for a Lorentz matrix but not for the color
one, and the color
indices
are contracted within each of two brackets separately. Up to
terms ${\cal O}(N_c^{-1})$ two brackets can be factorized.

To get an idea about the numerical value of $(1/2)f_B^2M_B$ we
should
substitute a numerical value for $f_B$ which is not measured so far.
Theoretical ideas about this fundamental constant will be discussed
later. Now let me say only that $(1/2)f_B^2M_B
\sim 0.1 \,\, {\rm GeV}^3$, with a significant uncertainty.
Those matrix elements which are due to nonfactorizable
contributions (see Eq. (\ref{67})) are essentially undetermined,
although they are expected to be suppressed compared to the
factorizable
matrix elements $(1/2)f_B^2M_B$.

As for the baryon matrix elements of the four-quark operators
next-to-nothing is known about them at the moment. Some very
crude
estimates within the naive quark model are available \cite{BB1}
  but
they are very
unreliable.

In conclusion of this section a remark is in order concerning
numerical estimates of the key parameter of the heavy quark theory,
$\bar\Lambda$. I postponed discussing the issue because its value
continues to be  controversial. QCD sum rules indicate
\cite{FQSR1,mupie,Nar}
that
$\bar\Lambda \sim 0.5$ GeV. This number is in full agreement
with the lower bound stemming from Voloshin's sum rule, see Sect.
3.6. However, some lattice calculations yield a factor of 2 lower
estimate. I am inclined to think that there is something wrong in the
lattice results. Perhaps, the lattice definition of $\bar\Lambda$ does
not fully correspond to that of the continuum theory. It is
inconceivable that such a low value of
$\bar\Lambda$ as 0.2 or even 0.3 GeV could be reconciled with the
lower bound implied by Voloshin's sum rule.

In the discussion above
we have totally disregarded
logarithmic dependence of the operators and their matrix elements
due to anomalous dimensions -- i.e. the issue of the normalization
point (including the normalization point of $\bar\Lambda$). This is
in line with that so far we pretend that hard gluons do
not exist. A brief excursion into this topic will be undertaken later;
here it is only worth  mentioning that all numerical estimates
presented above
refer to a low normalization point, of order of several units times
$\Lambda_{\rm QCD}$.

We are ready now to review classic problems of the heavy quark
theory. We will gradually move from simpler to more sophisticated
problems.

\subsection{Mass formula revisited}

In Sect. 1.5 we have found the first subleading term in the mass
formula for the heavy flavor hadrons. The parameter
$\bar\Lambda$
was related to the expectation value of the gluon anomaly,
see Eq. (\ref{defl}).  It is very easy to continue the expansion one
step further and find the next subleading term, of order $1/m_Q$.
One could have extended the derivation along the lines suggested in
Sect. 1.5. This was done in Ref. \cite{BSUV}. This is not the fastest
route, however. Instead, let us observe that the $1/m_Q$ term in the
Hamiltonian (\ref{hamil}) can be considered as the first order
perturbation;
the corresponding correction to the mass is merely the expectation
value of this perturbation,
$$
M_{H_Q} = m_Q +\bar\Lambda +
\frac{1}{2m_Q} (2M_{H_Q})^{-1}
\langle H_Q|{\vec\pi}^2 +\vec\sigma\vec B |H_Q\rangle + ... =
$$
\beq
m_Q +\bar\Lambda + \frac{(\mu_\pi^2 - \mu_G^2)_{H_Q}}{2m_Q}
+ ...
\label{massf2}
\eeq
The terms of order $1/m_Q^2$ are neglected. If we keep only the
terms up to  $1/m_Q$ it does not matter whether the state $H_Q$ we
average over is an asymptotic state (corresponding to $m_Q=
\infty$) or the real physical heavy flavor state. I remind that, unlike
HQET, we work with the physical states. The difference becomes
noticeable only at the level $1/m_Q^2$. In this order the mass
formula does not reduce any more to the expectation values of local
operators. A part of the $1/m_Q^2$ correction is due to non-local
correlation functions, see Ref. \cite{BSUV} for further details. Eq.
(\ref{massf2}) was first presented in Ref. \cite{FN}.

\subsection{The scaling law of the pseudoscalar and vector coupling
constants}

The pseudoscalar and vector meson constants $f_P$ and $f_V$ are
defined as
\beq
\langle 0|\bar Q \gamma_\mu\gamma_5 q |B_Q\rangle = i f_P
p_\mu
\,\,\,  \langle 0|\bar Q \gamma_\mu q |B^*_Q\rangle = i f_V
M_V\epsilon_\mu\, .
\label{defpv}
\eeq
An alternative definition of the pseudoscalar constant can be given
in terms of the pseudoscalar current,
\beq
\langle 0|\bar Q i\gamma_5 q |B_Q \rangle = f_P' M_B \, .
\label{defpvalt}
\eeq
The constant $f_B$ is one of the key parameters of the heavy quark
physics, just in the same way  the constant $f_\pi$ is a key
parameter of the soft pion physics. Below we will show that in the
limit $m_Q\rightarrow\infty$ all three constants,
$f_P, f_V$ and $f_P'$,  coincide with each other and scale as $m_Q^{-
1/2}$ modulo a weak logarithmic dependence on $m_Q$. (Needless to
say, that both masses, $M_P$ and $M_V$ also coincide in this limit.)
For definiteness let us consider $f_P'$. Two other constants can be
treated in a similar manner. The subscripts will be omitted in the
remainder of this section to avoid overloaded expressions.

Start from  the two-point function
\beq
{\cal A} (k) =i\int e^{ikx} d^4x
\langle {\rm T}\{ \bar Q (x)i\gamma_5 q(x) \, \,   \bar q(0)
i\gamma_5
Q(0) \}\rangle \, ,
\label{2pf}
\eeq
where $k$ is the external momentum.
The two-point function ${\cal A}(k)$ develops a pole at $k^2=M^2$,
the
position of the ground state pseudoscalar,
\beq
{\cal A} (k)  =\frac{f^2M^2}{k^2-M^2} +\mbox{excitations} ,
\label{pole}
\eeq
Of course, the
currents  produce from the vacuum not only the ground state mesons
but also all
excitations
in the given channel.  It is clear that to isolate the lowest-lying pole
we
should
keep $k^2$ close to $M^2$. Keeping in mind Eq. ({\ref{mf}) it is
natural
to represent $k$ as
$$
k_\mu =\{ m_Q +\epsilon, 0,0,0\}
$$
where $\epsilon$ scales like $\Lambda_{\rm QCD}$ while
$m_Q\rightarrow
\infty$.  With this parametrization of $k_\mu$ we merely separate
the
mechanical (uninteresting) part of the momentum. The pole is
achieved
at $\epsilon =\bar\Lambda$; near the pole
\beq
{\cal A}(\epsilon ) \approx \frac{f^2 M}{2(\epsilon -\bar\Lambda)}\,
{}.
\label{epole}
\eeq
The value of the coupling constant is obtained by amputating the
pole,
\beq
f^2 =\lim_{\epsilon\rightarrow\bar\Lambda}\left\{
\frac{2(\epsilon -\bar\Lambda )}{M} {\cal A}(\epsilon ) \right\}\, .
\label{f2}
\eeq

Let us now examine the theoretical expression for the same
two-point
function. In the  background field technique (which is already pretty
familiar, right?)
 we write
\beq
{\cal A}(k) = i \mbox{Tr}
\left\{ i\gamma_5\frac{1}{\not\!\!{\cal P}- m_q}i\gamma_5
\frac{1}{{\not\!\!  k}+\not\!\!{\cal P} - m_Q}\right\}\, .
\label{2pft}
\eeq
Superficially this expression  looks the same as if the quarks were
treated as
free;
they are not, however; the coupling to the background field is
reflected in the fact that
${\cal P}_\mu$  is the momentum operator, not just a
$c$-number four-vector.

Now we will
take advantage of the fact that $m_Q\rightarrow\infty$. As usual, we
close our eyes on any possible hard contributions, assuming that
${\cal P}$, the
momentum operator of the light quark, is soft, i.e. does not scale with
$m_Q$
in
the large mass limit but, rather ${\cal P}\sim\Lambda_{\rm QCD}$.
(This is the reason, by the way, why the large  external momentum
$k$ was directed through the heavy quark line in Eq. (\ref{2pft}).)
Intuitively
it is clear that the hard components of ${\cal P}$ should be irrelevant
for the lowest-lying state whose ``excitation energy" measured from
$m_Q$ is of order $\Lambda_{\rm QCD}$.

If ${\cal P}$ is soft and $m_Q\rightarrow\infty$ the heavy quark
Green function in the leading approximation takes the form
$$
\frac{1}{{\not\!\! k} +\not\!\!{\cal P}-m_Q}=
({\not\!\! k} +\not\!\!{\cal P}+m_Q)\,\,
\frac{1}{(k+{\cal P} )^2 - m_Q^2 +(i/2)\sigma G } =
$$
\beq
\frac{\gamma_0 + 1}{2}
\,\frac{1}{\epsilon + {\cal P}_0 }
\label{prop}
\eeq
where in the second line  all
$1/m_Q$ terms are omitted.
No explicit $m_Q$ dependence is left! This means that ${\cal
A}(\epsilon)$
scales as $m_Q^0$. Equation (\ref{f2}) immediately implies then
that $f$ scales as
\beq
f\sim m_Q^{-1/2}\, .
\label{fM}
\eeq

Equation (\ref{prop}) for the heavy quark Green function in the limit
$m_Q\rightarrow\infty$ is in one-to-one correspondence with the
leading term $\bar Q\pi_0 (1+\gamma_0)/2 Q$ in the low-energy
Lagrangian (\ref{HQLm}).
The analysis of the scaling law of the coupling constants presented
above
is a simplified version of that carried out many years ago by
Shuryak \cite{Shuryak}. Later it was established that the power
dependence on
$m_Q$ in Eq.
(\ref{fM}) is supplemented by a logarithmic dependence appearing
due to the
hard gluon exchanges \cite{hybrid}.

A few words about the numerical value of the coupling constants.
Although in principle $f_D$ and $f_B$ are measurable
experimentally, practically it is a very hard measurement, especially
for $B$. No experimental number for
$f_B$ exists so far. Its value was estimated in the
QCD sum
rules  and on lattices more than once. Leaving aside a dramatic
evolution of
the issue I will say only  that the recent and most reliable results
cluster around 160 MeV both, in the sum rules \cite{FQSR,FQSR1}
and  in the lattice calculations \cite{lattice}. It is curious to note
that in $(m_b)^{1/2}f_B$  the preasymptotic $1/m_Q$ correction
turned out to be unexpectedly  large and negative
\cite{FQSR,mupie,Pat,lattice}; at the same time in $(m_b)^{1/2}{f'}_B$
the preasymptotic $1/m_Q$ correction
 is much more modest \cite{FQSR,FQSR1}.

\subsection{Proof of the Isgur-Wise formula}

I return to my promise to prove the Isgur-Wise formula
(\ref{genIW}).
Consider the three-point function depicted on Fig. 6. The
sides of the
triangle are the Green functions of the quarks in the background
gluon field.
 The reduction theorems tell us that in order to get the
transition amplitudes
$\langle H_c | \bar c \Gamma b | H_b \rangle $  from this
three-point function
we must ``amputate" it:  multiply by $(p^2-M_B^2)$
and
 $(p'^2-M_D^2)$, tending $p^2$ to $M_B^2$ and $p'^2$ to $M_D^2$.
This singles
out the meson states we want to pick up.  For the  vector
mesons
we must also multiply the three-point functions by its polarization
vector
$\epsilon_\mu$. The last  step necessary for amputation is dividing
by the coupling constants
(residues)
connecting the currents $\bar b i\gamma_5 c $ and  $\bar b
\gamma_\mu c$ to the respective mesons. If
the
currents
are normalized  appropriately -- and we will always do that -- the
corresponding coupling constants
in the
pseudoscalar and
vector channels are the same, $fM$ (see Sect. 3.4). It is convenient
to
combine the pseudoscalar and
vector
channels together by introducing the currents
\beq
J = B\bar b i\gamma_5 q + B_\mu \bar b \gamma_\mu q\,\,\,
\mbox{and}\,\,\, J' = D^*\bar q i\gamma_5 c + D^*_\mu \bar q
\gamma_\mu c
\label{J}
\eeq
where $B$ and $D$ are external constants marking the annihilation
and creation of the initial and final $B$'s and $D$'s ($B_\mu$ and
$D_\mu$ denote the polarization vectors of $B^*$ and $D^*$,
respectively).

The expression for the three-point function of Fig. 6 takes the form
\beq
i\mbox{Tr}
\left\{ {\cal M}'\frac{1}{\not\!\! p' +\not\!\!{\cal P} -m_c}\Gamma
\frac{1}{{\not\!\! p} +\not\!\!{\cal P} -m_b}
 {\cal M}
\frac{1}{\not\!\!{\cal P} - m_q}\right\}
\eeq
where the matrices ${\cal M}' $ and  ${\cal M}$ are introduced in
Eq. (\ref{m}).  In Sect. 3.4 the $m_Q\rightarrow\infty$ limit of the
quark propagator was obtained in the rest frame. Here we have two
heavy flavor states, initial and final, and both can not be at rest
simultaneously. Therefore, we need  the very same propagator in the
arbitrary frame. Let $p_\mu = m_Qv_\mu + \epsilon_\mu$. Then a
trivial generalization of Eq. (\ref{prop}) is
$$
\frac{1}{{\not\!\! p} +\not\!\!{\cal P}-m_Q}=
({\not\!\! p} +\not\!\!{\cal P}+m_Q)\,\,
\frac{1}{(p+{\cal P} )^2 - m_Q^2 +(i/2)\sigma G } \rightarrow
$$
\beq
\frac{\not\! v + 1}{2}
\,\frac{1}{(\epsilon + {\cal P})v }
\label{propv}
\eeq

Using this propagator in Eq. (\ref{propv}) we rewrite the three-point
function
of Fig. 6 as follows
$$
\left({\cal M}'\frac{\not\!{v}'+1}{2} \Gamma \frac{\not\!{v}+1
}{2}{\cal M}
\right)_{\alpha\beta}\, \times
$$
\beq
i\mbox{Tr} \left\{
\frac{1}{(\epsilon ' +{\cal P})v'}\, \,
\frac{1}{(\epsilon  +{\cal P})v}\,
  \left(\frac{1}{\not\!\!{\cal P} - m_q}\right)_{\beta\alpha}\right\}\,
{}.
\label{three1}
\eeq
The expression in the braces is independent of the heavy quark
masses; moreover, it is proportional to the three-point function in
the theory with the scalar heavy quarks considered  in Sect. 1.4.  As
was
explained there, in this theory in the limit $m_Q\rightarrow\infty$
only one form factor survives.

One subtle point deserves discussing here.  When I speak about the
heavy flavor mesons I keep in mind  particles built from the  heavy
quark and a light antiquark, which is not always in line with the
accepted nomenclature. Say, they call $B$ meson a particle with the
$b$ antiquark, not quark. Since this distinction plays no role in my
lectures I will continue to ignore this linguistic nuance
referring to the $b\bar q$ states as $B$ mesons.  All equations
presented above assume that the $b$ quark in the initial state
annihilates to produce the $c$ quark in the final state.
Simultaneously a light antiquark in the initial meson is annihilated
and the same light antiquark reappears  in the final meson.

Return now to the model with spinless heavy quarks. The heavy
flavor hadrons we now deal with are spin-1/2 baryons. More exactly,
we have {\em anti}baryon in the initial state and {\em anti}baryon
in the final state.  This means that
near the mass shell the
expression in the braces in Eq. (\ref{three1}) takes the form
\beq
\left(\frac{-\not\!{v}+1}{2} \,\,  \frac{-
\not\!{v}'+1}{2}\right)_{\beta\alpha}
\, \sqrt{M_BM_D} f_Bf_D
\frac{1}{v'\epsilon ' -\bar\Lambda}\,\frac{1}{v\epsilon
-\bar\Lambda} \, \xi (y)\, ;
\label{sctri}
\eeq
the minus sign  between the unit term and the $v$ term in the
density matrices is due to the fact that we deal with the antibaryons.
Amputating the legs and combining Eqs. (\ref{sctri}) and
(\ref{three1}) we get the Isgur-Wise
formula (\ref{genIW}) since ${\cal M}(-\not\!{v}+1) =
(\not\!{v}+1){\cal M}$,  and so on.

\subsection{The Bjorken sum rule and all that}

The Isgur-Wise function $\xi (y) $ carries information about the
structure of the light cloud. Needless to say that the heavy quark
expansion {\em per se} does not help to calculate this function. One
has to rely on methods applicable in the strong coupling regime
which are outside the scope of my lectures (QCD sum rules, lattices,
...). Still, some interesting and important
relations emerge. Here we will discuss a sum rule for the slope
of the Isgur-Wise function and related topics.

We are already familiar with the sum rule technology in the heavy
quark theory.  In Sect 2.1
we dwelled on a simplified problem: inclusive decays
of a spinless heavy quark $Q$ into a lighter spinless quark $q$
and a fictitious spin-zero photon $\phi$. The ``photon" was assumed
to be on mass shell, $q^2 = 0$. The predictions obtained referred to
the
moments of the ``photon" energy. Now you are mature enough
to face actual problems from real life. We will concentrate on the
decays of a $b$ containing hadron into a $c$ containing hadron plus
the lepton pair $l\nu$. The four-momentum of the lepton pair is
a free parameter, in particular, $q^2\neq 0$. We can and will choose
the value of $q$ to our advantage.

Consider a transition $H_b\rightarrow H_c$ induced by some
particular current, say,  axial-vector.
At zero recoil $\xi = 1$. In  the SV limit where the velocity of
the recoiling hadron is small
\beq
\xi (y) = 1 - \rho^2 (y-1) + ... = 1-\rho^2\frac{{\vec v}^2}{2} + ...
\label{defrho}
\eeq
where $y=vv'$, $\vec v$ is the spatial velocity of $H_c$ in the $H_b$
rest frame,  and the slope parameter $\rho^2$ was introduced
in Ref. \cite{Bj}.  It plays the same role as, say, the charge radius of
pions.

To get  relations involving $\rho^2$ we start from consideration of
the transition operator similar to that of Eq. (\ref{trans}). The
expectation value of the transition operator over the $B$ meson state
yields the hadronic amplitude whose imaginary part is proportional
to the probability of the inclusive decay $B\rightarrow X_c l\nu$
with the fixed value of $q$, the
momentum carried away by the lepton pair $l\nu$. (Here
$X_c$ denotes an
inclusive hadronic state containing one $c$ quark.)
A new
element compared to the toy model of Sect. 2.1 is the heavy quark
spin.  Another distinction is the fact that, to achieve the SV limit,
we do not need now to assume that $m_c$ is close to $m_b$.
In the
semileptonic decay $B\rightarrow X_c l\nu$ one can fine-tune the
lepton pair momentum in such a way that
$q^2$ is close to its maximal value, $q^2_{\rm max} = (M_B-M_D)^2$;
then the $c$ containing hadronic state produced is almost at rest, and
we
are in the SV limit even though the charmed quark is significantly
lighter than the $b$ quark. In other words, for such values of $q^2$
the $c$ quark is always slow.

This transition operator describes the forward scattering of $B$ to
$B$ via intermediate states $D^*$ and ``excitations".
(We focus for definiteness on the
axial-vector
current, $\bar c\gamma_\mu\gamma_5 b$. The vector  current can
be treated in a similar way.)
The excitations can include, for instance, $D^*\pi\pi$, and so on. In
general, all intermediate states except the lowest-lying $D^*$ will be
referred to as excitations.  The transition
operator
\beq
\hat T_{\mu\nu}  = i\int d^4 x {\rm e}^{-iqx} T
\{\bar b (x) \gamma_\mu\gamma_5 c (x)\, , \,
\bar c\gamma_\mu\gamma_5 b\}
\label{transqcd}
\eeq
in the Born approximation
is given by the diagram of Fig. 1.   The  hadronic amplitude  obtained
by averaging  ${\hat T}_{\mu\nu}$ over the $B$ meson state,
\begin{equation}
h_{\mu\nu} =\frac{1}{2M_B} \langle B|{\hat T}_{\mu\nu}|B\rangle
\label{htdef}
\end{equation}
contains various kinematical factors.
 In the general case the
hadronic tensor $h_{\mu\nu}$ consists of five different covariant
structures
\cite{chay,koyrakh}:
\begin{equation}
h_{\mu\nu} = -h_1g_{\mu\nu}
+h_2v_\mu v_\nu -ih_3\epsilon_{\mu\nu\alpha\beta}v_\alpha
q_\beta
+h_4 q_\mu q_\nu + h_5 (q_\mu v_\nu + q_\nu v_\mu ).
\label{five}
\end{equation}
Moreover, the invariant hadronic functions $h_1$ to $h_5$ depend
on two
variables, $q_0$ and $q^2$, or $q_0$ and $|\vec q \,|$. For
$\vec q =0$
only one variable survives, and only two of five tensor structures in
$h_{\mu\nu}$ are independent.

Each of these hadronic invariant functions
satisfies a dispersion relation in $q_0$,
\begin{equation}
h_i(q_0) =\frac{1}{2\pi}\int\;
\frac{w_i ({\tilde q}_0)d{\tilde q}_0}{{\tilde q}_0 - q_0}
\,\, + \mbox{ polynomial}
\label{disp}
\end{equation}
where $w_i$ are observable structure functions,
$$
w_i =2\,{\rm Im}\, h_i \, .
$$

For our purposes it is quite sufficient to analyze only one function,
namely, $h_1$.  Moreover, for the time being we will disregard
all non-perturbative corrections ${\cal O}(\Lambda_{\rm QCD}^2)$
which means that operators in the expansion of $\hat T_{\mu\nu}$
other than $\bar
b b $ can be neglected, and the $B$ meson expectation  value of
$\bar b b $ can be replaced by unity. Calculating $h_1$ in this
approximation is a trivial problem ( it was a part of the exercise
suggested in Sect. 2.1).  Specifically,
\begin{equation}
-h_1^{AA}
=(m_b+m_c-q_0) \frac{1}{z} +{\cal O}(\Lambda_{\rm QCD}^2)
\label{hbare}
\end{equation}
 where
\beq
z= (m_b -q_0 - E_c)(m_b -q_0 + E_c)\, , \,\,\, E_c^2 = m_c^2 +{\vec
q}^2 \, .
\label{zdef}
\eeq
I remind that $\vec q$ is assumed to be fixed, and $\Lambda_{\rm
QCD}
\ll |\vec  q
|\ll M_D$, so that actually I will expand in $\vec  q$ keeping only the
terms up to second order. It is convenient to  shift $q_0$ by
introducing a new variable
\begin{equation}
\epsilon = q_{0max}-q_0
\end{equation}
where
\begin{equation}
q_{0max} = M_B- E_{D^*},\,\,\, E_{D^*}= M_{D^*}+\frac{{\vec
q}^2}{2M_{D^*}}\, .
\label{q0max}
\end{equation}
When $\epsilon$ is real and positive we are on the physical cut
where the actual intermediate states (e.g. $D^*$) are produced. Here
the
imaginary part of $h_1$ is given by the ``elastic" contribution of
$D^*$ plus inelastic  excitations.
For negative $\epsilon$ we are below the cut. The result for
$h_1$ above can be trusted if $-\epsilon \gg \Lambda_{\rm QCD}$
since the expansion actually runs in $\Lambda_{\rm QCD}/\epsilon $.
The expansion in the inverse heavy quark mass also requires, of
course, that $|\epsilon |\ll m_{c,b}$.
A bridge between the physical domain of positive $\epsilon$ and the
Euclidean domain of negative $\epsilon$ where the calculation is
done is provided by the dispersion relations.

At the next stage the amplitude $h_1$ is expanded
in powers of $\Lambda_{\rm QCD}/\epsilon$ and
$\epsilon /m_{b,c}$. Polynomials in $\epsilon$ can be discarded since
they have no imaginary part. We are interested only in negative
powers of $\epsilon$. The coefficients in front of $1/\epsilon^n$
are related, through dispersion relations, to the integrals over the
imaginary part of $h_1$ with the weight functions proportional
to the excitation energy to the power $n-1$.
Indeed,
$$
-h_1 (\epsilon ,{\vec q}^2) =
\frac{1}{2\pi}\int d\tilde\epsilon\,\frac{w_1(
\tilde\epsilon ,{\vec q}^2)}{\epsilon -\tilde\epsilon}=
$$
\begin{equation}
\frac{1}{\epsilon}\cdot\frac{1}{2\pi}\int
d\tilde\epsilon\,w_1(
\tilde\epsilon ,{\vec q}^2)+
\frac{1}{\epsilon^2}\cdot\frac{1}{2\pi}\int
d\tilde\epsilon\,\tilde\epsilon\, w_1(
\tilde\epsilon ,{\vec q}^2)+
\frac{1}{\epsilon^3}\cdot\frac{1}{2\pi}\int
d\tilde\epsilon\,{\tilde\epsilon}^2 w_1(
\tilde\epsilon ,{\vec q}^2)+...
\label{disp1}
\end{equation}
Thus, our immediate task is to built the $1/\epsilon$ expansion from
the amplitude (\ref{hbare}).

The theoretical expression for the amplitude $h_1$ above
 knows nothing, of course, about the
meson masses; it contains only the quark masses. Correspondingly,
it is very convenient to build first the expansion of $h_1$ in an
auxiliary quantity,
\begin{equation}
\epsilon_q = m_b- E_c - q_0,\,\,\, E_c =m_c +\frac{{\vec
q}^2}{2m_c}\, .
\label{quarke}
\end{equation}
Then, if necessary, we  reexpress the expansion obtained in this way
in terms of $\epsilon$,
\begin{equation}
\frac{1}{\epsilon_q} = \frac{1}{\epsilon} + \frac{(\epsilon
-\epsilon_q)}{\epsilon^2} +...
\label{epsq}
\end{equation}
 The difference between
$\epsilon_q$ and $\epsilon$ is ${\cal O}(\Lambda_{\rm
QCD}\cdot{\vec
q}^2/m_{b,c}^2)$ and ${\cal O}(\Lambda_{\rm QCD}^2/m_{b,c})$.

 After these introductory remarks, assembling all information in our
disposal, we get
\beq
-h_1 = \left( 1 - \frac{{\vec v}^2}{4} \right) \,\frac{1}{\epsilon}
+\bar\Lambda \frac{{\vec v}^2}{2}\frac{1}{\epsilon^2} + ...
\label{hexp}
\eeq
plus terms of higher order in ${\vec q}^2$ or $\Lambda_{\rm
QCD}$. In deriving Eq. (\ref{hexp}) I used the fact that
$\epsilon - \epsilon_q =\bar\Lambda{\vec v}^2/2$ plus terms of
higher order in ${\vec q}^2$ or $\Lambda_{\rm QCD}$.

This completes the theoretical aspect of the calculation. The
coefficients in front of $1/\epsilon$ and $1/\epsilon^2$ in $h_1$
are known; Eq. (\ref{disp1}) tells us that these coefficients
are equal to integrals over $w_1$, the spectral density. So, what
remains to be done is to express the spectral density in terms of the
contribution coming from the physical intermediate states.
Let us assume for simplicity that the spectrum of the intermediate
states is discrete.  Denote the mass of the $i$-th state by $M_i$ and
the energy by $E_i$; the lowest lying meson, $D^*$, corresponds to
$i=0$. Then the propagator of the $i$-th meson
$$
\frac{1}{(M_B-q_0-E_i)(M_B-q_0+E_i)}
$$
at positive $\epsilon$ has the imaginary part
$$
(2E_i)^{-1}\pi\delta (\epsilon - \delta_i)
$$
where $\delta_i$  is the excitation energy
(including the corresponding kinetic energy),
$$
\delta_i = E_i - E_{D^*} .
$$
For the ``elastic"  $B\rightarrow D^*$ transition  $\delta_0$ vanishes,
of course.

Now it is rather obvious that the structure function
$w_1$ reduces to
\begin{equation}
w_1 (\epsilon) =
\sum_{i=0}^{\infty}\frac{|f_{B\rightarrow i}|^2}{2E_i} \, 2\pi
\delta (\epsilon
- \delta_i ) ,
\label{sum}
\end{equation}
where the sum runs over all possible final hadronic states,
the term with $i=0$ corresponds to the ``elastic" transition
$B\rightarrow D^*$ while $i=1,2,\ldots$ represent excited states
with the energies $E_i = M_i +{\vec q}^2/(2M_i)$;
furthermore, $|f_{B\rightarrow i}|^2$ looks like  the square of a
form factor. Strictly speaking
 $|f_{B\rightarrow i}|^2$ is not exactly
the square of a form factor; rather this is the (appropriately
normalized) contribution to the
given structure function coming from the multiplet of the
degenerate states which includes summation over spin states as well.
By appropriate normalization I mean that we routinely  insert the
normalization factor $(2M_B)^{-1}$.
In the particular example considered (the axial-vector current) $D$ is
not produced
in the elastic transition, so that in the elastic part one needs
to sum only over polarizations of $D^*$.  Say, at zero recoil
$f_{B\rightarrow D^*}=
\sqrt{2M_{D^*}} F_{B\rightarrow D^*}$ where
$F_{B\rightarrow D^*}$ is the $B\rightarrow D^*$ form factor at zero
recoil, see Eq.
(\ref{defF}) below.

Let us examine in more detail the elastic contribution, $i=0$.
The form factor of the $B\rightarrow D^*$ transition generated by
the axial-vector current is given in Eq. (\ref{2}).  Using this
expression we readily obtain
\begin{equation}
(2E_{D^*})^{-1}|f_{B\rightarrow D^*}|^2= \frac{M_{D^*}}{E_{D^*}}
\left(
\frac{1+vv'}{2}
\right)^2 |\xi (vv')|^2 \approx
1 - \rho^2 {\vec v}^2\, .
\label{elastic}
\end{equation}

Comparing the $1/\epsilon$ coefficient in the dispersion
representation (\ref{disp1}) with that of Eq. (\ref{hexp}) we conclude
that
\begin{equation}
\frac{1}{2\pi}\int d\epsilon \; w_1(\epsilon ) =
1-\rho^2{\vec v}^2 +
\sum_{i=1}^{\infty} \frac{|f_{B\rightarrow i}|^2}{2E_i} =  1-
\frac{{\vec
v}^2}{4}\, .
\label{bjorken}
\end{equation}
which implies, in turn
\beq
\rho^2 = \frac{1}{4} + \sum_{i=1}^{\infty} \frac{|f_{B\rightarrow
i}|^2}{2M_i{\vec v}^2} \, .
\label{bjrho}
\eeq
In Sect. 1.3 we learnt that at zero recoil (i.e. $\vec v =0$) only the
elastic transition survives. As a consequence of the heavy quark
symmetry  for all non-elastic transitions $|F_{B\rightarrow i}|^2
\sim {\vec v}^2$. The ratio  $|f_{B\rightarrow i}|^2/{\vec v}^2$
stays finite in the limit of small $\vec v$. Eq. (\ref{bjrho}) is the
Bjorken sum rule proper.  Since the contribution of the excited states
on the right-hand side is obviously positive it tells us, in particular,
that $\rho^2 >1/4$. This inequality is not very informative, though,
since both, the QCD sum rule \cite{FQSR1,BSR}
and lattice calculations indicate that
$\rho^2$ is only slightly less than unity, perhaps,  close to 0.8.

Leaving technicalities aside let me summarize the physical meaning
of the result obtained. The coefficient in front of $1/\epsilon$ in Eq.
(\ref{hexp}) does not contain $\Lambda_{\rm QCD}$. This means that
we calculate the  probability of the decay $b\rightarrow c ``l\nu "$
with given value of ${\vec v}^2$ merely in the parton model;
this probability is equal to that of the physical decay
$B\rightarrow X_c ``l\nu "$; the latter is comprised of the elastic
transition $B\rightarrow  D^* ``l\nu "$ and the transition of $B$
into excitations. (The quotation marks are used to emphasize the fact
that the decays that are measured are induced by both, the
axial-vector and vector, currents, while we focus now only on the
transitions induced by the axial-vector current.)  All probabilities of
production of the excited states are proportional to ${\vec v}^2$
(at small ${\vec v}^2$), and so is the part of the elastic transition
containing $\rho^2$. The sum of these two contributions must
coincide with the ${\vec v}^2$ term obtained in the parton model.
The very same analysis, by the way, presents a  proof of the fact that
$\xi (y=1) = 1$. (Do you see this?)

Now, we make the next step, proceeding to the $1/\epsilon^2$ terms.
The $1/\epsilon^2$ term in Eq. (\ref{hexp})  is proportional to
$\bar\Lambda$, hence the result we are about to get evidently goes
beyond the parton model. Combining Eq. (\ref{hexp}) with the
dispersion representation (\ref{disp1})  we find
\beq
\bar\Lambda\frac{{\vec v}^2}{2} =
\sum_{i=1}^{\infty}\frac{|f_{B\rightarrow i}|^2}{2E_i} \, \delta_i =
\sum_{i=1}^{\infty}\frac{|f_{B\rightarrow i}|^2}{2M_i} \,
(M_i - M_{D^*})\, .
\label{vsr}
\eeq
The sum runs not from zero to infinity but from 1 to infinity since
$\delta_0 =0$. Moreover, since all $|F_{B\rightarrow i}|^2$ for $i=1,2,
...$
are proportional to ${\vec v}^2$, and we are interested only in the
${\vec v}^2$  term, it is legitimate to substitute, as I did, $E_i$ by
$M_i$ and $\delta_i$ by $M_i - M_{D^*}$.

Eq. (\ref{vsr}) is the optical (or Voloshin's) sum rule; superficially it
looks the same as in the toy model of Sect. 2.1.  Please, remember
this sum rule -- it gives a unique opportunity to {\em measure}
$\bar \Lambda$, one of the  key parameters of the heavy quark
theory.
To this end one has to measure the inelastic transition probabilities
in the semileptonic decays $B\rightarrow X_c l\nu$ in the SV limit.
This is a difficult measurement, but not impossible, at least in
principle. Before venturing into this noble task -- extraction of
$\bar\Lambda $ from experimental data -- I must warn you that
acceptable accuracy can be achieved only provided that the
perturbative corrections (hard gluons) as well as nonperturbative
ones,
of the next order in $\Lambda_{\rm QCD}$, are included in the sum
rules. We will briefly discuss the impact of the perturbative
corrections in Lecture 5.

Those who are anxious to get something practical from the optical
sum rule in the absence of the necessary measurements should not
be discouraged. We can still get a lower bound on
$\bar\Lambda$. Indeed, let us rewrite Eq. (\ref{vsr}) as follows:
$$
\bar\Lambda\frac{{\vec v}^2}{2} =
\sum_{i=1}^{\infty}\frac{|f_{B\rightarrow i}|^2}{2M_i} \,  (M_1-
M_{D^*})
+\sum_{i=2}^{\infty}\frac{|f_{B\rightarrow i}|^2}{2M_i} \,  (M_i
- M_1) =
$$
\beq
(M_1- M_{D^*}) \left(\rho^2 -\frac{1}{4}\right){\vec v}^2
+\sum_{i=2}^{\infty}\frac{|f_{B\rightarrow i}|^2}{2M_i} \,  (M_i
- M_1) \, ,
\eeq
where Eq. (\ref{bjrho}) is substituted.
Since the second term on the right-hand side is obviously positive we
conclude that
\beq
\bar\Lambda >2(M_1- M_{D^*})\left(\rho^2 -\frac{1}{4}\right)
\sim 500\,\,  \mbox{MeV}
\label{ineql}
\eeq
where $M_1$ is the mass of the first excited resonance with the
quantum numbers of $D^*$ ($M_1 -M_{D^*}\sim 0.5$ GeV).

Following the same line of reasoning one can derive the ``third" sum
rule relating
$\mu_\pi^2$ to an appropriately  weighted sum over excitations
\cite{Grozin}. The corresponding inequality analogous to Eq.
(\ref{ineql}) takes the form
\beq
\mu_\pi^2 > 3(M_1- M_{D^*})^2\left(\rho^2 -\frac{1}{4}\right)\sim
0.45 \,\, \mbox{GeV}^2\, .
\eeq

\subsection{Deviations of the $B\rightarrow D^*$ form factor from
unity at zero recoil}

The heavy quark theory began from the observation that
the $B\rightarrow D^*$ axial-vector form factor at zero recoil
is exactly unity in the limit $m_b\rightarrow\infty \, , \,\,\,
m_c\rightarrow\infty\, , \,\,\, m_b /m_c$ arbitrary, see Sect. 1.3.
This is  purely a symmetry statement, as usual, dynamics resides in
the corrections. In this section we will discuss deviations from unity.

When the heavy quark mass is infinite it is nailed at the origin,
both in the initial $B$ meson and in the final $D^*$. The light cloud
then does not notice the replacement of one quark by another,
the overlap is unity. If we make the quark masses finite
they start jiggling inside the mesons, and this motion is different
in $B$ and $D^*$ since the  heavy quark velocities are
different.  On top of this the difference in the relative spin
orientations
of the heavy quarks and light clouds shows up. These two effects
lead to deviations from unity. At a heuristic level there is no doubt
that the deviations are of order of (i) the square of the characteristic
heavy quark momentum ($\vec p$ itself can not enter since there is
no preferred orientation) or (ii) chromomagnetic correlation
$\vec \sigma\vec B$. In both cases dimensional arguments
prompt us that the deviation from unity at zero recoil
is proportional to $1/m_{c,b}^2$; linear effects in
$1/m_{c,b}$ are absent. The assertion was first formulated in Ref.
\cite{VS} and was cast in the form of a theorem (Luke's theorem)
in Ref. \cite{Luke}.  The proof presented below
is abstracted from the recent work \cite{vcb}.

Let us define the $B\rightarrow D^*$ form factor at zero recoil as
follows
\beq
\langle D^* |\bar c\gamma_\mu\gamma_5 b |B\rangle
=i \sqrt{4M_BM_D^*} F_{B\rightarrow D^*} D_\mu^* \, ,
\label{defF}
\eeq
to be compared with Eq. (\ref{BD*}).  Conceptually our present
derivation is very close to that leading to the Bjorken and Voloshin
sum rules (Sect. 3.6).  We will again consider the transition operator
induced by the axial-vector current limiting ourselves to the spatial
components of the current.  Technically it is simultaneously simpler
and more involved. Simpler -- because  at the point of zero recoil one
must put $\vec q = 0$, so that kinematics is trivial. In particular,
from the very beginning only one structure ($h_1$) survives in the
general
decomposition (\ref{five}). The calculation is more complicated on
the other hand
since now one  has to keep track of terms of order $\Lambda_{\rm
QCD}^2$. Those of order $\Lambda_{\rm QCD}$  are simply absent!

The quantity $\epsilon$ is defined now as
\beq
\epsilon = M_B-M_{D^*} - q_0
\eeq
and we continue to assume that $\Lambda_{\rm QCD}\ll |\epsilon |\ll
m_{c,b}$ and continue to  examine our old acquaintance, $h_1$,
expanded in powers of $1/\epsilon$ and $\epsilon /m_{c,b}$.
The result of a relatively simple calculation (which the reader is
encouraged to do) is
\beq
-h_1 = \left\{ 1 - \Delta \right\}\frac{1}{\epsilon} + {\cal
O}(\Lambda_{\rm QCD}^3)
\frac{1}{\epsilon^2} + ...
\label{zrh}
\eeq
where
$$
\Delta = \frac{1}{3}\frac{\mu_G^2}{m_c^2}
+\frac{\mu_\pi^2-\mu_G^2}{4}\left(
\frac{1}{m_c^2}+\frac{1}{m_b^2}+\frac{2}{3m_cm_b}
\right) \, .
$$

An explanatory remark is in order here concerning the
$1/\epsilon^2$ term in $h_1$.
The theoretical expression for $h_1$, as it naturally emerges
from the computations, depends on $\epsilon_q$, not on $\epsilon$
where $\epsilon_q$
is the energy measured from the  ``quark"  threshold, see
Eq. (\ref{quarke}). In the kinematics at hand,
when we are at the point of zero recoil,
$$
\epsilon - \epsilon_q = M_B - M_{D^*} - (m_b-m_c) =
$$
\beq
-(\mu_\pi^2 -\mu_G^2)\left(
\frac{1}{2m_c} - \frac{1}{2m_b} \right)
-\frac{2}{3m_c}\mu_G^2 + ...
\label{deltaa}
\eeq
where I invoked Eq. (\ref{massf2}). (Can you figure out why the
coefficients in this expression and in Eq. (\ref{deltaa}) are different?
Hint: The parameters $\mu_G^2$ and $\mu_\pi^2$ in Eq.
(\ref{deltaa}) are defined as the expectation values of the
corresponding operators over the {\em pseudoscalar} meson.
This is not the case in Eq. (\ref{massf2}).)

We first expand $-h_1$ in $1/\epsilon_q$ and then pass to the
physical variable $\epsilon$ and rearrange the expansion. In the
$1/\epsilon_q$ expansion the corrections of order ${\cal
O}(\Lambda_{\rm QCD})$ are absent from the very beginning
-- an obvious fact hardly requiring further comments -- while the
term ${\cal O}(\Lambda_{\rm QCD}^2/\epsilon_q^2)$ does appear
explicitly. This term, however, is killed in passing from
$1/\epsilon_q$ to $1/\epsilon$, see Eq. (\ref{deltaa}).

 That is why I assert
that the coefficient in front of $1/\epsilon^2$ is
${\cal O}(\Lambda_{\rm QCD}^3)$.  Moreover, as we will see
shortly this coefficient in Eq. (\ref{zrh}) is positive. In principle, it is
calculable (more exactly, expressible in terms of several
new phenomenological parameters) but this will be of no concern
to us in this lecture.

Repeating, step by step, the derivation of Sect. 3.6 we conclude that
\beq
|F_{B\rightarrow D^*}|^2 + \sum_{i=1,2,...}|F_{B\rightarrow excit}|^2
=1 -\Delta
\eeq
and
\beq
\sum_{i=1,2,...}|F_{B\rightarrow excit}|^2
(M_i - M_{D^*}) = {\cal O}(\Lambda_{\rm QCD}^3)\, .
\eeq
The latter sum rule, by the way, is the reason why we know that
the coefficient ${\cal O}(\Lambda_{\rm QCD}^3)$ in front of
$1/\epsilon^2$ is positive -- the left-hand side of the sum rule
is obviously positive-definite.
These two relations, taken together, plus positivity of
$|F_{B\rightarrow i}|^2$,  imply that
$|F_{B\rightarrow D^*}|^2$ is limited from below and from above,
\beq
1-\Delta - \frac{{\cal O}(\Lambda_{\rm QCD}^3)}{M_1 - M_{D^*}}
< |F_{B\rightarrow D^*}|^2< 1-\Delta
\label{1-F}
\eeq
where
$M_1$ is the mass of the first excited state produced by the
axial-vector current, $M_1 - M_{D^*} = {\cal O}(\Lambda_{\rm
QCD})$.

Not only is it  seen  that the deviation of $|F_{B\rightarrow D^*}|^2$
from unity starts from terms scaling like $1/m_{c,b}^2$,
with no $1/m_{c,b}$ corrections, but we understand now the
reasons lying behind this remarkable fact. Moreover, we have an
idea of how large the actual deviations are since Eq. (\ref{1-F})
establishes a lower limit for these deviations in terms of the
parameter $\Delta$ which is determined numerically  rather well.
In this aspect the derivation I present here goes beyond
the more conventional  analysis of Ref. \cite{FN,Mannel}. The reader
is
nevertheless  advised to consult the latter works to get a broader
perspective of the heavy quark theory -- the more approaches you
master the better for you.

Qualitatively it is quite clear why the deviation
of $|F_{B\rightarrow D^*}|^2$ from unity starts from
$\Lambda_{\rm QCD}^2/m_{c,b}^2$. Indeed, let us return to the
Bjorken formula (\ref{defrho}). In this formula it is assumed $\vec
v\gg \Lambda_{\rm QCD}/m_Q$ so that actually we do not
distinguish between  the velocity
of the recoiling final heavy hadron and that of the final quark. At
zero recoil the heavy hadron is nailed, but not the heavy quark. The
latter experiences a primordial motion inside the nailed hadron, with
the velocity ${\vec v}^2\sim \Lambda_{\rm QCD}^2/m_Q^2$.
So, a reasonable guess would be to extrapolate Eq. (\ref{defrho})
down to ${\vec v}^2\sim \Lambda_{\rm QCD}^2/m_Q^2$.  As we see,
this guess works.

It is worth emphasizing that our analysis need
not be confined to the transitions induced by the spatial components
of the axial-vector current. We could consider the temporal
components, or vector currents, or something else. Each time we
get additional information.  For instance, from the transition operator
induced by the vector currents we get a sum rule proving the
inequality $\mu_\pi^2 > \mu_G^2$  obtained in Sect. 3.2 from a
quantum-mechanical argument.

\newpage

\section{Lecture 4. Theory of the Line Shape}.

\renewcommand{\theequation}{4.\arabic{equation}}
\setcounter{equation}{0}

In this lecture I will discuss one of the most interesting and
practically important applications of the heavy quark theory,
the spectra in the end point domain in the inclusive decays.
Inclusive weak
decays of heavy flavors, in particular, semileptonic decays,
are close relatives of famous deep inelastic scattering
-- the processes where a highly virtual photon scatters off nucleons
to produce an inclusive multiparticle state.  The latter are related to
the former via channel crossing. Deep inelastic lepton-nucleon
scattering was in the focus of theoretical activity in the late sixties
and the beginning of seventies and
 was instrumental in discovering and
developing QCD  \cite{DISRev}. It is thus quite surprising
that for a long time there were hardly any attempts
to treat the  beauty decays in QCD proper along essentially the same
lines as it was done in deep inelastic scattering.
Realization of the idea that the $1/m_Q$ expansion
in the theory of the line shape can play the same role as the twist
$1/Q^2$ expansion in DIS came with the 20 years delay
\cite{JR,motion-,motion} --
I see absolutely no reasons why the corresponding theory was
worked out only recently and not  20 years ago.

The
theory of the
line shape in QCD resembles that of the M\"{o}ssbauer
effect.
To explain what I mean it is convenient to consider, for definiteness,
the transition $B\rightarrow X_s \gamma$ where $X_s$ denotes
the inclusive hadronic state with the $s$ quark. This decay has been
recently observed experimentally. (Description of $B\rightarrow
X_q l\nu$ is conceptually similar but is more technically involved).

Again, to avoid inessential technicalities  I will neglect the quark and
photon
spins. So we will consider the transition $Q\rightarrow q\phi$ where
all fields $Q$,  $q$ and $\phi$ are spinless. Thus, to begin
with, we will limit ourselves to the toy model described in Sect. 2.1,
see Eq. (\ref{Qqphi}).
The
mass of the final quark $m_q$   will be treated as a free
parameter which can vary from zero almost up to $m_Q$.
For our approach to be valid we still need
that $\Delta m \equiv m_Q-m_q\gg\Lambda_{\rm QCD}$ although
the
mass difference may be small compared to the quark masses.

 To warm up we will
 put the final quark mass to zero. At the level of
the
free quark decay the photon energy is then fixed  by the two-body
kinematics of the decay $Q\rightarrow q\phi$, namely, $E_\phi
=m_Q/2$. In other words, in the rest frame of the decaying $Q$
quark the photon energy
spectrum
is a monochromatic line at $E_\phi =m_Q/2$ (Fig. 7). On the other
hand,
in the actual hadronic decays $H_Q\rightarrow X_q\phi$
the kinematical boundary of the spectrum lies at
$M_{H_Q}/2$. Moreover, due to multiparticle final states
(which are, of course, present at the level of the hadronic decays) the
``photon" line will be smeared. In particular, the {\em window} --
a gap between $m_Q/2$ and $M_{H_Q}/2$ -- will be closed
(Fig. 7). There are two mechanisms smearing the monochromatic
line of the free-quark decay.  The first is purely perturbative: the
final quark $q$
can shake off a hard gluon, thus leading to  the three-body
kinematics. This mechanism tends to diminish the photon energy and
may be important at $E<m_Q/2$. We will defer its discussion till later
times. The second mechanism is due to the ``primordial"
motion of the heavy quark $Q$ inside $H_Q$ and is
non-perturbative.
Even if the decaying $B$ is nailed at the origin so that
its velocity vanishes, the $b$ quark  moves inside
the light cloud, its momentum being of order $\Lambda_{\rm QCD}$.
This is
the QCD analog of the Fermi motion of the nucleons in the nuclei.
It is quite clear that this motion affects the decay spectra.
Say, if the ``primordial" heavy quark momentum is parallel to that of
the photon, the photon produced gets more energy, and vice versa,
for the antiparallel momenta it gets less.
 It is quite clear that
this effect is preasymptotic (suppressed
by inverse powers of $m_b$): while typical energies of the
decay products are of order $m_b$ a shift due to the heavy quark
motion is of order $\Lambda_{\rm QCD}$.

Only the second
mechanism will be of interest for us in this lecture.

The window (i.e. the domain kinematically inaccessible for free
quarks) plus the adjacent domain below the window, of width
several units $\times \Lambda_{\rm QCD}$, taken together,
form what is called the end point domain.
Below I will outline the main elements of the theory allowing one to
translate
an intuitive picture of the $Q$ quark primordial motion inside $H_Q$
in QCD-based  predictions for the spectrum in the end point domain
\footnote{Basics of the theory of the line shape were worked out in
Refs. \cite{JR,motion-,motion}. Further crucial steps were undertaken
in Refs. \cite{motion1,motion2,motion3,motion4}. In my presentation
I follow mainly Bigi {\it et al.} \cite{motion}.}
{}.
The spectrum below the end point domain is the realm of the
perturbative physics (hard gluon emission).

\subsection{Formalism}

Let us return back to Sect. 2.1 and
consider the transition operator
defined there.  Since we are
interested in
the energy spectrum the ``photon" momentum $q$ must be fixed.
Let us assemble Eqs. (\ref{spectrum}) and (\ref{ImT}) together.
For convenience I will reproduce the result here again, taking into
account the fact that now $m_q$ is assumed to vanish,
\beq
\frac{d\Gamma}{dE} =\frac{m_Q}{M_{H_Q}}\Gamma_0
\left\{ \langle\bar QQ\rangle\delta \left( E-\frac{m_Q}{2}\right)
-
\frac{\langle{\vec\pi}^2\rangle}{4m_Q}\delta '\left( E-
\frac{m_Q}{2}\right)
+\frac{\langle{\vec\pi}^2\rangle}{24}\delta ''\left( E-
\frac{m_Q}{2}\right)+...\right\}
\label{grsing}
\eeq
If in the leading approximation the spectrum is just a delta
function, the corrections are  more and more singular!  The higher
the correction the stronger the singularity. Nonsense? No, this was to
be expected: the width of the $\phi$ line in
the transition $H_Q\rightarrow  X_q \phi$ is of order
$\Lambda_{\rm QCD}$.
We expand in the powers of $\Lambda_{\rm QCD} /m_Q$; hence we
must
expect the enhancement of the singularities in each successive order.
Equation (\ref{grsing}) gives all terms up to  $\Lambda_{\rm
QCD}^2$.
It is clear that to describe the shape of the line one needs to sum up
the infinite number of terms in this expansion.

Then in the approximation of Fig. 1 (no hard gluon exchanges)
the transition operator is given by Eq. (\ref{born}) with $m_q$ set
equal to zero.
To construct  the operator product expansion to all orders we
observe that the momentum operator
$\pi$ corresponding to the residual motion of the heavy quark
is $\sim\Lambda_{\rm QCD}$ and the expansion in $\pi /k$ is
possible.
Unlike the problem of the total widths, however, in the end point
domain $k^2$ is anomalously small, the expansion parameter is
of order unity, and there exists an infinite set of terms where all
terms are of the same order of magnitude.

To elucidate this statement let us examine different terms
in the denominator of the propagator,
$$
k^2 +2\pi k +\pi^2 .
$$
In the end point domain
\beq
|E-(1/2)m_Q|\sim \bar\Lambda
\eeq
It is quite trivial to find that in this domain
$$
k_0\sim|\vec k|\sim m_Q/2, \,\, k^2\sim m_Q\bar\Lambda ;
$$
in particular, at the kinematical boundary (for the maximal value
of the ``photon" energy)
$k_0 = M_Q/2$ and $k^2 = -m_Q\bar\Lambda$. Hence, in the
end point domain
$$
k^2\sim k\pi\gg \pi^2.
$$
In other words, when one expands the propagator of the final quark
in the transition operator,
\beq
\bar Q \frac{1}{k^2 +2\pi k +\pi^2 } Q\, ,
\eeq
in $\pi$, in the leading approximation all terms $(2k\pi /k^2)^n$
must be
taken into account while terms containing $\pi^2$ can be omitted.
The first subleading correction would contain one $\pi^2$ and
arbitrary number
of $2k\pi$'s, etc.

Thus, in this problem it is twist of the operators ($\equiv$ dimension
- Lorentz spin) in the operator product expansion,
not their dimension, that counts. For connoisseurs I will add that this
aspect makes the theory of the line shape in the end point domain
akin to that of deep inelastic scattering (DIS). Keeping only those
terms
in the expansion that do not vanish in the limit $m_Q\rightarrow
\infty$
(analogs of the twist-2 operators in DIS)
we get the following series for the transition
operator
\beq
\hat T = -\frac{1}{k^2}\sum_{n=0}^{\infty}\left(
-\frac{2}{k^2}\right)^n
k^{\mu_1}...k^{\mu_n} \left( \bar Q\pi_{\mu_1}...\pi_{\mu_n}Q
- {\rm traces}\right) .
\label{T18}
\eeq
Traces are subtracted by hand since they are irrelevant
anyway; their contribution is suppressed as
$k^2\pi^2 /(k\pi )^2\sim\Lambda_{\rm QCD} /m_Q$ to a
positive power. Another way to make
the same statement is to say that in Eq. (\ref{T18}) the four-vector
$k$ can be considered as {\em light-like}, $k^2=0$.

\subsection{The light cone distribution function}

After the transition operator is built the next step is averaging
of $\hat T$ over the hadronic state $H_Q$. Using only the
general arguments of the Lorentz covariance one can write
\beq
\langle H_Q|\bar Q\pi_{\mu_1}...\pi_{\mu_n}Q - {\rm
traces}|H_Q\rangle
=a_n\bar\Lambda^n (v_{\mu_1}...v_{\mu_n}  - {\rm traces})
\label{T19}
\eeq
where $a_n$ are constants parametrizing the matrix elements.
Their physical meaning will become clear momentarily. Right now it
is worth
noting that the term with $n=1$ drops out ($a_1=0$). Indeed,
$\langle H_Q| \bar Q\vec\pi Q|H_Q\rangle$ is obviously zero for
spinless $H_Q$ while $\pi_0$ through the equation of motion
reduces to ${\vec\pi}^2/(2m_Q)$ and is of the next
order in $1/m_Q$. Disappearance of $\bar Q\pi_\mu Q$ means
that there is a gap in dimensions of the relevant operators.

Let us write $a_n$'s as moments of some function $F(x)$,
\beq
a_n=\int dx x^n F(x) .
\label{T20}
\eeq
Then, $F(x)$ is nothing else than the primordial line-shape function!
(That is to say,
$F(x)$ determines the shape of the line before it is deformed
by hard gluon radiation; this latter deformation is
controllable by perturbative QCD). The variable $x$ is
related to the photon energy,
$$
x= \frac{2}{\bar\Lambda } \left( E - \frac{m_Q}{2}\right)\, .
$$
If this interpretation is accepted -- and I will prove
that it is correct --
it immediately implies that (i) $F(x) > 0$, (ii) the upper limit
of integration in Eq. (\ref{T20}) is 1, (iii) $F(x)$ exponentially
falls off at negative values of $x$ so that practically
the integration domain in Eq. (\ref{T20}) is limited from below at
$-x_0$ where $x_0$ is a positive number of order unity.

To see that the above statement is indeed valid we substitute
Eqs. (\ref{T19}), (\ref{T20})  in $\hat T$,
\beq
\langle H_Q| \hat T|H_Q\rangle
=-\frac{1}{k^2}\sum_n \int dy F(y) \, y^n
\left( -\frac{2\bar\Lambda kv}{k^2 +i0}\right)^n ,
\label{T21}
\eeq
and sum up the series. The $i0$ regularization will prompt us
how to take the imaginary part at the very end. In this way we
arrive at
\beq
\frac{d\Gamma}{dE} =-\frac{4}{\pi}
\Gamma_0 \frac{m_QE}{M_{H_Q}}{\rm Im}\,
\int dy F(y)
\frac{1}{k^2 + 2y\bar\Lambda kv +i0}
=(2/\bar\Lambda )\Gamma_0 F(x) ,
\label{T22}
\eeq
where the variable $x = k^2/(2\bar\Lambda kv)$ was written out
above in terms of the ``photon" energy, and $\Gamma_0$
is the total decay width in the parton approximation.
Corrections to Eq. (\ref{T22}) are of order $\bar\Lambda /m_Q$.

Thus, we succeeded in getting the desired smearing:
the monochromatic line of the parton approximation is replaced
by a finite size line whose width is of order
$\bar\Lambda$. The pre-asymptotic
effect we deal with is linear in $\Lambda_{\rm QCD} /m_Q$.

At this point you might ask me how this could possibly happen.
We have already learnt that there is a gap
in dimensions of the operators in the expansion -- no operators
of dimension 4 exist -- and the correction
to $\bar Q Q$ is also quadratic in $1/m_Q$ (the CGG/BUV theorem).
There are no miracles -- the occurrence of the effect linear
in $\Lambda_{\rm QCD} /m_Q$ became possible due to the
summation of the
infinite
series in Eq. (\ref{T18}); no individual term in this series gives
rise to $\Lambda_{\rm QCD} /m_Q$.

To avoid misunderstanding it is worth explicitly stating
that the primordial distribution function $F(x)$ is {\em not}
calculated;
rather $F(x)$ is related to the light-cone distribution function
 of the
heavy quark inside $H_Q$, namely $\langle \bar Q(n\pi
)^nQ\rangle$,
$n^2=0$, or more explicitly
\beq
F(x)\propto\int dt{\rm e}^{ixt\bar\Lambda}
\langle H_Q|{\bar Q}(x=0 ){\rm e}^{-i\int_0^t nA(n\tau )d\tau}
Q(x_\mu=n_\mu t)
|H_Q\rangle ,
\label{T23}
\eeq
where $n$ is a light-like vector\footnote{Similar light-cone
distributions for light quarks
are well known \cite{CS} in the theory of deep inelastic scattering,
see also \cite{BaBr}.}
$$
n_\mu = (1,0,0,1).
$$
Unfortunately, this primordial function is not the one that will be
eventually
measured from $d\Gamma /dE$; the actual measured line shape will
be essentially deformed by radiation of  hard gluons. I will say
a few words about this  in Lecture 5.

The primordial distribution function $F(x)$ which we defined here
can be called the {\em light-cone} distribution function.  This is clear
from the expression (\ref{T23}) which has a very transparent
physical meaning. The quark $q$ produced is massless and,
therefore, propagates along the light cone from the point of emission
to the
point of absorption in the transition operator defining the
distribution
function.

If we looked at the physical line shape sketched on Fig. 7 more
attentively, through a microscope, we would notice that a
smooth curve is obtained as a result of adding up many channels,
specific decay modes.
A typical interval in $E$ that contains already enough channels
to yield a smooth curve after summation is $\sim\Lambda_{\rm
QCD}^2/m_Q$.
Roughly one can say that
the spectrum of Fig. 7 covers altogether $m_Q/\Lambda_{\rm QCD}$
resonance
states produced in the $H_Q$ decays and composed
of $q$ plus the spectator (I keep in mind here
that the final hadronic state is produced through decays
of highly excited resonances, as in the multicolor QCD). These states
span the window between $m_Q/2$ and $M_{H_Q}$ and the adjacent
domain
to the left of the maximum at  $E=m_Q/2$.

\subsection{Varying the mass of the final quark}

So far I was discussing the transition into a massless final quark.
It is very interesting to trace what happens with the
line shape and the primordial distribution
function as the final quark mass $m_q$ increases.

Inspection of the transition operator shows that as long as
$m_q^2\ll m_Q\bar\Lambda$ nothing changes in our formulae
at all in the leading-twist approximation. Since the characteristic
values of $k^2$ in the end point domain are of order
$m_Q\bar\Lambda$ and $m_q^2\ll m_Q\bar\Lambda$ one can
merely
neglect the final quark mass altogether.

A more interesting regime is $m_q\sim (m_Q\bar\Lambda )^{1/2}$.
In this regime one can not neglect $m_q^2$ in the denominator.
It is not difficult to see, however \cite{motion3},  that, as in the
massless case,  all
traces
can be neglected since $k^2\pi^2 /(k\pi )^2\sim\Lambda_{\rm QCD}
/m_Q$. This means that the same light-cone distribution function
$F(x)$ that
emerged in the massless case describes the line shape
if $m_q\sim m_Q\bar\Lambda$ as well. The only change that occurs
is a shift of the end point spectrum, as a whole, to the left.
Indeed, if previously the variable $x$ was defined as
$(2/\bar\Lambda ) [E - (1/2)m_Q] $, now when $m_q\neq 0$
$$
x= \frac{2}{\bar\Lambda } (E - E_0)
$$
 where  $E_0 =(2m_Q)^{-1}(m_Q^2-m_q^2)$.  The maximum
of the distribution, in particular, shifts from $m_Q/2$ to $E_0
=m_q/2 -{\cal O}(\bar\Lambda )$.

What happens if one continues to increase $m_q$?
Increasing the quark mass further results in more drastic changes.
The trace terms can not be omitted any more, and the
light-cone function gives place to other distribution functions.
This is obvious already from a simple kinematical argument.
Indeed, with $m_q$ increasing
the window shrinks. When we eventually come to the SV limit
$$
\Lambda_{\rm QCD}\ll \Delta m \equiv m_Q-m_q\ll m_{Q,q}
$$
it shrinks to zero. In this limit
the photon energy in the two-body quark decay,
$\Delta m (1+ \Delta m(2m_Q)^{-1}) $ differs from the
maximal photon energy in the hadronic decay,
$\Delta M (1+ \Delta M(2M_Q)^{-1}) $,  only by a tiny amount
inversely proportional to $m_Q$ ($\Delta m$ and
$\Delta M$ stand for the quark and meson mass differences,
respectively).

Thus, the kinematical consideration prompts us that the line shape
must essentially change. Anticipating the results of the calculation
let me describe the situation pictorially.
 Simultaneously with the
shrinkage of the window the peak becomes more asymmetric
and develops a two-component structure (Fig. 8). The dominant
component of the peak, on its right-hand side, becomes
narrower and eventually collapses into a delta
function when $m_q$ becomes a finite fraction of $m_Q$. A
shoulder develops on the left-hand
side;  the number
of the hadronic states populating the end point domain becomes
smaller -- instead of $m_Q/\Lambda_{\rm QCD}$ states at $m_q = 0$
we are speaking of just several states at $m_q=$ a finite fraction of
$m_Q$.
When we approach the SV limit the
height of the shoulder corresponding to the production of the excited
states becomes very small, proportional to ${\vec v}^2\ll 1$ (Fig. 9).
This is the end of the evolution -- starting from the
light-cone distribution function at $m_q =0$ we
continuously pass to the {\em temporal}
distribution function in the SV limit. It is the temporal distribution
function that shapes   the inelastic shoulder  on Fig. 9.

This rather sophisticated picture, hardly reproducible in
 naive quark models, emerges from the operator product expansion
(in the leading approximation) if one follows along the same lines as
previously. The
transition
operator $\hat T$ for $m_q\neq 0$ is given in Eq. (\ref{S1});
I reproduce it here again for convenience,
\beq
\hat T = \frac{1}{m_q^2-k^2}\sum_{n=0}^\infty
\bar Q \left( \frac{2m_Q\pi_0 +\pi^2 - 2q\pi}{m_q^2-k^2}\right)^n Q\,
,
\label{T24}
\eeq
 Notice that
$2m_Q\pi_0 +\pi^2$ acting on Q yields zero (the equation of motion)
and in the SV limit $q$ must be treated as a small
parameter,
$$
q_0 /m_Q\equiv E/m_Q = v\ll 1;
$$
$v$ is the spatial velocity of
the heavy quark produced. Although $v$ is small the inclusive
description  is still valid provided that
$\Delta m \gg\Lambda_{\rm QCD}$.

In the zeroth order in $q$ the only term surviving in the
sum (\ref{T24}) is that with $n=0$, and we are left with the single
pole,
the elastic contribution
depicted on Fig. 9. This is the extreme realization of the
quark-hadron
duality. The inclusive width is fully saturated
by a single elastic peak. We have already discussed this phenomenon
in Lecture 2. What might seem to be a miracle at first
sight
has a symmetry explanation -- the phenomenon is explained by the
heavy
quark symmetry. The  fact that the parton-model
monochromatic line is a survivor of hadronization is akin
to the M\"{o}ssbauer effect.

If terms ${\cal O}(v^2)$ are switched on the transition operator
acquires an
additional part,
\beq
{\hat T}_{v^2} = \frac{4}{3}{\vec q}^2
\frac{1}{(m_q^2-k^2)^3}
\sum_{n=0}^\infty \left( \frac{2m_Q}{m_q^2-k^2}\right)^n\, \bar Q
\pi_i\pi_0^n\pi_i  Q.
\label{T25}
\eeq
{}From this expression it is obvious that the shape of the $v^2$
shoulder
is given  by the {\em temporal} distribution function $G(x)$
whose moments
are introduced through the matrix elements
\beq
\langle H_Q |\bar Q\pi_i\pi_0^n\pi_i Q|H_Q\rangle =
\bar\Lambda^{n+2}\int dx
x^n G(x) .
\eeq
Alternatively, $G(x)$ can be written as a Wilson line along the time
direction,
\beq
G(x)\propto \int dt{\rm e}^{ixt\bar\Lambda}
\langle H_Q|\bar Q (t=0,\vec x =0)
\pi_i {\rm e}^{-i\int_0^t A_0(\tau )d\tau}\pi_i Q(t, \vec x
=0)|H_Q\rangle .
\eeq

Intuitively it is quite clear why the light-cone
distribution function gives place to the temporal
one in the SV limit. Indeed, if the massless final quark propagates
along
the light-cone, for $\Delta m \ll m$  the quark $q$ is at rest in the
rest frame
of $Q$, i.e. propagates only in time.

In terms of $G(x)$ our prediction
for the line shape following from Eq. (\ref{T25}) takes the form
$$
\frac{d\Gamma}{dE}\propto
\left[ 1-\frac{v^2}{3}\int\left(\frac{1}{y^2}+\frac{\bar\Lambda
/E_{max}}{y}\right)
G(y)dy\right]\delta (x)+
$$
\beq
\frac{v^2}{3}\left(\frac{1}{x^2}+\frac{\bar\Lambda
/E_{max}}{x}\right) G(x) ,
\eeq
where $x=(E-E_{max})/\bar\Lambda$. The $v^2$ corrections affect
both, the
elastic peak
(they reduce the height of the peak) and the shoulder (they create
the
shoulder). The total decay rate stays intact, however: the
suppression of the elastic peak is compensated by the integral over
the inelastic contributions in the shoulder. This is the Bjorken sum
rule thoroughly considered in Lecture 3.
 It is important that we do not have to guess  or make {\em ad hoc}
assumptions --
 a situation typical for model-building -- QCD itself tells us
what distribution function enters in this or that case and in what
particular way.

\subsection{Real QCD: Inclusive semileptonic decays}

{}From the analysis presented above the following remarkable fact
should be clear.
The very same primordial distribution functions that determine the
line
shape in the radiative transitions appear in the problem of the
spectra in the semileptonic decays. In particular,
in $b\rightarrow ul\nu$ we deal with $F(x)$.

Of course, kinematical conditions are different. Now the hadronic part
of the process, $B\rightarrow X_u l\nu$ inclusive decay,
depends on two variables, for instance, $q_0$ and $q^2$, or
$q_0$ and $|\vec q|$.  The probability of the decay, in the free quark
approximation, is proportional to $\delta (m_Q - q_0 - |\vec q|)$
\cite{Paschos}.
In other words, in this approximation only a line on the
$q_0, |\vec q|$ plane is populated (Fig. 10).  (I assume that
we are not interested in the individual momenta of $l$ and $\nu$
and measure just the total momentum of the lepton pair. This is
quite a
fantastic formulation of the problem since
experimentally the neutrino energy and momentum are not
measured, of course; only the electron energy is usually measured.
Nevermind, let us keep in mind a {\em  gedanken} experiment.)
The end point domain is defined now as a band whose  width is
several units $\times \Lambda_{\rm QCD}$ adjacent to the above
quark line (Fig. 10).  Needless to say that in the physical decay
 the whole large triangle is populated; the inner part of the triangle,
to the left of
the end point band, is due to the hard gluon emission.  The smearing
of the
delta-like  spectrum in the band is due to the primordial motion of
$b$ inside $B$, and is described by the light cone distribution.

A
 trivial  modification compared to Sect. 4.3 is
the occurrence of several structure functions. All five  structure
functions are expressible, however, in terms of the same
light cone primordial distribution function $F(x)$ where,
as previously, $x = -\bar\Lambda^{-1}k^2/2k_0$. Since $q_0$ and
$\vec q$ are independent variables in the case at hand
\beq
x = -\bar\Lambda^{-1}k^2/2k_0 = -\bar\Lambda^{-1}k^2/(k_0+|\vec
k|) = -\bar\Lambda^{-1}(m_Q- q_0-
|\vec q|)
\label{xdis}
\eeq
where in the denominator
the difference between $k_0$ and $|\vec k|$ is neglected which is
perfectly
legitimate in the end point band.  In this band $k_0 -|\vec k | =
{\cal O}(\Lambda_{\rm QCD})$, and the difference between
$k_0$ and $|\vec k|$ becomes important only at the level of the
subleading twists which are not included anyway.

Thus, we observe a scaling behavior:
the structure functions that generally speaking could depend on two
variables, $q_0$ and $|\vec q |$, actually depend only on the single
light-cone combination (\ref{xdis}). This is the analog of the
Bjorken scaling in deep inelastic scattering! In the rest of the
phase space, outside the end point band,  the approximate equality
$k_0 \approx |\vec k|$ is not valid, of course, and
the above scaling is not going to take place. The primordial
distribution falls off -- presumably
exponentially -- outside the end point band.
The hard gluon emissions will populate the phase space outside this
domain creating long logarithmic tails. The primordial part is buried
under these tails. Therefore, outside the band
 one can not expect that the structure functions depend
on the single combination $q_0+|\vec q|$ anyway.

Guesses about
a scaling behavior in the inclusive semileptonic
decays are known in the literature \cite{Paschos}. Now
we are finally able to say for sure what sort of scaling takes place,
where it is expected to hold and where and how it will be violated.

I will not go into further details which are certainly
important if one addresses the problem of extraction of $V_{ub}$
from experimental data.  Some of them are discussed
in the  literature , others still have to be
worked out.
Applications of
the theory to data analysis is a separate topic going beyond the scope
of this lecture.

What can be said about the light cone distribution function
$F(x)$?  This function depends on the structure of the light cloud of
the $B$ meson
and,  thus, belongs to the realm of the soft physics. The moments of
this function
are related to the expectation values of the operators
$\bar Q \pi_{\mu_1} ... \pi_{\mu_n} Q$ (see Eq. (\ref{T19}));
in  real QCD the properly normalized matrix elements on the
left-hand side include the factor $(2M_B)^{-1} $). The knowledge of
the infinite set of these expectation values would be equivalent to
the knowledge of the structure of the light cloud. Needless to say
that this is beyond our abilities at present. Still, we know a few first
moments of $F(x)$ and have a general idea of the shape of this
function.
It must be  positive everywhere in the
physical domain, vanish at $x=1$  and  have exponential
fall-off at large negative $x$. The latter property ensures the
existence of
all moments. Moreover,
$$
a_0 = \int dx F(x) = 1\, ,
$$
$$
a_1 = \int dx x F(x) = 0\, ,
$$
and
$$
a_2=\int\; dx \,x^2F(x)=\frac{\mu_{\pi}^2}{3\bar{\Lambda}^2}\, .
$$
Estimates of the third moment also exist in the literature
\cite{motion,Mannel}. I can not
dwell on this issue now and will only mention that $a_3$ is
constrained by exact inequalities, i.e. \cite{motion4}
\begin{equation}
a_2 < \frac{1}{4}+\sqrt{\frac{1}{4}-a_3}\; ,   \;\;\; a_3 <
\frac{1}{4}-
\left(a_2-\frac{1}{2}\right)^2\, .
\label{ineq}
\end{equation}
To derive these inequalities  one merely observes that for any $t$
the integral from $-\infty$ to $1$ over $x$ over the function $(1-
x)(x-t)^2 F(x)$ is
positive; on the other hand this integral  is a second order polynomial
in $t$ and,  hence, its discriminant must be  negative.

A sketch of a function satisfying all these requirements is given on
Fig. 11.

A natural desire to extend the formalism
described above
to the semileptonic inclusive transitions $b\rightarrow c
l \nu$ encounters  serious technical difficulties. The essence of
the problem is as follows.
The final quark $c$ can be  treated as heavy, although at
the same time , $m_c^2\ll m_b^2$. The ratio $m_c^2/m_b^2
\approx 0.07$ is a small parameter while $m_c^2/(\bar\Lambda
m_b)
\sim 1$.  Under the circumstances the type of the distribution
function
describing  the primordial motion of  $b$  inside  $B$
and determining the measurable structure functions $w_1$ to $w_5$
will depend on
the value of  $|\vec q|$, and the scaling property
-- dependence on one particular combination of variables -- is lost.

The $q_0, |\vec q|$ plane is shown on Fig. 12. In the free quark
approximation the transition probability is proportional to
$\delta (m_b-q_0 - E_c)$ where $E_q = (m_c^2 +{\vec q}^2)^{1/2}$,
and all events are concentrated along the line indicated on Fig. 12.
At the hadronic level the phase space consists of the full triangle,
with one   side curved. The end point band is also curved.

The fact that one side of the triangle is distorted compared to
$b\rightarrow u l\nu$ is not crucial. What is important is the change
of dynamics as we move from the upper left corner to the lower
right one. In the case of $b\rightarrow  u l\nu$ moving along the
end point band in this direction does not affect the measured
structure functions (apart from the extreme domain of soft $u$
-- the exclusive resonance domain --  where our description fails
altogether). The situation is different in the
$b\rightarrow c l\nu$ transition.

If $|\vec q|^2\gg m_c^2$ one recovers \cite{motion3}
the same
light-cone
function $F(x)$ as in the transition $b\rightarrow ul\nu$ or
$b\rightarrow s\gamma$.
Modifications  are  marginal. First, some extra terms
explicitly proportional to $m_c/m_b$ are generated in the structure
functions due to the fact that
$\not\!\!{\cal P} +m_c$ replaces $\not\!\!{\cal P}$
in the numerator of the quark Green function. Moreover,
if in the $b\rightarrow ul\nu$ transitions the scaling variable in the
end-point domain is
\beq
x=\bar\Lambda^{-1}(q_0 +|\vec q| -m_b) ,
\eeq
in the $b\rightarrow cl\nu$ transition it is shifted by a constant term
of
order 1,
\beq
x=\bar\Lambda^{-1}(q_0 +|\vec q| -m_b) +\frac{m_c^2}{\bar\Lambda
m_b} \, .
\label{MNsv}
\eeq
To see how this shift occurs \cite{motion3} and to reveal limitations
of the approximation let us start from the parton model variable
$m_b -q_0 -E_c$. In the limit $|\vec q |^2/m_c^2\rightarrow\infty$
inside the end point band formally one may substitute $E_c$ by
$$
E_c\rightarrow  |\vec q | +\frac{m_c^2}{2|\vec q |}
\rightarrow |\vec q | + \frac{m_c^2}{m_b}\left(
1+{\cal O}(m_c^2/m_b^2)\right) \, .
$$
If $m_c^2 = {\cal O}(\bar\Lambda m_b)$ formally one may discard
the ${\cal O}(m_c^2/m_b^2)$ correction. In this way we arrive at the
scaling variable (\ref{MNsv}).

It is worth emphasizing that the occurrence of the light-cone
distribution function in this regime, the same as in the $b\rightarrow
u$ transition,  is a remarkable fact.  Indeed, if we could examine the
measured structure functions ``in the microscope" in these two cases
we would see that their microstructure is quite different.  As was
already mentioned, in the $b\rightarrow u$ transition the end point
band is saturated by the production of $\sim m_b/\bar\Lambda$
states, with the spacing between the individual states of order
$\bar\Lambda^2/m_b$. In the $b\rightarrow c$ transition, even if
we are in the upper left corner of the phase space where the
light-cone distribution is relevant, the number of the states produced
is
$\sim m_b/m_c\sim m_c/\bar\Lambda$ and the spacing is of order
$\bar\Lambda^2/m_c$.  We deal with a much coarser structure in
the latter case, and still  all resonance contributions being summed
up must
add up to produce the light-cone distribution, formally the same one
that is created  by a much larger number of resonances in the
$b\rightarrow u$ transition. (Purely theoretically we can not predict
fine grain versus coarse grain composition of the structure functions
if we limit ourselves to the leading twist. Only analysis of all twists
could resolve these details, would this analysis be possible).

The word ``formally" is used above three times, not accidentally.
Practically
in the $b\rightarrow c l\nu$ transition $|\vec q|$ can never be much
larger than $m_c$. Indeed, the {\em maximal} value
of $|\vec q|$, corresponding to $q^2 = 0$ is $(m_b^2 - m_c^2)/2m_b
\sim 2$ GeV, that  is only $\sim 1.5 m_c$. Therefore, only by
stretching a point and only in a narrow domain near $q^2 = 0$,
one can expect that the light cone function of the variable
(\ref{MNsv}) is, perhaps, more or less relevant.

As $|\vec q|$ decreases and becomes  less than
$m_c$ (this regime takes place in a large part of the phase space)
the light-cone distribution function
becomes irrelevant. The measurable structure functions are
determined by a different distribution -- the light-like vector
$n_\mu$ in Eq. (\ref{T23})  is replaced by
$$
w_\mu = \left( 1, \frac{\vec q}{E_c}\right)\, ,
$$
 and it becomes clear that the would be scaling variable
$x=\bar\Lambda^{-1}(q_0 +E_c -m_b)$ fails to represent all
dependence of the structure functions on $q_0$ and $|\vec q|$.
When $q^2$ approaches its maximal value,
$$
q^2_{max} = (M_B-M_D)^2 ,
$$
$|\vec q|$ tends to zero and we eventually approach the SV regime
which I have
already
discussed, with fascination, in the toy example above. In the SV
limit the velocity of $H_c$ produced is small,
and the structure functions probe the primordial motion  described
by the temporal distribution function $G(x)$ where now
$$
x\approx \bar\Lambda^{-1}(q_0 -\Delta m),
$$
$\Delta m $ is the quark mass difference coinciding, to the
leading order with $M_B-M_D$.

Thus, changing $q^2$ from zero to $q^2_{max}$ results in an
evolution of the distribution function appearing in theoretical
formulae for $d\Gamma (B\rightarrow X_cl\nu)$, from
light-cone to  temporal, through a series of intermediate
distributions. The
physical reason for this evolution is quite clear -- what distribution
function is actually measured depends on the parton-model velocity
of the quark produced in the $b$ decay.
In the limiting cases of very large recoil and very small recoil
the problem is solved in the sense that the structure functions
are expressed in terms of the light-cone and temporal distribution
functions, respectively.  The intermediate case $|\vec q |\sim m_c$
is not worked out in detail so far. It is beyond any doubt, however,
that the parton-model type scaling will not take place.

\newpage

\section{Lecture 5. Including Hard Gluons. Generalities of the
Operator Product Expansion.}

\renewcommand{\theequation}{5.\arabic{equation}}
\setcounter{equation}{0}

Finally the time comes when I can not ignore any more the existence
of
hard gluons. Hard gluons are mere nuisance from the point of
view of the theory of hadrons since they play no, or very little, role
in the structure of the low-lying hadronic states. Yet, if we want to go
beyond purely academic exercises, however beautiful they might
look, and descend down into a messy world of real hadronic physics,
hard gluons can not be forgotten about since they ``contaminate"
nearly every experimentally measurable quantity. To make contact
with the real world we have to consider interplay between
the soft and hard physics.

The hard gluons manifest themselves in many ways. They contribute
to the coefficient functions in the effective Lagrangian
(\ref{N2}) obtained by integrating out all degrees of freedom with
the characteristic frequencies down to $\mu$. They show up in the
calculations of the total decay rates and spectra discussed in Lectures
3 and 4 resulting in perturbative  corrections
which, in some instances, change the answer quite drastically. They
result in the fact that all basic parameters of the
heavy quark physics -- the heavy quark mass, $\bar \Lambda$,
$\mu_\pi^2$ and so on -- generally speaking, become $\mu$
dependent
and can not be treated as universal constants. Here we will address
some of these issues in brief.

\subsection{Calculation of the effective Lagrangian}

I have already started discussing this topic in Sect. 1.2.
The original QCD Lagrangian (\ref{lagr})
is formulated at very short distances.
In principle, it codes all information necessary for calculation of all
observable amplitudes. We just do the functional integral and ...
Alas, there are very few functional integrals that can be calculated
analytically; numerical evaluation on lattices may take years,
and I even dare to assert that some amplitudes will never be
calculated that way.  So, we take the original Lagrangian and start
evolving it down, integrating out all fluctuations with the frequencies
$\mu <\omega < M_0$ where $M_0$ is the original normalization
point, and $\mu$ will be treated, for the time being, as a current
parameter. In this way we get the Lagrangian which has the
form
\beq
{\cal L} =
\sum_n  C_n(M_0;\mu ){\cal O}_n(\mu )\, .
\label{10}
\end{equation}
The coefficient functions $C_n$ represent the contribution of virtual
momenta from $\mu $ to $M_0$. The operators ${\cal O}_n$ enjoy
full rights of the Heisenberg operators with respect to all field
fluctuations with frequencies less than $\mu$. The sum in Eq.
(\ref{10}) is infinite --
it runs over all possible Lorentz singlet gauge invariant operators
with the appropriate quantum numbers; for instance, if $CP$ is
conserved, only $CP$-even operators will appear in (\ref{10}).
If, say,  the electromagnetic processes are included,
the operators in the Lagrangian (\ref{10}) may contain the
photon and electron fields, and  so on. All operators can be ordered
according to their dimension; moreover, we can use the equations
of motion stemming from the original QCD Lagrangian to get rid  of
some of the
operators in the sum. Those operators that are reducible to full
derivatives give vanishing contributions to the physical (on mass
shell) matrix elements and can thus be  discarded as well.

If one just abstractly writes the expression (\ref{10}) one is free
to take any value of $\mu$; in particular, $\mu= 0$ would mean
that {\em everything} is calculated and we have the full $S$ matrix,
all conceivable amplitudes, at our disposal.  Nothing is left to be done.
In this case Eq. (\ref{10}) is just a sum of all possible amplitudes.
This sum  then must be written in terms of the physical hadronic
states,
of course, not in terms of the quark and gluon operators
since the latter degrees of freedom are simply non-existent at large
distances.

This is day-dreaming, of course. Needless to say that in our explicit
calculation of the coefficient functions we have to stop somewhere,
at such virtualities that the quark and gluon degrees of freedom are
still relevant, and the coefficient functions $C_n(M_0, \mu )$ are still
explicitly calculable.  On the other hand, for obvious reasons it is
highly desirable to have $\mu$ as low as possible.  In the heavy
quark theory there is an additional requirement that $\mu$ must be
much less than $m_Q$.  The process of calculating the coefficients
$C_n(M_0, \mu )$ is called {\em matching} in the more standard
presentation of HQET. Actually we see that this procedure is nothing
else than a generalization of Wilson's idea of the renormalization
group and the (Wilsonean) operator product expansion. Using the
standard OPE language has an evident advantage: all well-studied
elements of the latter approach can be immediately adapted in the
environment of the
heavy quark expansions. In particular all parameters one
can read off from the Lagrangian (\ref{10}) depend on $\mu$
(including, say, the heavy quark mass).  Let us assume that
$\mu$ is large enough so that $\alpha_s (\mu )/\pi \ll 1$,
on the one hand, and small enough so that there is no large gap
between $\Lambda_{\rm QCD}$ and $\mu$. The possibility to make
such a choice of $\mu$ could not be anticipated {\em apriori}
and is an extremely fortunate feature of QCD, a gift from the Gods.
Quarks and gluons with the offshellness larger than $\mu$
chosen that way are called hard.

Needless to say that the parameter $\mu$ is in our minds, not in
Nature. All observable amplitudes must be $\mu$ independent.
The $\mu$ dependence of the coefficient functions $C_n$ must
conspire with that of the matrix elements of the operators ${\cal
O}_n$ in such a way as to ensure this $\mu$ independence of the
physical amplitudes.

What can be said about the calculation of the coefficients $C_n$ ?
Since $\mu$ is sufficiently large, see above, the main contribution
comes from  perturbation theory. We just draw all relevant Feynman
graphs and calculate them, generating an expansion in $\alpha_s
(\mu )$
which for brevity I will denote by $\alpha_s$, with the argument
omitted,
$$
C_n= \sum_l a_l \alpha_s^l \, .
\label{asum}
$$
 Sometimes some graphs will contain not only powers of $\alpha_s
(\mu )$ but powers of $\alpha_s\ln (m_Q /\mu )$. This happens if
the anomalous dimension of the operator ${\cal O}_n$ is
nonvanishing -- quite a typical situation -- or if a part of a
contribution to $C_n$ comes from characteristic momenta of order
$m_Q$ and is, thus, expressible in terms of $\alpha_s (m_Q)$,
and we rewrite it in terms of $\alpha_s (\mu )$. Nevermind,
this is a trivial technicality. You are supposed to know how to sum up
these logarithms.

As a matter of fact the  expression (\ref{asum})
is not quite accurate theoretically. One should not forget that,
in doing the loop integrations, in $C_n$ we {\em must} discard the
domain
of virtual momenta below $\mu$, by definition of $C_n(\mu )$.
Subtracting
this domain from the perturbative loop integrals we introduce in
$C_n$
power corrections of the type $(\mu /m_Q)^n$ by hand.
In principle, one should recognize the existence of such corrections
and try to learn how to deal with them. The fact that they are there
was realized long ago (see e.g. V. Novikov {\em et al}, Ref.
\cite{Wilson}) and then largely ignored.
If it is possible
to
choose $\mu$ sufficiently small these corrections may be
insignificant
numerically and can be omitted. This is what is actually done in
practice. This is one of the elements of a simplification of the
Wilsonean operator product expansion.  The simplified version is
called the {\em practical version of OPE}, see below. Certainly, at the
modern stage of the theoretical development it is desirable to return
to the issue to engineer a better procedure than just discarding these
$\mu /m_Q$ terms in the coefficient functions.
Attempts in this direction are under way \cite{Ji}.

Even if perturbation theory  dominates in the coefficient functions
they still
contain also  nonperturbative terms coming from short distances.
Sometimes  they  are referred to as noncondensate nonperturbative
terms. An
example is provided by the so called direct instantons with the sizes
of order $m_Q^{-1}$.
These contributions  fall off as high powers of $\Lambda_{\rm
QCD}/m_Q$ and are very poorly controllable theoretically.
Since the fall off of the
noncondensate nonperturbative corrections is extremely steep,
basically
the  only thing we need to know is a critical value of $m_Q$. For
lower values of $m_Q$
 no reliable theoretical predictions are possible at
present. For higher values of $m_Q$  one can  ignore the
noncondensate
nonperturbative contributions. There are good reasons to believe
that the $b$ quark, fortunately, lies above the critical point.
Again, I must add that the noncondensate
nonperturbative contributions are neglected in the practical version
of OPE.

(Do we see  seeds of the nonperturbative contribution in Eq.
(\ref{asum})? Yes, we do.
At any finite order the perturbative contribution
is well-defined. At the same time, if
the coefficients in the series (\ref{asum})  grow factorially with $l$ --
and this is actually the case --
the tail of the series,
$l> 1/\alpha_s$,   must be regularized which may bring in terms
of order
\begin{equation}
\exp {(-C/\alpha_s (m_Q))} \sim \left(
\frac{\Lambda_{QCD}}{m_Q}\right)^\gamma
\label{12}
\end{equation}
where $C$ is some positive constant and the exponent $\gamma$
need not be integer. In a sense, one may say that contributions to
$C_n$ of this
type are vaguely related to diagrams with $1/\alpha_s$ hard gluon
loops.)

Thus, two sources of nonperturbative corrections in the physical
amplitudes are indicated. Those due to nonperturbative terms
in the coefficient functions are systematically ignored (and, perhaps,
rightly so, as I tried to convince you)
in these lectures and in all works based on the practical version of
OPE which constitute the overwhelming majority of all works
devoted to the $1/m_Q$ expansions. The second source is operators
of  higher dimensions in the Lagrangian (\ref{10}), the so called
condensate corrections. The latter
were in the center of our attention; they generate   the $1/m_Q$
expansions discussed above. One new element which I would like to
add here, is that the series of $1/m_Q$ terms generated by
higher-dimensional operators is also asymptotic and divergent in
high orders \cite{Shifconf}. Of course, we always calculate only one,
at best two,
first  $1/m_Q$ corrections, truncating the series.
If, however, one would ask what the impact of the high-order tail of
the power series is, the answer would be: this tail is reflected in
exponentially small terms $\sim\exp (-m_Q)$. This type of
contribution is certainly not seen in OPE truncated at any
finite order.
A transparent example is again
provided by instantons. This time one has to fix the
size of the instanton $\rho$ by hand,  $\rho_0\sim
 \Lambda^{-1}$. Then their contribution to physical amplitudes is
${\cal O}(\exp (-m_Q\rho_0
))$.
 The relation between exp$(-m_Q\rho_0)$ piece and the high-order
terms of the
 power series is conceptually akin  to the connection between
$\exp (-1/\alpha_s)$ terms and $l\sim 1/\alpha_s$ orders
in the perturbative expansion.

Summarizing, Wilsonean OPE (\ref{10}) leads
to expansions in different parameters.
Purely logarithmic terms
$(\ln m_Q)^{-l}$ are due to ordinary perturbation theory.  Terms of
the
type $(m_Q^2)^{-k} (\ln m_Q)^{-\gamma}$ reflect higher-dimension
operators and direct instantons. In the
former case the values of $k$ are  integer,  the latter case
may produce non-integer values of $k$.
In the practical version of OPE we calculate the coefficient functions
perturbatively. All
non-perturbative terms come from condensates within this
approximation. The condensate power series is truncated: only those
operators whose dimension is smaller than some number are
retained.

The practical version of
OPE was heavily used in connection with the QCD sum rule method.
It was checked \cite{SVZ} that in the
majority of channels this is a valid approximation allowing one to
calculate  in the Euclidean domain down to $\mu$ as low as
0.6 or 0.7 GeV.
The validity of this
approximation is an element of luck; it relies, among other things, on
the fact that $\Lambda_{QCD}$ is significantly smaller than 1 GeV,
and $\alpha_s{\rm (1 GeV)}/\pi$  is already a small parameter.

I hasten to add that  some
exceptional channels where the practical version of OPE
fails at much larger values of $\mu$ were detected in the analysis of
glueballs \cite{Hadrons}. It would be interesting to explore the issue
in the context
of the heavy quark theory. The existing theory
gives no
clues for establishing the domain of validity of the practical version
of OPE
from first principles, neither does it tell us about when the
exponential
terms, not visible by standard methods, become negligibly small. At
this point
we have to rely on indirect methods and phenomenological
information.

\subsection{Untangling hard gluons from soft ones}

The coefficient functions $C_n$ in  Eq. (\ref{10}) contain, generally
speaking,  an infinite number of  perturbative terms, and
non-perturbative contributions of different types.  Practically we
often calculate them to the first nontrivial order. For instance,
in Lecture 3 we treated the transition operator in the Born
approximation; thus, all coefficients in OPE were found to order
$\alpha_s^0$. For a number of purposes (although not always, of
course) such a calculation, ignoring the hard gluon exchanges
altogether, is quite
sufficient. Let me remind that by hard gluons I mean those with
offshellness from $\mu$ up to $m_Q$. Let us ask a question  -- can
one find a theoretical parameter which would justify the
approximation of no hard gluon exchanges? In other words, does a
parameter exist that would allow one to switch the hard gluons
on/off ?

Each extra hard loop contains the running gauge coupling $\alpha_s
(\mu )$,
\beq
 \frac{\alpha_s (\mu ) }{\pi} = \frac{2}{b} \left[
\ln\left(\frac{\mu}{\Lambda_{\rm QCD}}\right)\right]^{-1}
\label{alphas}
\eeq
where $b$ is the first coefficient of the Gell-Mann-Low function,
$$
b = \frac{11}{3} N_c - \frac{2}{3}n_f\, .
$$
If we could make $b$ very large the running law of $\alpha_s$
would be very steep effectively switching off all hard gluons.
Indeed, once $\mu$ is bigger than, say, 2$\Lambda_{\rm QCD}$
and $b\rightarrow\infty$ the gauge coupling constant $\alpha_s
(\mu )
\rightarrow 0$. The first idea which immediately comes to one's
mind is to make $b$ large by tending the number of colors $N_c$
to infinity.  Alas, this idea does not work. It is known from the early
days of QCD that the expansion parameter in all planar diagrams is
$N_c\alpha_s$, not $\alpha_s$ itself \cite{tHooft}. Thus, the diagram
of Fig.
13 is of the same order in $N_c$ as the Born graph of Fig. 1.
So, we have to rely on numerical smallness of $1/b$. For instance, in
the theory with three light flavors and three colors $b=9$,
quite a large number.  This is not the first time in physics we have to
deal with numerical enhancements. It is true that it is always better
to have an adjustable parameter, which could be sent to infinity at
will, than to deal with just a large fixed number. It is quite
unfortunate that we do not have such a parameter at our disposal in
the real world QCD.  If one still wants to have $b$ as an adjustable
parameter one could try a trick. Let us assume that, apart from
quarks and gluons, our theory contain quark {\em ghost} fields.
These ghost fields are perfectly the same as the quark fields,
with a single exception -- each ghost loop has an extra minus sign.
The quark ghost fields may or may not have a mass term.
Let us say that they do have a mass term $m_{gh}$ equal to
$\Lambda_{\rm QCD}$.  Then  they would automatically decouple in
the soft contributions.
The action of such a crazy theory has the form
\beq
iS = iS_{\rm QCD}
+ \sum_{q}\bar q_{gh} (i\not\!\!{D}- m_{gh}) q_{gh}
=
$$
$$
iS_{\rm QCD}
-N_{gh}\ln\,{\rm det}\, \left\{ i\not\!\! D -m_{gh}\right\}
\label{fQCD}
\eeq
where $S_{\rm QCD}$ is the action of quantum chromodynamics, see
Eq. (\ref{lagr}), and $N_{gh}$ is the number of the quark ghosts, a
free parameter
assumed to be large. Notice the ghostly minus sign in front of the
logarithm of the determinant. After some thinking one may conclude
that, perhaps, this theory is not so crazy.
Let us postulate that the initial particles we consider belong to our
world -- $B$, $D$ and so on --
i.e.  they do not carry these quark ghosts. Of course, if $B$ decays
the quark ghosts do appear in the final state, and the probability of
their emission is negative. This does not mean, however, that
the total amplitude is not unitary, as one could suspect from the fact
that we introduced the fields with a wrong metric.
Indeed, it is obvious that the only role of  the quark ghosts is to
switch off all hard gluons in the limit $N_{gh}\rightarrow\infty$
since in this limit
$$
b= 11-\frac{2}{3}N_f + \frac{2}{3}N_{gh} \rightarrow\infty
$$
and $\alpha_s (\mu )\rightarrow 0$, according to Eq. (\ref{alphas}).
In particular, the diagram of Fig. 13 where the gluon line is dressed
with the bubble insertions vanishes. All soft contributions
with $\mu \lsim \Lambda_{\rm QCD}$ remain intact, however,
and the positivity of the forward scattering amplitudes is not
violated.

If there exists a stringy representation of QCD it should refer to
the fake ``QCD", Eq. (\ref{fQCD}), rather than to the real one since in
the
string amplitudes there is no place for hard gluons.

The idea of treating $b$ as a numerically large parameter is not new
in QCD. In the purely perturbative calculations it constitutes the basis
of the so called BLM approach \cite{BLM}. Originally the BLM
approach was
engineered as a scale-setting procedure intended as a substitute for
full computations of ${\cal O}(\alpha_s^2 )$ corrections. Assume that
${\cal O}(\alpha_s )$ corrections in some amplitude are known
exactly. In order $\alpha_s^2$ typically one has to deal with a large
number
of graphs. The idea is to pick up only those which contain
a ``large parameter", $b\alpha_s^2$, presuming that the  graphs
without $b$ are numerically suppressed. Typically there are very
few graphs producing $b\alpha_s^2$. By doing so we can
approximately
determine the scale $\mu$ in the ${\cal O}(\alpha_s )$ term without
labor and time-consuming
calculation of a large number of all $\alpha_s^2$ contributions.
Later,  it was suggested \cite{BRN, BRN1} to extend the prescription
of the ``$b$ graph dominance" to even  higher orders, a more
extremist and dangerous approach. In both cases the limit of large
$b$ is used to get some information about perturbation theory.
I  use this limit in order to switch off the perturbative hard gluons in
the first place pushing the theory to the mode where only the soft
gluons
survive, hopefully providing a more transparent picture of the
infrared dynamics determining the regularities of the hadronic
world.

\subsection{Impact of hard  gluons}

Having said all that let us return to the real world where
$b$ is fixed, not infinity, and examine several examples
of corrections due to  hard gluons.
An instructive example to begin with is the calculation of the
coefficient in front of the chromomagnetic operator ${\cal O}_G$ in
the effective Lagrangian ${\cal L}_{\rm heavy} (\mu )$,
see Eq. (\ref{N2}) \footnote{In the limit $b\rightarrow\infty$
the coefficient given in this expression does indeed vanish, in full
accord with the argument of the previous section.}
This coefficient takes into account virtual gluons with offshellness
from $\mu$ to $m_Q$.

The line of reasoning is as follows. Our starting point is
$\mu = m_Q$. At this normalization point the Lagrangian we deal
with is the QCD Lagrangian (\ref{lagr}) with the coupling constant
and heavy quark mass normalized at $m_Q$. We then descend a little
further, down to $\mu$ equal to a finite fraction of $m_Q$, say,
$m_Q/5$. This is sufficient to make the $Q$ quark nonrelativistic and
make all nonrelativistic expansions work. Being interested only in
logarithms of $m_Q$ we ignore any nonlogarithmic $\alpha_s$
corrections that may appear at this stage. The nonrelativistic
expansion of the Lagrangian $\bar Q (i\not\!\!{D} - m_Q) Q $
implies that the operator $\bar Q (i/2)\sigma G Q$ appears with the
coefficient $C_0=1/(2m_Q)$. Further evolution down to $\mu =$
several units $\times\Lambda_{\rm QCD}$ will change $C$;
in particular, at one loop
\beq
C_0 \rightarrow C(\mu ) = C_0 \left( 1
+\gamma\frac{\alpha_s}{4\pi}\ln
\frac{m_Q^2}{\mu^2} + \mbox{non-log terms}\right)
\label{defgam}
\eeq
where $\gamma$ is a number.
Our goal is to find $\gamma$ and then sum up all leading logarithms.
This is not the end of the story, however, if one wants
to represent the result in the form (\ref{N2}), where the sum over
the operators includes only  the Lorentz invariant ones. The leading
operator  is
${\cal L}^0_{\rm heavy}(\mu )$. The coefficient $C(\mu )$
should be represented as
$$
C(\mu ) = C_0 +\left( C(\mu ) - C_0\right) \, ;
$$
then $C_0$ is swallowed back in the definition of ${\cal L}^0_{\rm
heavy}(\mu )$ while the expression in the brackets represents $c_G$
in Eq. (\ref{N2}).

The relevant one-loop graphs are depicted on
Fig. 14. At first sight the number of diagrams is rather large,
and the computation might seem rather cumbersome. My task is to
reduce it to a back-of-the-envelope calculation by using several
smart observations and the background field technique.

First of all, as it was already mentioned, we will be hunting only for
the terms containing
$\alpha_s \ln m_Q/\mu$ omitting all $\alpha_s$ terms without
logarithms.
 The logarithms $\ln m_Q/\mu$ have a dual nature
-- they appear from the loop integrations where the integrands
presents an infrared limit with respect to heavy quarks $Q$ while
presenting simultaneously the ultraviolet limit with respect to
gluons. That is why they were called hybrid in Ref. \cite{hybrid}, the
 paper
where these logarithms were discovered. In the language of HQET
they are referred to as matching logarithms.

Secondly, in this perturbative calculation we will naturally discard all
$1/m_Q$ corrections.

The closed circle on the diagrams of Fig. 14 denotes the vertex
$(i/2)\sigma^{ij}G_{ij} = -\vec\sigma\vec B$. Let us consider for
definiteness only one term with $i,j =1,2$, i.e. $-\sigma_z B_z$
keeping in mind that other terms will give the same.

It is absolutely obvious that the graph of Fig. 14$e$ gives
no contribution in our approximation.  Indeed, the very existence of
this graph is due to the nonlinear term $[A_1 A_2]$
in the definition of  $G_{12}$. However, neither $A_1$ nor $A_2$
interact with the heavy quark in the leading in $1/m_Q$
approximation, as it is clear from Eq. (\ref{Lzerore}), only $A_0$.
(We work in the rest frame of the heavy quark $Q$.)

Next, let us analyze the diagrams $c$ and $d$. To this end it is
convenient to write the gluon Green function in the background field.
For a detailed exposition of the technique the reader is referred to
the review paper \cite{NSVZ}. For our purposes we need so little that
it is quite in order to carry out all necessary derivations here.
Let us split the four-potential $A_\mu$ in two parts --
the external field $(A_\mu )_{\rm ext}$ and the quantum part
$a_\mu$ which will propagate in loops,
\beq
A_\mu = (A_\mu )_{\rm ext} + a_\mu
\label{split}
\eeq
As explained in Ref. \cite{NSVZ} the gauge conditions on $(A_\mu
)_{\rm
ext}$ and $a_\mu$ may be different, for instance,
the Fock-Schwinger gauge with respect to the background field and
the Feynman gauge with respect to the quantum field.
Here we do not need to discuss the gauge condition on $(A_\mu
)_{\rm ext}$. The quantum field $a_\mu$ will be treated in the
Feynman gauge. The definition of the gluon propagator in the
background field is standard:
\beq
D_{\mu\nu}^{ab} = \langle T\{ a_\mu^a (x) , a_\nu^b (0)\}\rangle \, .
\eeq
The Lagrangian of the quantum gluon field in the Feynman gauge has
the form
\beq
{\cal L} = -\frac{1}{2} (D_\mu^{\rm ext} a_\nu^a )^2
+ ga_\mu^a (G^b_{\mu\nu})_{\rm ext} a_\nu^c f^{abc}
\label{gllagr}
\eeq
plus cubic and higher order terms in $a_\mu$ plus the ghost terms
--
all irrelevant for the calculation at hand. Here
$$
D_\mu^{\rm ext} a_\nu^a = \partial_\mu a_\nu^a
+gf^{abc}(A_\mu^b)_{\rm ext} a_\nu^c\, .
$$
The second term in the Lagrangian (\ref{gllagr}) describes the
interaction of the
magnetic moment of the  gluon quantum with the background field.
If we switch off this magnetic terms for a short while we
immediately
observe that both graphs, Fig. 14$c$ and $d$, vanish. Indeed, the
Lorentz structure of the first term in Eq. (\ref{gllagr}) is such that
the Green function generated by it is
obviously proportional to $g_{\mu\nu}$. Hence the loops
displayed on Figs. $c$ and $d$
can not be
formed. Say, the diagram $c$ requires converting the $a_i$ quantum
leaving the vertex into
the $a_0$ quantum coupled to the heavy quark. Let us now switch
on the magnetic term and take into account the fact that the
background field is chromomagnetic, not chromoelectric
(I remind that we are interested in the vertex $-\vec\sigma\vec B$.)
This means that the graph $c$ still vanishes since the
conversion of $a_i$ into $a_0$ can only take place
in the chromoelectric background ($G_{0i}^{\rm ext}$). The diagram
$d$ is not vanishing, however, and is readily calculable.
We start from the vertex
$$
(i/2) \bar Q\sigma^{12}G_{12}^ct^c Q\rightarrow (i/2)
\bar Q\sigma^{12} g a_1^a a_2^b f^{abc}t^c Q \ ,
$$
make one insertion of  the magnetic term in the Lagrangian
(\ref{gllagr}),
$$
i ga^{\rho \tilde a} (G_{\rho\phi}^{\tilde b})_{\rm ext} a^{\phi\tilde
c}
f^{\tilde a\tilde b\tilde c}\, ,
$$
and after that one can take the free  gluon propagators, which yields
\beq
(i/2)
2g^2\bar Q\sigma^{12}f^{abc} t^c Q f^{a\tilde b b} G_{12}^{\tilde b}
(-i)^2 i \int\frac{1}{k^4}\frac{d^4k}{(2\pi )^4}
\label{equlo}
\eeq
where the factor 2 comes from two different ways of pairings.
The integral over $dk$ is evidently logarithmically divergent both at
the upper and lower ends and should be cut off at $m_Q$ from above
and at $\mu$ from below.   This logarithmic divergence should be
welcome since  in this way we are going
to get the desired hybrid logarithm.
Equation (\ref{equlo}) immediately leads to
$$
-2 \frac{N_cg^2}{16\pi^2}\,  (i/2) \, \bar Q \sigma^{12}G_{12}^c t^c Q
\ln \left(\frac{m_Q^2}{\mu^2}\right)
$$
where $N_c$ is the number of colors ($N_c=3$).
In other words the factor produced by  the one-loop graph of Fig.
14$d$ is
\beq
\gamma\mid_{{\rm Fig.\,\, 14}d} =
-2 N_c\, .
\label{fact1}
\eeq

The last step in our exercise is calculation of the diagrams
of Fig. 14$a$ and $b$. The Feynman integral for the diagram $a$
is quite trivial,
\beq
g^2\int \frac{d^4 k}{k^4} \bar Q \frac{1}{k_0}
t^a (-\sigma_z B_z^b t^b) t^a \frac{1}{k_0}
Q (-i)\frac{1}{k^2}
\label{zerqed}
\eeq
where $k$ is the virtual gluon momentum. Now, a minute reflection
shows that in the Abelian theory (i.e. if the gluons were photons and
the diagram of Fig. 14$a$ was considered in QED) this contribution
must be exactly canceled by that coming from diagrams $b$.
This assertion can be traced back to the nonrenormalization of
the $\bar Q\gamma_\mu Q$ vertex in QED. (We should take into
account the fact that the hybrid logarithms do not depend on the
Lorentz structure of the vertex at all \cite{hybrid} and are the same
for
$\gamma_\mu$ and $\sigma_{\mu\nu}$.) This observation implies
that in QCD the net effect of the two diagrams 14$b$ reduces to
replacing
Eq. (\ref{zerqed}) by
$$
g^2\int \frac{d^4 k}{k^4}\, \frac{1}{2} \bar Q \left( t^a [t^bt^a]
+[t^at^b]t^a\right) (-\sigma_z B_z^b )
Q \frac{-i}{k_0^2}\frac{1}{k^2} =
$$
\beq
\frac{N_cg^2}{2}\, \int \frac{d^4 k}{k^4}\,\bar Q (-\sigma_z B_z^b t^b)
Q\frac{i}{(k_0+i\epsilon)^2}\frac{1}{k^2+i\epsilon}\, .
\label{k3}
\eeq
The $i\epsilon$ prescription indicated explicitly defines the
integration contour (Fig. 15).  We  first do the $k_0$ integration
using the residue theorem, then the remaining $d^3k$ integration
and arrive at
\beq
\frac{N_cg^2}{16\pi^2}\,  (i/2) \, \bar Q (-\sigma_z B_z^c t^c) Q
\ln \left(\frac{m_Q^2}{\mu^2}\right)
\eeq
leading to the following ``dressing" factor due to diagrams 14$b$
and $c$:
\beq
\gamma\mid_{{\rm Fig.\,\, 14}a+b} =
N_c\, .
\label{fact2}\, .
\eeq

The diagram 14$f$ must be discarded in the background field
calculation -- it merely renormalizes the gauge coupling constant
included in the definition of ${\cal O}_G$.

The overall one-loop dressing factor is obtained by adding up Eqs.
(\ref{fact1}) and (\ref{fact2}),
\beq
\gamma =
-N_c = -3\, .
\label{fact3}
\eeq

Now
the renormalization group  allows us to sum up all  leading
log terms, in a standard manner;  the summation leads
to Eq. (\ref{N3}).
The same result can be rephrased as follows: in the $1/m_Q$
expanded effective Lagrangian ${\cal L}_{\rm heavy} (\mu)$
 the overall
coefficient in front of ${\cal O}_G$
is
\beq
C(\mu ) =
\left(\frac{\alpha_s(\mu)}{\alpha_s(m_Q)}\right)^{-
\frac{3}{b}}\, .
\label{ad}
\eeq
This is nothing else than the reflection of the hybrid anomalous
dimension of the
operator ${\cal O}_G$  found in Ref. \cite{Falk}.

It is curious to note that ${\cal O}_\pi$, the second operator of
dimension
5 (see Eq. (\ref{kinop})),  has vanishing anomalous dimension which
can be proven with no calculations  in no time.

To see that this is
indeed the case
we merely repeat the argument preceding and following Eq.
(\ref{defgam}). Let us assume for a short while that the hybrid
anomalous dimension of the operator ${\cal O}_\pi$ is non-vanishing.
Then after evolving to a low normalization point its coefficient
gets renormalized, and there is no way one could absorb ${\cal
O}_\pi$ back into a Lorentz invariant expression
$\bar Q(\not\!\!{\cal P} - m_Q)$  in the effective
Lagrangian.  Needless to say that ${\cal L}_{\rm heavy} (\mu )$
(before the $1/m_Q$ expansion)
must be expressible in terms of the Lorentz invariant structures.

A close line of reasoning leading to the same conclusion takes
advantage of the
expansion (\ref{N9}),
\beq
\bar Q Q - \frac{1}{2 m_Q^2}
\bar Q\frac{i}{2}\sigma G Q
=\bar Q\gamma_0Q -
\frac{1}{2 m_Q^2}
\bar Q{\vec\pi}^2Q + ...
\label{zad1}
\eeq
where the dots denote terms of the higher order in
$1/m_Q$. The left-hand side is Lorentz scalar while the right-hand
side
is written as a sum of terms that are {\em not} Lorentz scalars
individually.  The matrix element of $\bar Q\gamma_0 Q $ has the
meaning of
energy $E$ (which at small velocities reduces to $m + {\vec
p}^2/2m$) while that of the second term on the right-hand side
has the meaning of $-{\vec p}^2/2m$. The first term is not
renormalized by the gluon dressings, of course. If the coefficient
of the second term was distorted by the anomalous dimension, the
cancellation of the Lorentz noninvariant part
would be ruined, and the right-hand side could not be equal to the
left-hand side.

Concluding this section let us discuss
the impact of the hard gluons on the scaling law of, say, pseudoscalar
coupling $f_P$ defined in Sect. 3.4. In this section it was shown that
$f_P \sim m_Q^{-1/2}$ modulo logarithmic corrections. Now we
address the issue of the logarithmic corrections due to the hybrid
anomalous dimension of the current $\bar Q\gamma_\mu\gamma_5
q$. Let us add to the original QCD Lagrangian the term $\Delta{\cal
L}={\cal A}_\mu
\bar Q\gamma_\mu\gamma_5 q$ where ${\cal A}_\mu$
is an auxiliary $c$-number field and evolve $\Delta{\cal L}$ down to
$\mu$. The result of this evolution is the anomalous dimension
 $$
(\bar Q\gamma_\mu\gamma_5 q)_{m_Q} =
\left(\frac{\alpha_s(\mu)}{\alpha_s(m_Q)}\right)^{
\frac{2}{b}}\, (\bar Q\gamma_\mu\gamma_5 q)_\mu\, ;
$$
the subscript here indicates the normalization point.
The corresponding calculation is even simpler than that of the
anomalous dimension of ${\cal O}_G$ and will not be discussed here.
The interested reader is referred to Ref. \cite{hybrid} or to review
papers \cite{HQETRev} \footnote{The wording in these reviews is
somewhat different. You will read about the matching logarithms of
HQET for the axial current. Technically this is perfectly the same as
the anomalous dimension within our approach.}.
Correspondingly the complete asymptotic scaling law of
$f_P$ is $f_P \sim m_Q^{-1/2}(\alpha_s (m_Q))^{-2/b}$.

\subsection{$\mu$ dependence of the basic parameters of the heavy
quark theory. Measuring $\Lambda (\mu )$}.

The Lagrangian (\ref{N2}) summarizes the evolution from a high
normalization point down to $\mu$. Since all operators in this
Lagrangian are normalized at $\mu$ it is perfectly natural
that their matrix elements are also $\mu$ dependent.
In particular, the matrix elements of ${\cal O}_G$ and ${\cal O}_\pi$
denoted  by $\mu_G^2$ and $\mu_\pi^2$ in Lecture 3 depend on
$\mu$. Actually, $\mu_G^2$ depends on $\mu$ rather strongly,
through logarithms of $\mu$ -- this is explicitly demonstrated
by the fact that $({\cal O}_G)_{m_Q} = (\alpha_s (\mu ) /
\alpha_s (m_Q))^{-3/b} ({\cal O}_G)_{\mu }$.  In this case hardly
anybody would even think about
tending $\mu \rightarrow 0$.  As for the operator ${\cal O}_\pi$,
the situation here is trickier. As we saw,
it has no diagonal anomalous dimension, still some $\mu$
dependence appears through mixing with $\bar Q Q$, see \cite{BSUV}
for
details.

Let us discuss now $\bar\Lambda$, another basic parameter of the
heavy quark theory. The issue of its $\mu$ dependence was at the
epicenter of a heated debate recently.  By itself  $\bar\Lambda$
never appears in  ${\cal L}_{\rm heavy}$; moreover  the quark mass
$m_Q$
appears in the $1/m_Q$ expanded effective Lagrangian (\ref{HQLm})
only through
$1/m_Q$ corrections. Therefore, in the limit $m_Q\rightarrow\infty$
(which is often identified with HQET) it is quite tempting to say that
$\bar\Lambda = M_{H_Q} -m_Q$ is a universal constant.
For a few years it was taken for granted that such a constant
exists. Within the framework of our approach based on the
Wilsonean treatment of full QCD it is perfectly clear that this is not
the case. The quark mass in Eq. (\ref{N2}) explicitly depends on
$\mu$
resulting in a $\mu$ dependence of $\bar\Lambda$.

Since the issue is of importance let us rephrase this
statement as follows. Since quarks are permanently confined the
notion of the heavy quark mass becomes ambiguous.
To eliminate this
ambiguity one must explicitly specify the procedure of measuring
``the heavy quark mass''. The definition through the effective
Lagrangian (\ref{N2}) is consistent. Other definitions are certainly
conceivable; {\em any} consistent procedure will necessarily
involve a cut-off parameter $\mu$ , and then $\bar{\Lambda} (\mu
) = M_Q-
m_Q(\mu )$ \footnote{In the literature you can find assertions that
an ``absolute" heavy quark mass, or the so-called pole mass, can be
defined
and can be shown to be a universal number independent of any
cut-offs. These assertions are false. The notion of the pole mass exists
only to a given {\em finite} order of perturbation theory. No
consistent
definition of the pole mass can be given already at the level of the
leading nonperturbative corrections ${\cal O}(1/m_Q)$. The notion of
the pole mass is absolutely foreign to the approach I present here,
therefore, I do not want to go into details, see \cite{Pole}. I will only
say that it
assumes it is possible to separate perturbative contributions
from nonperturbative (?!) in contradiction with our approach
which separates soft contributions from hard.}.

The effective Lagrangian (\ref{N2}) is not something you directly
measure, neither is $m_Q$. Defining $\bar\Lambda$ or $m_Q$ is
equivalent to saying
how they are measured, how the parameters in the effective
Lagrangian are related to measurable quantities. To this end one
can use any suitable prediction of the heavy quark theory, in
particular, Voloshin's sum rule (\ref{I1e}). To avoid inessential
technicalities I will discuss the issue in the framework of the toy
model of Sect. 2.1. All results can be immediately extended to the
real QCD.  Equation (\ref{I1e}) gives a nice definition of
$\bar\Lambda$ in terms of a measurable quantity,  the average
value of $E_0^{phys} - E$ where $E$ is the energy of the $\phi$
quantum. The problem is that in Sect. 2.1 we discussed the question
switching off all hard gluons,
so that the above average value looked like a $\mu$ independent
number. To see where the $\mu$ dependence comes from we
must include  hard gluon corrections.

If the gluon field is treated only as a soft medium
the spectrum of the decay $H_Q\rightarrow X_q + \phi $
looks roughly as on Fig. 9.
The shoulder to the left of the elastic peak arises due to
production of the excited states. It is important
that in this approximation the spectrum rapidly (exponentially)
decreases outside the end point domain, so that the entire region
of $E$ from zero up to $E_0^{phys} - $ several units $\times
\Lambda_{\rm QCD}$ remains unpopulated. The average
$$
\int_0^{E_0^{phys}} dE \, (E_0^{phys} - E)\,
\frac{1}{\Gamma_0}
\frac{d\Gamma}{dE}
$$
is then a well-defined number independent of $m_Q$ or any cut-offs.

The situation drastically changes once we include  hard gluon
emission. In calculating radiative gluon correction we can disregard,
in the leading approximation,
nonperturbative effects, like the difference between $m_Q$
and $M_{H_Q}$ or the  motion of the initial quark inside ${H_Q}$.
Thus we deal
with the  decay of the free quark $Q$ at rest into $q +\phi +$
gluon.  The virtual gluon contribution merely
renormalizes the constant $h$  in the analysis  presented above
and is irrelevant.

The effects from real gluon emission are most
simply calculated in the
Coulomb gauge, where only the graph shown in Fig. 13 contributes.
A straightforward computation yields \cite{BSUV,Volopt} to leading
order in
$\vec v\,^2$
\begin{equation}
\frac{d\Gamma^{(1)}}{dE}=\Gamma_0\frac{8\alpha_s}{9\pi}
\frac{E^3}{E_0m_Q^2}\frac{1}{E_0-E}\, .
\label{rad}
\end{equation}

Hard gluon  emission obviously contributes to the spectrum in
the
entire interval $0<E<E_0^{phys}$
creating a long ``radiative" tail to the left of the end point domain.
(note that in this calculation
one can put $E_0^{phys}$ to $E_0$). In the first order calculation
$\alpha_s$ does not
run, of
course.
Its scale dependence shows up only in  the  two-loop
calculation; it is quite evident, however, that
it is $\alpha_s (E-E_0)$ that enters. Therefore, strictly speaking,
one cannot apply Eq. (\ref{rad}) too close to $E_0$. Even
leaving aside the blowing up of $\alpha_s (E-E_0)$, there exists
another
reason not to use Eq. (\ref{rad}) in the vicinity of $E_0$: if
$E$ is close to
$E_0$, the emitted gluon is soft; such gluons are to be treated
as belonging to the soft gluon medium in order to avoid double
counting.

The separation between soft and hard gluons
is achieved by explicitly
introducing a normalization point $\mu$. The value of $\mu$ should
be large enough to justify a small value for $\alpha_s(\mu)$.
On the other hand
we would like to choose $\mu$ as small as possible. We then  draw a
line: to the left of $E_0-\mu$ the gluon is considered
to be hard, to the right  soft. At $E<E_0 -\mu$ the experimentally
measured spectrum must follow the one-loop formula (\ref{rad}),
see Fig. 16.

Let us return now to Voloshin's  sum rule, i.e.   the first moment
of $ E_0^{phys} -E$, with radiative corrections  included.  A
qualitative
sketch of how
$d\Gamma/dE$ looks now is presented in Fig. 16. Because of the tail
to the left of the end point domain we can not define $\bar\Lambda$
as the
 value of $E_0^{phys} - E$ averaged over the entire range of the
$\phi$ energy, $0<E<E_0^{phys}$. The integral would be proportional
to $\alpha_s m_Q$ because of the domain of small $E$. Besides, this
would contradict the physical meaning of what we want to define.
By evolving the effective Lagrangian down to $\mu$ we include
all gluons harder than $\mu$ in $m_Q$, thus excluding them from
$\bar\Lambda$.  Thus, we must accept that
\begin{equation}
\bar{\Lambda} (\mu ) = \int_{E_0^{phys}-\mu}^{E_0^{phys}}\,
\frac{2}{v_0^2}\,\frac{1}{\Gamma_0}
\frac{d\Gamma}{dE} (E_0^{phys}-E)\, dE .
\label{lambdaphys}
\end{equation}
Since the explicit form of the tail to the left of the end point domain
is
known (for small $\alpha_s (\mu )/\pi$ the physical spectrum
is supposed to tend to the  perturbative result)  the $\mu$
dependence of
$\bar\Lambda$ becomes obvious,
\begin{equation}
\delta\overline{\Lambda} =\delta\mu
\frac{16}{9}\frac{\alpha_s(\mu
)}{\pi}
{}.
\label{della}
\end{equation}

Equation (\ref{lambdaphys})   provides us with one
possible
physical
definition of $\bar{\Lambda} (\mu)$
(among others) relating this quantity to an integral over
a physically  measurable spectral density.
The
pole-mass based
definition, being applied to our example, would involve three
steps:
(i) Take the radiative perturbative
tail to the left of the shoulder and extrapolate it all the way to
the point $E=E_0$; (ii) subtract the result from the measured
spectrum; (iii) integrate the difference over $dE$
with the weight function $(E-E_0^{phys})$. The elastic peak
drops out and the remaining integral is equal to $\Gamma_0
(\vec v\,^2/2)\bar{\Lambda}$. It is quite clear that this
procedure
cannot be
carried out consistently -- there exists no unambiguous way to
extrapolate the perturbative tail too close to $E_0^{phys}\,$,
the end point of the spectrum.
Our procedure, with the normalization point $\mu$
introduced explicitly, is free from this ambiguity.

In practice, the $\mu$ dependence of $\bar{\Lambda} (\mu )$
may
turn out to
be rather weak.
This is the case if the spectral density is such as shown in Fig. 16,
where the contribution of the first excitations (lying within
$\sim \Lambda_{\rm QCD}$ from $E_0^{phys}$) is numerically much
larger
than the
radiative tail representing  high
excitations. It is quite clear that if the physical spectral density
resembles that of Fig. 16 and $\mu$ = several units
$\times \Lambda_{\rm QCD}$,
the running $\bar{\Lambda} (\mu)$ is rather insensitive to the
particular choice
of
$\mu$.

It remains to be added that a similar  definition of $\bar\Lambda
(\mu )$ works in  real QCD. Here it may be  defined through an
integral over the spectrum in the decay $B\rightarrow X_c l\nu$
measured in the domain where the recoil of the hadronic system is
small, $|\vec q|\ll m_c$, i.e. in the SV limit.

\subsection{Hard gluons and the line shape}

Actually we have already started considering this question in the
previous section  where the radiative correction to the spectrum in
the decay $H_Q\rightarrow X_q + \phi$ was found in the SV limit.
In this case the impact of the hard gluons is mild -- they provide a
long but squeezed  tail outside the end point domain which could be
evaluated in the leading (one-loop) approximation. The reason why
there are no violent distortions of the spectrum is simple:
the $q$ quark produced is slow, and slow quarks do not like to emit
hard gluons. If the final quark was fast it would produce gluons
like crazy through bremsstrahlung, and the impact of such
bremsstrahlung on the line shape would be much more drastic.
As a matter of fact, if $m_q \rightarrow 0$ one can not limit oneself
to
any finite number of gluons -- an infinite sequence of the so called
Sudakov (or double-log) corrections must be summed over.

By definition the Sudakov corrections are those in which each power
of $\alpha_s $ is accompanied by two powers of logarithm
$\ln m_Q/(m_Q -2E)$. When one approaches the end point domain
  the logarithm inevitably becomes large, and overcompensates
the smallness of the gauge coupling constant $\alpha_s$. So, the
more gluons  emitted the higher  the probability. The phenomenon
is classical in nature and has a transparent physical interpretation.
Indeed, in the initial state $H_Q$ the color field in the light cloud
corresponds to a static source. The final quark produced is very
fast. The stationary state of the color field corresponding to a
fast-moving color charge is strongly different from that of the
stationary charge. Therefore, the excess of the color filed is just
shaken off in the form of the multiple emission of gluons. If you
forbid to emit a large number of gluons and insist that the final state
is just ``one quark" (this would correspond to the two-body decay
kinematics and the delta-function-like narrow spectrum) then the
probability of such an improbable event is terribly
suppressed. This explains why for the massless final quarks the
narrow peak in the end point domain obtained in Sect. 4.2 will be
drastically distorted, and a well-developed tail to the left
of the end point domain will appear.

The theory of the Sudakov corrections constitutes a noticeable part of
the perturbative QCD, and here, of course, I have no possibility even
to scratch the surface. I will give just a few hints referring the
interested reader to the original papers and textbooks \cite{PQCD}.

The
first order probability of emission of the massless gluon in the
$b\rightarrow s\gamma$ decay
 is
\begin{equation}
\frac{1}{\Gamma}\frac{d^2\Gamma_{g}}{d\omega d\vartheta^2} =
\frac{2\alpha_s}{3\pi\omega(1-\cos{\vartheta})}
\label{firsto}
\end{equation}
where $\omega$ is the  gluon momentum and $\vartheta$ is its
angle relative
to
the momentum of the $q$ quark. In this expression it is assumed
that
$\omega\ll
m_Q$. As a matter of fact it is perfectly legitimate to make this
assumption since the double logarithm comes only from this domain
of integration. The $\phi$  energy in the presence of a  gluon in the
final state
is given by
\begin{equation}
E = \frac{m_b^2-2\omega (m_b-\omega) (1-\cos{\theta})}
{2m_Q-2\omega(1-\cos{\theta})} \simeq
\frac{m_Q}{2}-
\frac{k_\perp^2}{4\omega}\;  ,
\label{kinem}
\end{equation}
$$
k_\perp \approx \omega\vartheta \, .
$$
One starts from
computing the  (first order) probability $w(E)$ for the
gluon to be emitted with such momentum that the $\phi$ quantum
gets
energy below given $E$. This probability is obtained by integrating
the
distribution (\ref{firsto}) with
the constraint that $ (m_Q/2) -(k_\perp^2 /4\omega )$ is less than
the given $E$,
\begin{equation}
w(E)=
\int d\omega\, d\vartheta^2\;
\frac{1}{\Gamma}\frac{d^2\Gamma_{g}}{d\omega\,d\vartheta^2} \:
\theta \left(\frac{k_\perp^2}{4\omega}-\left(\frac{m_Q}{2}-
E\right)\right) =
\frac{2\alpha_s}{3\pi} \ln^2\frac{m_Q}{m_Q-2E} \; .
\label{w}
\end{equation}
I integrated over $\vartheta^2$ first; the upper limit of integration is
of order one, the lower limit is determined from the $\theta$
function
in Eq. (\ref{w}).  Then we can carry out the $\omega$ integration.
The upper limit is $m_Q/2$ while the lower limit is seen from the
same $\theta$ function,
$$
\omega \lsim \frac{m_Q}{2} - E \,  .
$$

The function $w(E)$ has the meaning  of the probability of  emission
of a sufficiently hard
gluon lowering  the $\phi$ energy below $E$. The all-order
summation of
double logs  amounts then to  merely exponentiating  this
probability \cite{PQCD},
\begin{equation}
S(E)={\rm e}^{-w(E)}\, .
\label{exp}
\end{equation}
The spectrum then takes the form
\begin{equation}
\frac{1}{\Gamma}\frac{d\Gamma}{dE}= -\frac{dS}{dE}\, .
\label{spectrS}
\end{equation}
We see that as $E$ approaches the end point, $E$ close to $m_Q/2$,
the spectrum gets suppressed, in full accord with our expectations,
since the presence of the very hard $\phi$ does not allow, purely
kinematically,
the gluon shower to develop and the color field to restructure itself.
Notice that the Sudakov corrections merely redistribute the
probability, since the full integral over the spectrum remains
unchanged.
They pump events out from the end point domain to lower values of
$E$.

The double log approximation {\em per se} does not allow us to
determine the
scale of $\alpha_s$ in the Sudakov exponent $S(E)$. For practical
purposes the scale setting  is of course very important
since $\exp (-w(E))$ is a steep function. The question goes far
beyond the scope of this lecture. Some partial answers can be found
in Refs. \cite{scse,smilga}, see also \cite{motion4}; suffice it to
mention here
that $\alpha_s$ in Eq. (\ref{w}) turns out to be \cite{smilga}
$\alpha_s (\sqrt{(m_Q-2E)m_Q})$.   Another point deserving
stressing  is that with the
classical Sudakov  formula  one can not travel over the energy axis
too
close to the end point $E=m_Q/2$ (even after the scale setting).
Indeed, if $E > (m_Q/2) -\mu$
the gluons emitted become too soft; such gluons constitute the soft
gluon medium and have nothing to do with the perturbative
calculation; they have to be referred to the primordial distribution
function. Equation (\ref{w}) is applicable provided that
$$
\bar\Lambda\ll m_Q- 2E\ll m_Q\, .
$$
If we come closer to the end point domain the classical Sudakov
factor must be modified by cutting off and discarding the
contribution of the soft gluons. This idea gained recognition only
recently; it is obviously premature to further immerse into this topic
for the time being.

If the effect of the hard (perturbative) gluon emission is known
the full physical spectrum is obtained by convoluting the
perturbative one with the primordial distribution function,
\begin{equation}
\frac{d\Gamma (E)}{dE}\; = \;\theta(E)\, \int dy F(y)
\frac{d\Gamma^{pert}_Q(E-(\bar \Lambda /2) y)}{dE}\;\;.
\label{convolution}
\end{equation}
Integration over $y$ runs from $-\infty$ to 1 (more exactly, the
lower limit of
integration is $y_0=-m_Q/\bar\Lambda$ but this difference can be
ignored). One should keep in mind that
$d\Gamma^{pert}_Q/dE$ is nonvanishing only in the interval
$(0, m_Q/2)$. The convolution formula above is legitimate  only as
long as
one does
not apply it to the
very low energy part, $E\sim \bar\Lambda$; further details are
presented in Ref.  \cite{motion4}.

%********
{\bf ACKNOWLEDGMENTS:} \hspace{.4em} I am grateful to Prof. D.
Soper and Prof. K. Mahanthappa for their kind invitation to deliver
these lectures at TASI-95. The first two lectures were prepared
during my visit to Nagoya University in November and December
1994. The visit became  possible due to Research Fellowship
provided by JSPS. It is my pleasure to thank Prof. S. Sawada, Prof. T.
Sanda, Prof. K. Yamawaki and other members of the Theory Group of
the Nagoya University for kind hospitality. Countless fruitful
discussions on all topics touched upon in these lectures with N.
Uraltsev, A. Vainshtein and M. Voloshin are gratefully acknowledged.
This work was supported in part by DOE under the grant
number DE-FG02-94ER40823.

%********
\newpage
\section{Figure Captions}

Fig.1. The forward scattering amplitude $Q\rightarrow Q$. The
dashed
line denotes the $\phi$ quantum. The solid line connecting two
vertices is the $q$ quark Green function in the background gluon
field.
The thick solid lines describe the $Q$ quarks
in the background field.

\vspace{0.5cm}

Fig. 2. The transition operator relevant to the total semileptonic
width
of the heavy mesons. The dashed lines denote the leptons, $l$ and
$\nu$. Other  notations are the same as on Fig. 1.

\vspace{0.5cm}

Fig. 3. The two-point function (\ref{diftr}). The wavy line is the
external
scalar or pseudoscalar current; the quark propagator is in the
background field.

\vspace{0.5cm}

Fig. 4. The two-point function of two scalar currents in the scalar QCD.

\vspace{0.5cm}

Fig. 5. The four-point function appearing in Eq. (\ref{difsc}).
The dashed line denotes the current $Q^\dagger G_{\alpha\beta}Q$.

\vspace{0.5cm}

Fig. 6. The three-point function relevant to the proof of the
Isgur-Wise formula.

\vspace{0.5cm}

Fig. 7. The ``photon" spectrum in the decay $H_Q \rightarrow
X_q\phi$. The final quark is assumed to be massless. The thick line
represents the delta-function spectrum of the free quark
approximation. The solid line is a sketch of the actual hadronic
spectrum in the end point domain (possible radiation of hard gluons
is neglected).

\vspace{0.5cm}

Fig. 8.  Evolution of the spectrum of Fig. 7 as the mass of the final
quark increases (the schematic plot refers to $m_q\sim m_Q /2$.
The effects of the hard gluon bremstrahhlung are not included.

\vspace{0.5cm}

Fig. 9. The photon spectrum in the SV limit, ${\vec v}^2\ll 1$.
The dashed line shows the would be spectrum of the free quark
decay.

\vspace{0.5cm}

Fig. 10. Kinematically allowed domain in the transition $B\rightarrow
X_u l\nu$. The thick line indicates the populated phase space in the
free quark decay. The shaded area of width $\sim \bar\Lambda$
is the end point domain populated due to  the primordial motion of
the $b$ quark inside $B$.  The shaded square in the lower right
corner is the exclusive resonance domain where the inclusive
approach developed here is inapplicable.

\vspace{0.5cm}

Fig. 11. More or less realistic light-cone primordial distribution
function versus $x$ (borrowed from Ref. \cite{motion4}).

\vspace{0.5cm}

Fig. 12. Kinematically allowed domain in the decay $B\rightarrow
X_c l\nu$. Two large circles show the domains where  description
should be based on the light cone and temporal distribution
functions, respectively.

\vspace{0.5cm}

Fig. 13. Correction of the first order in $\alpha_s$ to the transition
operator of Fig. 1. Shown is the only graph contributing to the
imaginary part in the Coulomb gauge. The gluon is denoted by the
curly line.

\vspace{0.5cm}

Fig. 14. One-loop diagrams determining the coefficient $c_G$. The
wavy line denotes the gluon quanta, dashed line background glion
field.

\vspace{0.5cm}

Fig. 15. Integration contour in the $k_0$ plane.

\vspace{0.5cm}

Fig. 16.  A sketch of the photon spectrum in the SV limit with the
hard gluon radiation included.

\end{document}